\documentclass{aa}

\usepackage{booktabs}

\usepackage{graphicx}
\usepackage{txfonts}

\usepackage[separate-uncertainty = true,multi-part-units=single,range-units = single]{siunitx}
\usepackage{tabularx}
\usepackage{caption}
\usepackage{subcaption}
\usepackage{lscape}
\usepackage{afterpage}
\usepackage{xltabular}

\usepackage{natbib}
\usepackage[colorlinks, citecolor=blue, linkcolor=blue]{hyperref}
\makeatletter
\renewcommand*\aa@pageof{, page \thepage{} of \pageref*{LastPage}}
\makeatother

\newcommand{\AS}[3]{$#1\,^{+#2}_{-#3}$}

\newcommand{\serval}{\texttt{serval}}
\DeclareSIUnit\gauss{G}
\DeclareSIUnit\day{d}
\DeclareSIUnit\year{yr}
\DeclareSIUnit\msol{\mathnormal{M_\odot}}
\DeclareSIUnit\rsol{\mathnormal{R_\odot}}
\DeclareSIUnit\mearth{\mathnormal{M_\oplus}}
\DeclareSIUnit\arcsec{arcsec}
\let\oldsim\sim 
\renewcommand{\sim}{{\oldsim}}

\usepackage{etoolbox}           
\renewcommand{\bfseries}{\fontseries{b}\selectfont} 
\robustify\bfseries             
\newrobustcmd{\B}{\bfseries}    
\newcolumntype{Y}{>{\raggedleft\arraybackslash}X}

\AtBeginDocument{\mathcode`v=\varv}

\begin{document} 

   \title{The CARMENES search for exoplanets around M dwarfs}
   \subtitle{The impact of rotation and magnetic fields on the radial velocity jitter in cool~stars}
    
   \author{
   H.\,L.~Ruh\inst{1} \and
   M.~Zechmeister\inst{1} \and
   A.~Reiners\inst{1} \and
   E.~Nagel \inst{1} \and
   Y.~Shan \inst{2,1} \and
   C.~Cifuentes \inst{3} \and
   S.\,V.~Jeffers \inst{4} \and
   L.~Tal-Or \inst{5,1} \and
   V.\,J.\,S.~B\'ejar \inst{6,7} \and
   P.\,J.~Amado \inst{8} \and
   J.\,A.~Caballero \inst{3} \and
   A.~Quirrenbach \inst{9} \and
   I.~Ribas \inst{10,11}
   J.~Aceituno \inst{12}
   A.\,P.~Hatzes \inst{13}
   Th.~Henning \inst{14} \and
   A.~Kaminski \inst{9} \and 
   D.~Montes \inst{15} \and
   J.\,C.~Morales \inst{10,11} \and
   P.~Sch\"ofer \inst{8} \and
   A.~Schweitzer \inst{16} \and
   R.~Varas \inst{8}
          }
   \authorrunning{H.~L.~Ruh et al.}
   \titlerunning{Radial velocity jitter in cool stars}

   \institute{
              Institut f\"ur Astrophysik und Geophysik, Georg-August-Universit\"at,
              Friedrich-Hund-Platz 1, D-37077 G\"ottingen, Germany
        \and
              Centre for Earth Evolution and Dynamics, Department of Geosciences, Universitetet i Oslo, 
              Sem Sælands vei 2b, 0315 Oslo, Norway
        \and
              Centro de Astrobiolog\'ia (CSIC-INTA), Campus ESAC, 
              Camino Bajo del Castillo s/n, 28692 Villanueva de la Ca\~nada, Madrid, Spain 
        \and
              Max Planck Institute for Solar System Research, 
              Justus-von-Liebig-Weg 3, 37077 G\"ottingen, Germany
        \and
              Department of Physics, Ariel University, Ariel 40700, Israel
        \and
              Instituto de Astrof\'sica de Canarias (IAC), 38200 La Laguna, Tenerife, Spain
        \and
              Departamento de Astrof\'isica, Universidad de La Laguna, 38206
              La Laguna, Tenerife, Spain
        \and
              Instituto de Astrof\'isica de Andalucía (IAA-CSIC), 
              Glorieta de la Astronom\'ia s/n, 18008 Granada, Spain
        \and
              Landessternwarte, Zentrum für Astronomie der Universit\"at Heidelberg, 
              K\"onigstuhl 12, 69117 Heidelberg, Germany    
        \and
              Institut de Ci\`encies de l'Espai (ICE, CSIC),
              Campus UAB, c/~Can Magrans s/n, E-08193 Bellaterra, Barcelona, Spain
        \and
              Institut d'Estudis Espacials de Catalunya (IEEC),
              c/ Gran Capit\`a 2-4, E-08034 Barcelona, Spain 
        \and 
              Centro Astron\'onomico Hispano en Andaluc\'ia,
              Observatorio de Calar Alto, Sierra de los Filabres, 04550 G\'ergal, Almer\'ia, Spain
        \and
              Th\"uringer Landessternwarte Tautenburg, 
              Sternwarte 5, 07778 Tautenburg, Germany
        \and
              Max-Planck-Institut für Astronomie, 
              K\"onigstuhl 17, 69117 Heidelberg, Germany
        \and
              Departamento de F\'sica de la Tierra y Astrof\'sica and IPARCOS-UCM (Instituto de F\'sica de Part\'culas y del Cosmos de la UCM),
              Facultad de Ciencias F\'sicas, Universidad Complutense de Madrid, 28040, Madrid, Spain
        \and 
              Hamburger Sternwarte, Universit\"at Hamburg, 
              Gojenbergsweg 112, 21029 Hamburg, Germany
             }

   \date{Received 22 May 2024 / Accepted 30 September 2024} 

  \abstract
   {Radial velocity (RV) jitter represents an intrinsic limitation on the precision of Doppler searches for exoplanets that can originate from both instrumental and astrophysical sources.}
   {We aim to determine the RV jitter floor in M dwarfs and investigate the stellar properties that lead to RV jitter induced by stellar activity.}
   {We determined the RV jitter in \num{239} M dwarfs from the CARMENES survey that are predominantly of mid to late spectral type and solar metallicity. We also investigated the correlation between stellar rotation and magnetic fields with RV jitter.}
   {The median jitter in the CARMENES sample is \SI{3.1}{\meter\per\second}, and it is \SI{2.3}{\meter\per\second} for stars with an upper limit of \SI{2}{\kilo\meter\per\second} on their projected rotation velocities. We provide a relation between the stellar equatorial rotation velocity and RV jitter in M dwarfs based on a subsample of \num{129} well-characterized CARMENES stars. RV jitter induced by stellar rotation dominates for stars with equatorial rotation velocities greater than \SI{1}{\kilo\meter\per\second}. A jitter floor of \SI{2}{\meter\per\second} dominates in stars with equatorial rotation velocities below \SI{1}{\kilo\meter\per\second}. This jitter floor likely contains contributions from stellar jitter, instrumental jitter, and undetected companions. We study the impact of the average magnetic field and the distributions of magnetic filling factors on the RV jitter. We find a series of stars with excess RV jitter and distinctive distributions of magnetic filling factors. These stars are characterized by a dominant magnetic field component between \SIrange{2}{4}{\kilo\gauss}.}
   {An RV jitter floor can be distinguished from RV jitter induced by activity and rotation based on the stellar equatorial rotation velocity. RV jitter induced by activity and rotation primarily depends on the equatorial rotation velocity. This RV jitter is also related to the distribution of magnetic filling factors, and this emphasizes the role of the magnetic field in the generation of RV jitter.}
 

   \maketitle

%

\section{Introduction} \label{sec:introduction}
    High-precision radial velocity (RV) searches of extrasolar planets often target low-mass stars, where it is possible to detect low-mass rocky planets \citep[e.g.][]{zechmeister2019,ribas2023} because the planet-to-star mass ratio is favorable. However, the active nature of these stars often prevents a definitive detection of companions \citep{wright2005,kossakowski2022}. Variability in stellar spectra might produce an apparent shift, and thus, it represents a source of excess noise in RV time series. This excess noise is termed stellar RV jitter \citep{saar1998,saar1997}. 
    
    In addition to stellar RV jitter, photon noise and instrumental noise contribute to the observed variability. The photon noise along with the spectral information in observed spectra poses a fundamental limit to the precision of a single RV measurement, the photon-limited precision \citep{bouchy2001}. For the CARMENES\footnote{Calar Alto high-Resolution search for M dwarfs with Exoearths with Near-infrared and optical Échelle Spectrographs} \citep{ribas2023,quirrenbach2016} sample of M dwarfs, \citet{reiners2018} reported a photon-limited precision down to $\sim$\SI{1}{\meter \per \second} for observations with the CARMENES visual channel (VIS) spectrograph and gave a signal-to-noise ratio ~(S/N) of 150. The instrumental stability of high-resolution spectrographs typically lies at a few \si{\meter \per \second} \citep{fischer2016} and reaches sub-\si{\meter\per\second} level in extreme-precision RV (EPRV) \citep{pepe2021,seifahrt2022}. The long-term instrumental precision of CARMENES lies at a few meters per second \citep{ribas2023}.
    
    In the regime of high instrumental stability, stellar jitter dominates, and stellar activity has to be addressed in order to obtain robust planet detections. Stellar activity is causally connected to the presence of magnetic fields. The plethora of activity signatures such as spots, faculae, and plages can induce various forms of RV variability \citep{hudson1988, crass2021}. 
    These surface inhomogeneities, which rotate along with the stellar surface, can result in a quasi-periodic signal \citep{desort2007}. The RV amplitude of a single equatorial spot on an edge-on star is thus expected to vary linearly with the area covered by the spot and the stellar rotation velocity \citep{boisse2012,desort2007,saar1997}. 

    The correlation of RV jitter with activity indicators and rotation has been explored extensively in the past: \citet{saar1998} studied jitter as a function of the projected rotation velocity $v \sin i$, the stellar rotation period $P_{\mathrm{rot}}$, the $B-V$ color index, and $\log R'_{\mathrm{HK}}$, \citet{wright2005} in terms of $B-V$, $F_{\mathrm{Ca~II}}$, and $\mathrm{\Delta}M_V$, and \citet{isaacson2010} studied it in terms of the excess in emission in the Ca~{\sc ii} H \& K lines, \citet{luhn2020b} in terms of  $\log R'_{\mathrm{HK}}$, stellar luminosity $L$, and absolute G magnitude, $M_{\rm G}$, \citet{suarez2017} in terms of $\log R'_{\mathrm{HK}}$ and the Mount-Wilson $S$-index, \citet{moutou2017} in terms of $v \sin i$, the average magnetic field, the absolute longitudinal magnetic field, and the total chromospheric emission, and \cite{talor2018} in terms of $v \sin i$ and H$\alpha$ luminosity.

    The magnetic field and stellar activity are both causally connected to rotation \citep[e.g.][]{skumanich1972,2012AJ....143...93R,2018A&A...614A..76J}. Convection and rotation are fundamental to the generation of magnetic fields in stars \citep[e.g.][]{brandenburg2005}. As a consequence, the average magnetic field strength is a function of the Rossby number, $Ro$, which is a measure of the relative strength of convection compared to rotation. The Rossby number is often computed as $Ro=P_{\mathrm{rot}}/\tau$, where $\tau$ is the convective overturn time. In M dwarfs, the magnetic field increases with decreasing $Ro$ up to several \si{\kilo\gauss} and saturates in fast-rotating stars \citep{reiners2009,reiners2022}.

    The suppression of convective shifts by magnetic fields in active regions can induce RV variability because active regions rotate into and out of the visible part of the stellar disk and the granulation pattern evolves. In the case of the Sun, a convective blueshift of $\sim$\SI{100}{\meter\per\second} is observed in individual spectral lines \citep{liebing2021,ellwarth2023}. Models predict lower convective velocities in M dwarfs than in the Sun, with root mean square (RMS) vertical velocities still of several hundred \si{\meter\per\second} compared to several \si{\kilo\meter\per\second} in the Sun  \citep{beeck2013,ludwig2002}. The resulting net convective line shifts are smaller than in the Sun and may turn into a convective redshift \citep{kuester2003,baroch2020,liebing2021}.
    For the Sun, the remaining short-term variations due to the changing granulation pattern are $\sim$\SIrange{1}{2}{\meter\per\second} \citep{dravins2023, almoulla2022b}. The lifetime of granules in the Sun is about a few minutes and is expected to be shorter for M dwarfs \citep{beeck2013}. Because RV measurements in M dwarfs have typical integration times longer than a few minutes, short-term variations in the granulation pattern are expected to average out. Supergranulation, that is, large-scale granulation patterns, can persist up to a timescale of a few days in the Sun, at least, and they induce a similar RV amplitude as the short-term variations \citep{almoulla2022b}.

    The correction of activity-induced RV variations using magnetic field measurements has been demonstrated for the Sun \citep{haywood2022} and is a promising avenue for correcting the RVs of other stars \citep[e.g.][]{lienhard2023, donati2023}. Magnetic fields can be directly measured from the Zeeman broadening of spectral lines, but they are notoriously difficult to measure and require high-resolution spectra with a high S/N \citep{2012LRSP....9....1R}. \cite{reiners2022} measured the average magnetic field in \num{292} stars of the CARMENES sample, which enables an investigation of the relation between RV variability and magnetic fields in the CARMENES sample. The measurement of magnetic fields includes fitting the components of different magnetic field strengths. The contribution from the magnetic components is reflected in the distribution of the magnetic filling factors and contains information about the structure of the stellar  magnetic field \citep{shulyak2019}.
    
    We determine the RV jitter in the CARMENES sample of M dwarfs. We investigate the jitter-rotation relation and the transition from slow to fast rotation. We then study the connection of RV jitter with the stellar average magnetic field. Last, we investigate the distributions of magnetic filling factors and discuss a series of stars with increased jitter and distinctive distributions of the magnetic filling factors.

\section{Observations}
    \label{sec:obs}
    
        CARMENES is a high-precision RV instrument that encompasses two echelle spectrographs and is located at the Calar Alto observatory in Spain\footnote{\url{https://carmenes.caha.es/}}. The visual channel CARMENES VIS operates at \SIrange{520}{960}{\nano\meter} and has a spectral power of \num{94500}, and the the near-infrared CARMENES NIR operates from \SIrange{960}{1710}{\nano\meter} and has a spectral power of \num{80400} \citep{quirrenbach2016}. The two spectrographs are optimized for Doppler searches of exoplanets around M dwarfs. 
        
        The CARMENES survey of M dwarfs started in 2016 and observed 361 stars during its Guaranteed Time Observations (GTO) phase \citep[2016-2020;][]{ribas2023}. CARMENES mainly observed early- and mid-type M dwarfs of solar metallicity \citep{marfil2021,reiners2018,alonsofloriano2015}. 
        The RV time series from the CARMENES GTO observations are publicly available\footnote{\url{http://carmenes.cab.inta-csic.es/gto/jsp/dr1Public.jsp}}. In addition to the GTO targets described in \cite{ribas2023}, we considered a further \num{40} CARMENES targets and \num{1793} RV measurements that were mostly added after the GTO phase. The radial velocities for the CARMENES survey are uniformly computed with the {\serval} pipeline \citep{zechmeister2018}. We applied telluric absorption correction (TAC) to the CARMENES spectra \citep{nagel2023} and nightly zero point (NZP) correction to the \serval \ radial velocities \citep{ribas2023}. This study is confined to CARMENES VIS radial velocities.
        
\begin{figure}
          \center
          \includegraphics[width=1\linewidth]{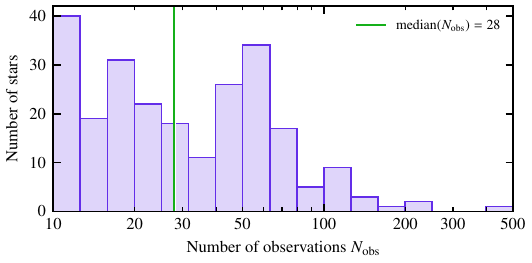}
          \caption{Distribution of observations of \num{239} stars in the CARMENES jitter sample. The sample contains stars with at least ten RV measurements. The survey aims to obtain at least \num{50} observations per star. Stars with fewer than \num{50} measurements have not reached completion or were discontinued due to strong activity-related RV scatter \citep{ribas2023}.}
          \label{fig:nrvs}
        \end{figure}

        In the following, we restrict our analysis to M dwarfs that have at least ten observations. This reduces the sample of CARMENES targets to \num{361} stars. Even though a minimum number of observations is required, the heterogeneous number of observations per star (Fig.~\ref{fig:nrvs}) and nonuniform sampling  limit our study because the RV variability is potentially underestimated in stars with a short observation timescale or an incomplete sampling of their activity signal. We removed \num{21} stars that are part of spectral binaries, spectral triples, or close resolved binaries (angular separation smaller than \SI{5}{\arcsec}), as indicated in Table~\ref{tab:excluded_stars}. We excluded \num{78} planet host stars and an additional \num{10} stars with planet candidates that are under investigation, as indicated in Table~G.1.
        Last, we removed \num{29} objects with significant 5$\sigma$ trends. The trends and their uncertainties were derived with a least-square fitting and an error rescaling. The best-fit value of the slope is given in Table~G.1. Overall, the sample then includes \num{239} unflagged objects. Removing binaries and stars with known planets allowed us to study the activity-related contribution of the RV jitter in isolation. We studied a subsample of \num{129} stars with available equatorial rotation velocities in detail (Sect.~\ref{sec:jitter_and_rotation}) and also studied a subsample of \SI{89} with additional average magnetic field measurements (Sect.~\ref{sec:jitter_and_bfield}).
         
        The stellar parameters considered in this work are mainly taken from the CARMENES input catalog \citep[Carmencita;][]{caballero2016}. We used rotation periods from \citet{2024A&A...684A...9S}. The stellar radii for the CARMENES survey sample were computed from effective temperatures and luminosities by \citet{schweitzer2019}. The average magnetic fields of CARMENES stars were determined by \cite{reiners2022}.  \\

\section{Method}
    \label{sec:method}
    We computed the RV jitter for the CARMENES sample. RV jitter refers to the excess noise in an RV time series \citep{baluev2009}. The uncertainty $\sigma_i$ in each RV data point $\mathrm{RV_i}$ is then given by the quadratic sum of its internal error $\sigma_{\mathrm{internal},i}$ and the excess noise $\sigma_{\mathrm{jitter}}$, that is, $\sigma_i = \sqrt{\sigma_{\mathrm{internal},i}^2 + \sigma_{\mathrm{jitter}}^2}$. When we model the probability density function of an RV measurement with a Gaussian distribution with a standard deviation $\sigma_i$ and constant expected value $\mu$, the log-likelihood function $\ln \mathcal{L}$ of an RV time series is given by  
    \begin{equation}
        \begin{aligned}
            \ln \mathcal{L} &= - \frac{1}{2} \sum \ln{\left(2 \pi \left[ \sigma_{\mathrm{internal},i}^2 + \sigma_{\mathrm{jitter}}^2\right]\right)} \\
            &\quad - \frac{1}{2} \sum{\frac{(\mathrm{RV}_i-\mu)^2}{\sigma_{\mathrm{internal},i}^2 + \sigma_{\mathrm{jitter}}^2}}.
        \end{aligned}
        \label{eq:jitter_lh}
    \end{equation}

     We computed $\sigma_{\mathrm{jitter}}$ (and the time-series mean $\mu$) by maximizing the log-likelihood function. To maximize the log-likelihood function in Eq.~(\ref{eq:jitter_lh}), we used the open-source script {\tt mlrms}\footnote{\url{https://github.com/mzechmeister/python/blob/master/wstat.py}.}. In Eq.~(\ref{eq:jitter_lh}), we made the intrinsic assumption of normally distributed data. Although signals such as sinusoids are non-Gaussian, their variance is captured by the jitter defined in Eq.~(\ref{eq:jitter_lh}). The jitter variance $\sigma_{\mathrm{jitter}}^2$ can be negative when the internal RV uncertainties are overestimated. For a negative jitter variance, the jitter values become imaginary. For formal reasons, we computed this imaginary jitter, but did not analyze them. To estimate the uncertainties of the derived jitter values, we applied the nonparametric bootstrapping method \citep{efron1979}. Therein, we drew 1000 samples with a replacement from each RV time series and computed the jitter in each sample. The standard deviation of the sample values represents our uncertainty estimate. 
    
\begin{figure}
      \center
      \includegraphics[width=1\linewidth]{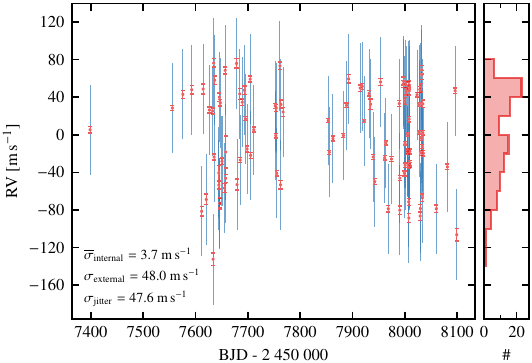}
      \caption{Radial velocity time series of EV~Lac (J22468+443). The radial velocities and internal uncertainties $\sigma_{\mathrm{internal},i}$ are displayed in red. The blue error bars indicate the quadratic sum of the internal uncertainties and the RV jitter $\sqrt{\sigma_{\mathrm{internal},i}^2 + \sigma_{\mathrm{jitter}}^2}$. The panel to the right shows a histogram of the RV measurements. The radial velocities are centered on their mean ($\mu=\SI{0}{\meter\per\second}$).}
      \label{fig:rvs_evlac}
    \end{figure}

    As an example, we show the CARMENES RV time series of EV~Lac in Fig.~\ref{fig:rvs_evlac}. The internal uncertainties cannot explain the scatter of the data points. We call this scatter the external uncertainty $\sigma_{\mathrm{external}}$ and computed it as weighted RMS with weights $w_i = \frac{1}{\sigma_{\mathrm{internal},i}}$. For EV~Lac, the RV jitter is close to the external uncertainty. EV~Lac is a known active star, so that the high RV jitter value for EV~Lac can be attributed to stellar activity \citep[e.g.][]{jeffers2022}, although the precise origin of the underlying signal is not relevant for the computation of the jitter.
    
\begin{figure}
      \center
      \includegraphics[width=1\linewidth]{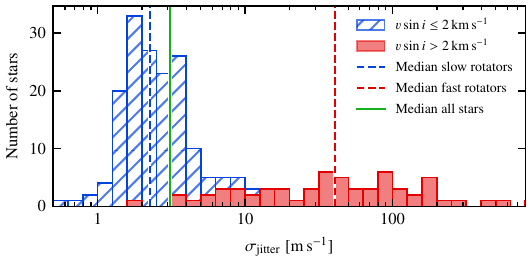}
      \caption{Histogram showing the jitter of \num{233} CARMENES stars. Slowly rotating stars ($v \sin i \leq \SI{2}{\kilo\meter\per\second}$; striped blue bars) have a median jitter of \SI{2.3}{\meter \per \second} (dashed blue line), and fast-rotating stars ($v \sin i > \SI{2}{\kilo\meter\per\second}$; filled red bars) have a median jitter of \SI{37}{\meter \per \second} (dashed red line). The median jitter for all stars is \SI{3.1}{\meter\per\second} (solid green line). Six stars with negative jitter variances are not shown.}
      \label{fig:histogram_jitter}
    \end{figure}

    A histogram of the resulting jitter values for slowly and fast-rotating stars as well as for the total sample is shown in Fig.~\ref{fig:histogram_jitter}. The distribution peaks at approximately \SI{2}{\meter\per\second}, and the median jitter lies at \SI{3.1}{\meter \per \second}. For \num{65} objects, we measure a jitter value lower than \SI{2}{\meter \per \second}. For \num{6} stars, we compute negative jitter variances indicating overestimated internal RV uncertainties. The star J19573$-$125 has the highest jitter value with \SI{1120 \pm 250}{\meter\per\second}. The median RV jitter of only the slowly rotating stars ($v \sin i \leq \SI{2}{\kilo\meter\per\second}$) lies at \SI{2.2}{\meter \per \second}. The lower median jitter in slow rotators signifies a correlation between jitter and rotation. We study this in detail in Sect.~\ref{sec:jitter_and_rotation}.

\section{Analysis and results}
    \subsection{Jitter-rotation relation}
        \label{sec:jitter_and_rotation}
        
        On the disk of a rotating star, each surface element has an RV that depends on its projected distance to the rotation axis and on the stellar rotation period. The light emitted by each surface element is Doppler-shifted, and we observe a broadened line profile in the light that is integrated over the projected stellar disk. When a spot blocks a part of the stellar disk, the line becomes asymmetric, and an apparent velocity shift is induced. The variability of these shifts can lead to the RV variability that we recognize as jitter.
    
\begin{figure}
          \center
          \includegraphics[width=1\linewidth]{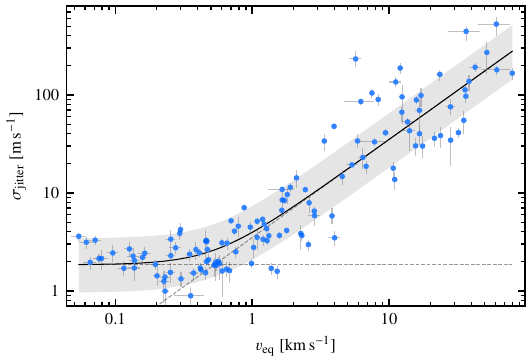}
          \caption{Jitter-rotation relation for \num{129} CARMENES M~dwarfs with known rotation periods. The RV jitter is fit as a function of the stellar rotation velocity $v_{\rm eq}$. The solid line displays the best fit, and the shaded region indicates the prediction interval. The jitter floor and the linear trend (dashed lines) correspond to parameters $\alpha$ and $\beta$ in Eq.~(\ref{eq:jitter_vrot}).}
          \label{fig:jitter_vrot_all}
        \end{figure}

        In Fig.~\ref{fig:jitter_vrot_all}, we show the relation between RV jitter and the equatorial rotation velocity $v_{\mathrm{eq}}$. We computed $v_{\mathrm{eq}}$ from the stellar radius $R_{\star}$ and from the rotation period $P_{\mathrm{rot}}$ as $v_{\mathrm{eq}} = 2 \pi\, R_{\star}/P_{\mathrm{rot}}$. The plot contains \num{129} CARMENES targets with  stellar radii from \citet{schweitzer2019} and rotation periods from \citet{2024A&A...684A...9S}, who estimated the uncertainties of the rotation period using Eq.~(4) of \citet{2024A&A...684A...9S}. Stars with known planets were excluded (see Sect.~\ref{sec:obs}). 
        The jitter approaches a limit for slow rotators and increases for fast rotators, in agreement with a trend in $v_{\mathrm{eq}}$. The observed increase in jitter with $v_{\mathrm{eq}}$ in the regime of fast rotation is expected for corotating star spots. A corotating star spot generates an RV signal with a semi-amplitude $K \propto v_{\mathrm{eq}}$.   
        
        In addition, the amplitude of the spot signal is a function of the stellar inclination and limb darkening as well as of the spot parameter latitude, spot size, and contrast. The spot parameters are expected to vary with time, and hence, the spot amplitude is expected to vary with time. The spot configuration of individual stars is generally unknown and might vary significantly. Consequently, the dependences on spot parameters and inclination can be expected to smear out the jitter-rotation relation. All these effects may explain the variance around the general jitter trend in Fig.~\ref{fig:jitter_vrot_all}. The variance might also be a function of the sampling, namely the observation time-span and number of observations with respect to the rotation period. The jitter floor seen for slow rotation is caused by other sources of intrinsic stellar noise and instrumental effects. We discuss the jitter floor in detail in Sect.~\ref{sec:jitter_floor}. 

        The relation between rotation and RV excess noise has traditionally been investigated using $v \sin i$ instead of $v_{\mathrm{eq}}$ \citep[e.g.][]{saar1997}. We used $v_{\mathrm{eq}}$ here because it can be computed for stars with very long rotation periods, allowing us to extend our analysis to stars that have only an upper limit on $v \sin i$. Additionally, the uncertainties on  $v \sin i$ are typically large compared to the formal uncertainties on $v_{\mathrm{eq}}$. It might be expected that $v \sin i$ has a stronger correlation to $\sigma_{\mathrm{jitter}}$ and thus represents a better parameterization. For \num{53} stars in our dataset with both $v \sin i$ and $v_{\mathrm{eq}}$ detected (i.e., $v \sin i>\SI{2}{\kilo\meter\per\second}$), the Pearson correlation coefficient between $v \sin i$  and $\sigma_{\mathrm{jitter}}$ is \num{0.62}, and it is \num{0.64} between $v_{\mathrm{eq}}$ and $\sigma_{\mathrm{jitter}}$. This shows that the correlation coefficients are similar for the fast-rotating stars.
            
\begin{figure}
          \center
          \includegraphics[width=1\linewidth]{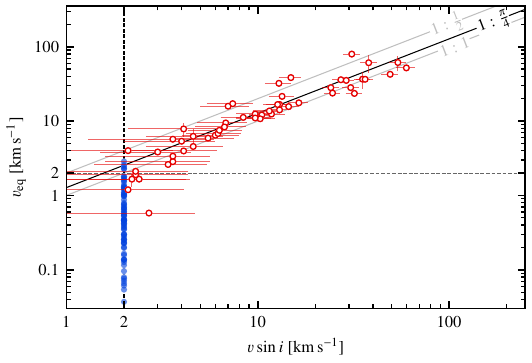}
          \caption{Equatorial rotation velocity vs. projected rotation velocity. $v_{\mathrm{eq}}$ is derived from the stellar radius and rotation period, and $v \sin i$ originates from fitting the rotational line broadening. The blue dots display stars with an upper limit on $v \sin i$, and red circles show stars with nominal values for $v \sin i$. The dashed lines indicate \SI{2}{\kilo\meter\per\second} in $v \sin i$ and $v_{\mathrm{eq}}$. The linear $1:\frac{4}{\pi}$ relation (solid black line) corresponds to the expected value of $\sin i$ for an isotropic distribution of the inclination angles (see Appendix~\ref{sec:sin_expectancy}). The relations $1:1$ and $1:\frac{1}{2}$ corresponding to $i=\SI{90}{\degree}$ and $i=\arcsin \frac{1}{2}= \SI{30}{\degree}$ are shown for comparison (solid gray lines).}
          \label{fig:vsini_vs_vrot}
        \end{figure}

        The correlation between $v \sin i$ and $v_{\mathrm{eq}}$ is shown in Fig.~\ref{fig:vsini_vs_vrot}. Following \citet{2024A&A...684A...9S}, we mostly adopted $v \sin i$ values from \citet{reiners2022} and assumed an uncertainty of \SI{10}{\percent} of their $v \sin i$ value, but a minimum uncertainty of \SI{2}{\kilo\meter\per\second}. An upper limit of \SI{2}{\kilo\meter\per\second} was used for $v \sin i$ values below this limit because $v \sin i$ values below this limit cannot be obtained reliably \citep{reiners2018}. One star, J06574+740, has inconsistent values of $v_{\mathrm{eq}}$ and $v \sin i$; the values deviate by more than $2\sigma$. The star shows an unusual beating pattern that might be related to binarity or to a complex spot pattern and stellar differential rotation \citep[see discussion by][]{2024A&A...684A...9S}. Out of \num{159} stars with an upper limit on $v \sin i$, \num{78} have values for $v_{\mathrm{eq}}$ that extend to $\sim$\SI{40}{\meter\per\second}. Above  $v \sin i=\SI{2}{\kilo\meter \per \second}$, the data follow the expected trend for an isotropic distribution of the inclination angles. 
    
        In order to describe the jitter-rotation relation, we assumed that the jitter variance ${\sigma}_{\mathrm{jitter}}^2$ is the quadratic sum of a constant jitter floor, parameterized by $\alpha$, and a jitter term linear in $v_{\mathrm{eq}}$, parameterized by $\beta$, such that
        \begin{equation}
            {\sigma}_{\mathrm{jitter}}^2 = \alpha^2 + \left(\beta \cdot  v_{\mathrm{eq}} \right)^2.
            \label{eq:jitter_vrot}
        \end{equation}
        We fit the relation in logarithmic space using a maximum likelihood estimation (MLE) with {\tt emcee} \citep{emcee}. The uncertainties of the jitter values were obtained by bootstrapping the RV data as described in Sect.~\ref{sec:method}. They do not necessarily represent the statistical uncertainties, but may rather result from the intrinsic variability of the stellar RV signal. We thus included the variance of the jitter as a nuisance parameter in our fit of the jitter-rotation relation (see Appendix~\ref{sec:fitting}). The fit relation is displayed in Fig.~\ref{fig:jitter_vrot_all}, and the fit parameters are listed in Table~\ref{table:fits}. A corner plot of the fitting parameters is shown in Fig.~\ref{fig:corner_jitter_rotation_relation}. The jitter floor lies at $\alpha=$ \AS{1.8}{0.2}{0.2}~\si{\meter\per\second}, and the best-fit value for $\beta$ is \AS{3.5}{0.3}{0.2}~$\cdot10^{-3}$. The jitter floor and the rotation term intersect (i.e., $\alpha= \beta \cdot v_{\mathrm{eq}}$) at  $v_{\mathrm{eq}} = \SI{0.5}{\kilo \meter\per\second}$.  The RMS around the jitter relation, which corresponds to the $\Delta$ in Eq.~(\ref{eq:jitter_vrot}) and Fig.~F.1, is \SI{2.7 \pm 0.2}{\deci\bel\squared}, corresponding to a factor of \num{1.9} of the upper and lower bounds of the prediction interval with respect to the expected jitter (Fig.~F.1).

        \begin{table}
        \centering
        \caption{Best-fit parameters of the relations between jitter, equatorial rotation velocity, and average magnetic field strength.}
        \begin{tabularx}{\columnwidth}{@{}X c c c@{}} 
            \toprule
            \toprule
            ${\sigma}_{\mathrm{jitter}}^2$   & $\alpha$ [m/s] & $\beta$ & $\gamma$ \\ 
            \midrule
            $\alpha^2 + \left(\beta \cdot  v_{\mathrm{eq}} \right)^2 $ &  \AS{1.8}{0.2}{0.2}  &  \AS{3.5}{0.3}{0.2}~$\cdot10^{-3}$ &  \\
            $\alpha^2 + \left(\beta \cdot \left(\langle B \rangle \cdot v_{\mathrm{eq}} \right)^{\gamma} \right)^2$ &  \AS{2.3}{0.3}{0.3} & \AS{3.9}{0.7}{0.6}~\si{\meter\per\second}  &  \AS{0.67}{0.05}{0.04}  \\
            \bottomrule
        \end{tabularx}
        \label{table:fits}
        \end{table}

        The relation provides a good fit to the data in general. A deviation from the relation is visible for fast-rotating stars with $v_{\mathrm{eq}}$ higher than \SI{3}{\kilo\meter\per\second}. J04198+425, an M8.5-type star, is the most prominent outlier. It has a jitter value of \SI{232 \pm 43}{\meter\per\second} at an equatorial rotation velocity of $v_{\mathrm{eq}}=\SI{5.70 \pm 0.57}{\kilo\meter\per\second}$. The high-jitter bump also includes the well-known active stars YZ~CMi and EV~Lac. These high-jitter stars are likely young and posses strong average magnetic fields \citep{contreras2024,reiners2022}. Additional information on the high-jitter stars (stars above the prediction interval displayed in Fig.~\ref{fig:jitter_vrot_all}) is provided in Appendix~\ref{sec:high-jitter stars}. 

    \subsection{Average magnetic fields}
        \label{sec:jitter_and_bfield}

        In this section, we focus on the connection between average magnetic fields and RV jitter. The average magnetic fields of \num{292} CARMENES stars were measured by \citet{reiners2022} by comparing polarized radiative transfer calculations with a selection of suitable observed spectral lines in the CARMENES coadded, high-S/N spectra. The lines were selected to represent a broad range of Landé g-factors, that is, sensitivities to the magnetic field, in order to differentiate from other broadening mechanisms. This allows for an accurate measurement of the magnetic field regardless of the rotational broadening. Upper limits were computed for stars with a weak magnetic field, which also have low rotation velocities ($v \sin i \leq \SI{2}{\kilo\meter\per\second}$).
        
\begin{figure}
          \center
          \includegraphics[width=1\linewidth]{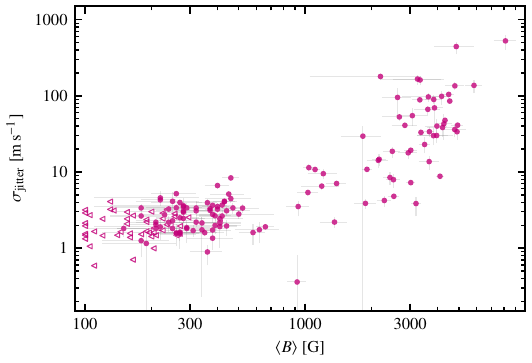}
          \caption{Radial velocity jitter vs. mean magnetic field. The upper limits on the average magnetic field are indicated by triangles.}
          \label{fig:jitter_bfield}
        \end{figure}

        A total of \num{178} stars of the \num{292} stars with a magnetic field measurement are part of our sample of stars with a sufficient number of observations and without indications of companions (see Sect.~\ref{sec:obs}). For an additional 6 stars, the literature reports mean magnetic field measurements \citep{cristofari2023,2021A&ARv..29....1K}, but we did not include these to sustain the homogeneity of the sample. Figure~\ref{fig:jitter_bfield} shows the RV jitter as a function of average magnetic field strength $\langle B \rangle$. For stars with an average magnetic field below \SI{1}{\kilo\gauss}, the RV jitter stays below \SI{10}{\meter\per\second}. Between \SIrange{1}{3}{\kilo\gauss}, the RV jitter takes values from a few \si{\meter\per\second} to tens of \si{\meter\per\second}. Above \SI{3}{\kilo \gauss}, the RV jitter assumes high values up to several hundred \si{\meter \per\second}. This regime of a high average magnetic field is also characterized by a weakened dependence of the magnetic field on rotation (see Fig.~\ref{sec:jitter_and_rotation}).
        
        We incorporated the average magnetic field into the jitter-rotation relation and tested different relations in order to minimize the scatter around the best-fitting model. A total of \num{89} stars have precise measurements of their average magnetic fields and equatorial rotation velocities. We allowed for additive and multiplicative terms of the average magnetic field and the equatorial rotation velocity to represent dependent and independent jitter generation. We also tested relations that only included one of the parameters, either the average magnetic field or the equatorial rotation velocity. The fit relations are listed in Table~\ref{tab:fit_results}. We applied the same fitting procedure as described in Sect.~\ref{sec:jitter_and_rotation} and Appendix~\ref{sec:fitting}.

        \begin{table*}
        
        \centering
        \caption{Stellar parameters, RV jitter, and magnetic field properties of the series of high-jitter stars with concentrated magnetic field distributions.}
        
        \begin{tabularx}{\linewidth}{@{}X X c S[table-format=.1(1)] S[table-format=.3(1),separate-uncertainty] S S[table-format=.2] @{}}            
            \toprule
            \toprule
            Karmn &                     Name &     SpT & $\sigma_{\mathrm{jitter}}$~[\si{\meter\per\second}] &      $v_{\mathrm{eq}}$~[\si{\kilo\meter\per\second}] &   $ B $~[\si{\gauss}] & $z$ \\
            \midrule
             J06574+740 &  2MASS J06572616+7405265 &  M4.0 V &    161(23) &  23.5(13) &   3330(830) & 14   \\
             J07446+035 &                   YZ CMi &  M4.5 V &   85.3(65) &  6.23(56) &   4540(150) & 7.7  \\
             J08536$-$034 &    LP 666-009, GJ~3517 &  M9.0 V &    135(16) &  11.2(11) &   4790(470) & 5.3  \\
             J18131+260 &                LP 390-16 &  M4.0 V &    104(11) &  7.50(29) &   4490(270) & 5.3  \\
             J22468+443 &                   EV Lac &  M3.5 V &   47.6(25) & 3.964(79) &   4320(110) & 9.0  \\
             J23548+385 &          RX J2354.8+3831 &  M4.0 V &   33.8(64) & 3.367(44) &   4900(130) & 9.3  \\
            \bottomrule
        \end{tabularx}
        \label{tab:weird}
        \end{table*}
        
        To compare the models, we employed the Akaike information criterion \citep[AIC;][]{akaike1974} and the Bayesian information criterion \citep[BIC;][]{schwarz1978}. According to \citet{kass1995}, an AIC or BIC difference smaller than \num{10} means that the comparison is indecisive, and an AIC or BIC difference smaller than \num{6} demonstrates only weak evidence in favor of the better-performing model. In our comparison, the three best-fitting relations produce a change in the BIC as well as AIC that is smaller than \num{6}. This only suggests a weak preference for the best-fitting relation. The best-fitting relation with the lowest BIC and AIC values includes the average magnetic field strength. In this relation, the RV jitter is a function of the product of the equatorial rotation velocity $v_{\mathrm{eq}}$ and the average magnetic field strength $\langle B \rangle$, such that
        \begin{equation}
            {\sigma}_{\mathrm{jitter}}^2 = \alpha^2 + \left(\beta \cdot \left(\frac{\langle B \rangle}{\SI{1}{\kilo\gauss}} \cdot \frac{v_{\mathrm{eq}}}{\SI{1}{\kilo\meter\per\second}} \right)^{\gamma}\right)^2.
            \label{eq:jitter_bvrot}
        \end{equation}

\begin{figure}
          \center
          \includegraphics[width=1\linewidth]{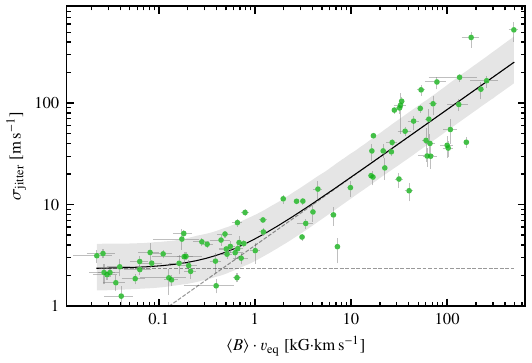}
          \caption{Similar to Fig.~\ref{fig:jitter_vrot_all}, but showing the RV jitter as a function of the product of the average magnetic field strength and stellar rotation velocity for \num{89} CARMENES stars. Equation~(\ref{eq:jitter_bvrot}) is the functional form of the model.}
          \label{fig:vrot_b_jitter}
        \end{figure}

        A single exponent for the product of $\langle B \rangle$ and $v_{\mathrm{eq}}$ instead of different exponents for both parameters is weakly preferred by our model comparison (see Table~G.3). The model performs better than the jitter-rotation relation we fit in Sect.~\ref{sec:jitter_and_rotation} (Eq.~(\ref{eq:jitter_vrot})) with a difference in BIC of \num{17.2}. The best-fit relation (Eq.~(\ref{eq:jitter_bvrot})) is shown in Fig.~\ref{fig:vrot_b_jitter}. The best-fit parameters are given in Table~G.3, and the respective corner plot is shown in Fig.~\ref{fig:corner_jitter_rotation_bfield_relation}. With $\alpha =$ \AS{2.3}{0.3}{0.3}~\si{\meter\per\second}, the $\alpha$-parameter takes a slightly higher value than $\alpha$ in the jitter-rotation relation without the average magnetic field (Eq.~(\ref{eq:jitter_vrot})). The slope parameter $\beta$ is not directly comparable between Eq.~(\ref{eq:jitter_vrot}) and Eq.~(\ref{eq:jitter_bvrot}) because the average magnetic field and the additional exponent $\gamma$ are included. The high-jitter stars lie closer to the expectation of the relation than in Fig.~\ref{fig:jitter_vrot_all}. J04198+425, which deviates most in Fig.~\ref{fig:jitter_vrot_all}, is not included because no measurement of the average magnetic field is available. 
        
    \subsection{Magnetic field components}
        \label{sec:jitter_outliers}
        
        We previously showed in Sect.~\ref{sec:jitter_and_rotation} that the observed RV jitter is related to the equatorial rotation velocity in fast rotators, and we showed that the RV jitter appears to be sensitive to the average magnetic field in Sect.~\ref{sec:jitter_and_bfield}. In this section, we investigate the impact of the distribution of the magnetic field filling factors.
        
        The average magnetic field is the sum of the magnetic field components $B_k$, weighted by their filling factors $f_k$, $\langle B \rangle = \sum_k f_k B_k$. The filling factors are derived from line profiles, which represent the sum of profiles emitted at different magnetic strengths and hence with different Zeeman broadening. Simplistically, the filling factors might be thought of as the relative area of the stellar disk that is penetrated by a magnetic field of a given magnetic field strength. The magnetic filling factors $f_k$ in the sum stated above are degenerate. In other words, stars with different magnetic filling factor distributions can have the same average magnetic field strength. We investigated the filling factor distributions of the stars in our sample based on the distributions fit by \citet{reiners2022}\footnote{The component distributions of all stars with average magnetic field measurements are available in their supplement (\url{http://carmenes.cab.inta-csic.es/gto/jsp/reinersetal2022.jsp}).}. 
        
\begin{figure}
          \center
          \includegraphics[width=1\linewidth]{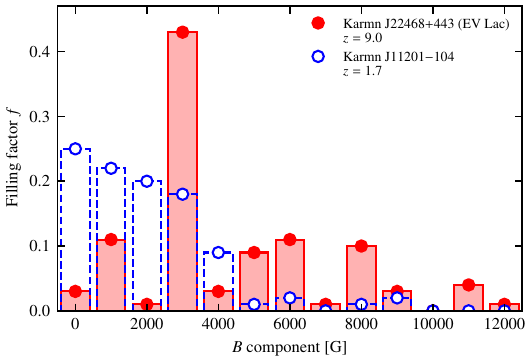}
          \caption{Distribution of magnetic filling factors for EV~Lac (solid red bars) and J11201$-$104 (dashed blue bars). The distribution index $z$ indicates the concentration of the distribution of filling factors (see Eq.~(\ref{eq:z-index})).}
          \label{fig:fprofiles}
        \end{figure}

        The filling factor distributions show different patterns, similar to the patterns described by \cite{shulyak2019}: Stars with small average magnetic field strengths have strong zero-field magnetic components. Stars with strong average magnetic field strengths display smooth distributions or concentrated distributions of the magnetic field components. In Fig.~\ref{fig:fprofiles}, we compare the filling factor distributions of EV~Lac with those of J11201$-$104, which shows a broad distribution of magnetic field components ranging from \SIrange{0}{4}{\kilo\gauss}. Its zero component, that is, the contribution of regions with a magnetic field strength below \SI{1}{\kilo\gauss}, has a filling factor of 0.25. In contrast, the filling factors in EV~Lac follow a narrow distribution that peaks at \SI{3}{\kilo \gauss} and has a maximum filling factor of \num{0.43}. The zero component is weak with a filling factor of \num{0.03}. With $v_{\mathrm{eq}}=\SI{4.55 \pm 0.04}{\meter\per\second}$, J11201$-$104 has a similar rotation velocity as EV~Lac. The average magnetic field strength is smaller in J11201$-$104 with \SI{2170 \pm 160}{\gauss}.
        
        To characterize the shape of the filling factor distribution, we computed the significance of the largest peak other than $\langle B_0 \rangle = \SI{0}{\gauss}$ in relation to all other field strengths. That is, we quantified to which extent the star carries predominantly one particular field strength in contrast to a broad distribution of field strengths. For this purpose, we computed the mean and the standard deviation of the filling factors $f_k$ excluding the peak component. We defined a distribution index
            \begin{equation}
            z = \frac{f_{\mathrm{peak}} - \mathrm{mean} f_{k,k \neq k_{\mathrm{peak}}}}{\mathrm{std} f_{k, k \neq k_{\mathrm{peak}}}},\label{eq:z-index}
        \end{equation}
        where $f_{\mathrm{peak}} = \mathrm{max} f_{k,k\neq0}$.
        The computed value serves as an indicator for the shape of the distribution: A high $z$ value corresponds to a narrower distribution that is more strongly peaked, whereas a low $z$ value corresponds to a more uniform distribution of the filling factors. When a negative value was obtained because a large zero component dominates the mean, we set the index to zero.
                
\begin{figure}
          \center
          \includegraphics[width=1\linewidth]{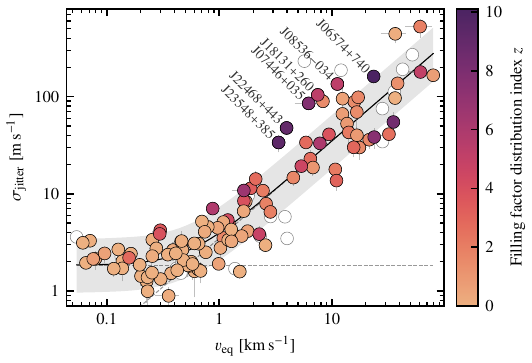}
          \caption{As Fig.~\ref{fig:jitter_vrot_all}, but color-coded with the filling factor distributions index and only for stars with measured average magnetic field. The labels mark the series of high-jitter stars with concentrated magnetic field distributions. The empty circles display stars without measurements of the average magnetic field.}
          \label{fig:bfield_structure}
        \end{figure}

        Figure~\ref{fig:bfield_structure} is a variation of our plot of the jitter-rotation relation, where the index of the filling factor distribution is now shown in the color-mapping. Stars with rotation velocities below \SI{1}{\kilo\meter\per\second} have index values around zero, which is expected because their magnetic fields are weak. A series of stars between \SIrange{4}{10}{\kilo\meter\per\second} display high $z$-index values $\geq 5$ as well as high jitter. These stars have similar narrow distributions with peaks between \SIrange{2}{4}{\kilo\gauss}. Their maximum filling factors lie between \numrange{0.35}{0.55}. J06574+740 is a possible extension of the series as it also shows a high $z$ value, but a jitter value closer to the prediction by the jitter-rotation relation. The high-jitter stars with high $z$ values are listed in Table~\ref{tab:weird}. These stars are mid-M dwarfs, except for J08536$-$034, which is of spectral type M9V. All of these high-jitter stars are RV-loud by the definition of \citet{talor2018} since their $v \sin i$ values are higher than \SI{2}{\kilo\meter\per\second} and their RV jitter values exceed \SI{10}{\meter\per\second}. The stars have H$\alpha$ pseudo-equivalent widths $\mathrm{pEW(H\alpha)} \leq \SI{-0.5}{\angstrom}$ and can be considered as $\mathrm{H\alpha}$-active following \citet{2018A&A...614A..76J}.
        
        The series of stars with excess jitter and a high $z$ value also have a strong average magnetic field strength of \SIrange{3.3}{4.9}{\kilo\gauss} and are located at the upper end of the magnetic fields measured for CARMENES targets (Fig.~\ref{fig:jitter_bfield}).
        Additionally, four of the stars, J07446+035, J18131+260, J22468+443, and J23548+385, produce magnetic fields  slightly more efficiently, as indicated by their Rossby numbers. Figure~F.4 shows the average magnetic field as a function of the Rossby number. We include the relations fit by \citet[][see their Fig.~9]{2024A&A...684A...9S}. J07446+035, J18131+260, J22468+443, and J23548+385 lie significantly above the relation at an $Ro$ of around \num{0.3}. J06574+740 and J08536$-$034 have lower Rossby numbers and lie close to the relation. 
        
        Figure~\ref{fig:bfield_structure} also displays stars with low $z$ values and high jitter, and vice versa. Individual high-jitter stars are discussed in Appendix~\ref{sec:high-jitter stars}. The most prominent high-$z$ stars with jitter below the prediction by the jitter-rotation relation are J05062+046 and J19511+464. Both stars occupy a similar region in the jitter-rotation plot around an equatorial rotation velocity of \SI{30}{\kilo\meter\per\second} and a jitter value around \SI{50}{\meter\per\second}. The two stars are likely very young and are members of young stellar kinematic groups \citep{contreras2024}. Both stars have relatively few observations, \num{12} and \num{13}.
        
        Stars with high jitter and low $z$ and vice versa might in general be explained by the following causes: (i) The variability of the RV signal of stars might not be completely sampled; many active stars only have about ten RV data points. Additionally, unaccounted-for trends might be present in these stars. (ii) The proposed index does not capture the full complexity of the magnetic field structure. (iii) The magnetic field measurements have large uncertainties, and the measurement reliability might decrease toward the regime of extreme rotation because the rotational line broadening increases.

\section{Discussion}
        With our version of the jitter-rotation relation, we provide an extension of previous work toward slowly rotating stars on the basis of a large sample of well-characterized M dwarfs. In this section, we discuss the jitter floor,  including possible contributions from instrumental effects, the implications of the jitter-rotation relation, and the connection of excess jitter to the magnetic field.
        
    \subsection{Jitter floor}
        \label{sec:jitter_floor}
        Following previous analysis, we characterize RV jitter as excess noise, where we count any residual instrumental or stellar signal as noise. This has the advantage of a unified description of noise sources without major assumptions on the nature of the signal. With the jitter, we describe the regimes of slow stellar rotation, where the instrumental or stellar noise floor dominates, and fast rotation, where the rotation-related activity signal dominates. 
        
        In this framework, we exclude time series with fewer than ten data points and stars with known planets or trends. The minimum number of required data points was chosen arbitrarily. Consequently, the jitter can be over- or underestimated when only a part of the signal is sampled. This especially affects active stars, for which observations have often been deferred after their active nature was established. After completion, CARMENES will attain 50 observations for its survey stars \citep{ribas2023}, which will allow for a more robust analysis in the future. 
        For our analysis, we used data that were corrected for telluric absorption and NZPs. The telluric absorption correction \citep{nagel2023} has a large effect mainly on fast rotators, where the RV information content is relatively low because of rotational line broadening \citep[e.g.][]{reiners2010}. Hence, we reduced systematic telluric and instrumental effects to the best possible extent.
        
        For stellar equatorial rotation velocities below $\sim\SI{1}{\kilo\meter\per\second}$, we show that a jitter floor of approximately \SI{2}{\meter\per\second} exists. The jitter floor might be caused by a baseline of stellar activity that is not related to rotation or instrumental instability. The signals of hidden companions also contribute to the observed jitter floor. \citet{ribas2023} estimated that a fraction of \AS{0.89}{0.08}{0.11} of M~stars host planets when the planets in a mass bin of $\SI{1}{\mearth}\leq M_{\mathrm{pl}} \sin i \leq\SI{10}{\mearth}$ and periods of \SIrange{1}{1000}{\day} are considered. Based on their assumptions (planet mass, orbital period distribution, and circular orbits), the median semi-amplitude of model planets in this parameter space lies at approximately \SI{1.3}{\meter\per\second} (see Appendix~\ref{sec:hidden_planets}).
        
        Instrumental instability might concern the accuracy of the wavelength solution, which is taken at the beginning of a night, or the accuracy of the drift measurements during the night \citep{2018schaefer}. Nightly zero point shifts affect the accuracy of the wavelength solution and are identified by measuring the average Doppler shifts of RV-quiet stars in each night \citep{talor2019,ribas2023}. Uncertainties on the NZPs are propagated to the corrected internal RV uncertainties used in this work. Therein, the NZP errors are typically $\lesssim \SI{1}{\meter\per\second}$. Since NZP uncertainties are included in the internal RV uncertainties, nightly shifts should not contribute significantly to the jitter. 

        \citet{ribas2023} reported a median reweighted RMS of \SI{3.9}{\meter\per\second} for the NZP-corrected CARMENES RVs and a jitter contribution of \SI{3.5}{\meter\per\second}. Computing the jitter from spectra that were corrected for telluric absorption and excluding known planet hosts, we find a slightly lower median jitter of \SI{3.1}{\meter\per\second}. 
        
        In the following, we compare the jitter floor in CARMENES to the jitter floor in other RV surveys. \citet{fischer2016} discussed the RV precision of the major Doppler search programs at the time. Their comparison was mostly based on the internal precision. The two largest programs in their comparison, executed with HARPS\footnote{High Accuracy Radial velocity Planet Searcher} \citep{2003Msngr.114...20M} and HIRES\footnote{High Resolution Echelle Spectrometer} \citep{1994SPIE.2198..362V}, provide an internal precision of \SI{0.8}{\meter\per\second} and \SI{1.5}{\meter\per\second}, where the precision is scaled to a hypothetical observation with an S/N of \num{200} for a \SI{3}{\kilo\meter\per\second} spectral bin at \SI{550}{\nano\meter}. The median internal precision of all targets of the CARMENES survey lies at \SI{1.6}{\meter\per\second}, with a median S/N of \num{95}. When we scale the precision to the hypothetical observation with $\mathrm{S/N} =\num{200}$, the CARMENES precision lies at \SI{0.8}{\meter\per\second}, which is similar to the precision of HARPS.
        \citet{fischer2016} also provided RMS radial velocities of programs running on the various instruments. The California Planet Search appears to reach an RMS floor at \SIrange{2}{4}{\meter \per \second}. The RMS in HARPS reaches \SIrange{1}{2}{\meter \per \second}. Therefore, HARPS demonstrates the highest precision of the instruments compared by \citet{fischer2016}. However, a general comparison is only of limited use because internal as well as external precision also depends on the spectral type, among other parameters. 
        \citet{wright2005} studied RV jitter in stars with F to M spectral types observed with HIRES. In inactive M dwarfs, they reported a jitter value of approximately \SI{5}{\meter\per\second}, which exceeds the median jitter in our sample by \SI{2}{\meter\per\second}. This might be related to the better instrumental stability of CARMENES, but it might also be due to differences in the samples.
        For the HARPS sample of M dwarfs, \citet{bonfils2013} reported an internal uncertainty of down to \SI{0.7}{\meter\per\second} and an external uncertainty between \SIrange{1.5}{2.0}{\meter\per\second} as a lower threshold for the brightest targets. The excess noise should consequently lie at \SIrange{1.3}{1.8}{\meter\per\second}. \citet{perger2017} reported an RV jitter of \SI{2.3}{\meter\per\second} in their HARPS-N\footnote{High Accuracy Radial velocity Planet Searcher for the Northern hemisphere} \citep{2012SPIE.8446E..1VC} survey of early-type M dwarfs. \citet{tuomi2019} combined RV data of 426 nearby M dwarfs from HARPS, HIRES, UVES\footnote{Ultraviolet and Visual Echelle Spectrograph} \citep{2000SPIE.4008..534D}, and PFS\footnote{Subaru Prime Focus Spectrograph} \citep{2016SPIE.9908E..1MT}. They obtained a median jitter of \SI{2.3}{\meter\per\second}.
        In a sample of \num{200} inactive mid- to late-type M dwarfs, \citet{pass2023a} reported internal uncertainties between \SIrange{5}{18}{\meter\per\second} for their most precise observations, and they measure a weighted standard deviation of \SI{20}{\meter\per\second} and \SI{21}{\meter\per\second} for the two instruments TRES\footnote{Tillinghast Reflector Echelle Spectrograph} \citep{2007RMxAC..28..129S} and CHIRON\footnote{CTIO HIgh ResolutiON spectrograph} \citep{2013PASP..125.1336T} that they employed. When we consider the internal uncertainty of \SI{5}{\meter\per\second}, the jitter floor in their survey lies at \SIrange{9}{11}{\meter\per\second}, which suggests that their results are limited by instrumental noise.
        
        The literature values as well as the jitter floor of \AS{1.8}{0.2}{0.2}~\si{\meter\per\second} determined in this work is overall compatible with a jitter floor of \SIrange{1}{3}{\meter\per\second} in M dwarfs. The question remains which contribution to the jitter floor dominates: instrumental, stellar jitter, or signals by unaccounted-for companions. We can attempt to estimate the contributions of instrumental noise, granulation jitter, and unaccounted companions. \citet{bauer2020} investigated the instrumental stability of CARMENES in detail. They reported an instrumental jitter in CARMENES VIS of \SI{1.2}{\meter\per\second}. To approximate the granulation component of RV jitter, we used the scaling law by \citet[see their Eq.~9]{luhn2020a}, which approximates the granulation-induced jitter as a function of stellar mass, luminosity, and effective temperature. For the stellar parameters of slowly rotating stars in our sample ($v\sin i \leq \SI{2}{\kilo\meter \per\second}$), we computed a granulation RV jitter up to \SI{0.3}{\meter\per\second}. We approximated the noise contribution caused by unaccounted-for companions with \SI{0.9}{\meter\per\second} (see Appendix~\ref{sec:hidden_planets}). The combined jitter from the quadratic sum of these contributions is \SI{1.5}{\meter\per\second}, which agrees with the observed jitter floor established above. 

        The jitter floor does not represent a hard limit because planet searches have succeeded to detect planets with RV semi-amplitudes below \SI{1}{\meter\per\second}. \citet{faria2022} claimed the detection of a planet with an RV semi-amplitude of \SI{39 \pm 7}{\cm\per\second} around the slowly rotating mid-type M dwarf Proxima Centauri with ESPRESSO\footnote{Echelle SPectrograph for Rocky Exoplanets and Stable Spectroscopic Observations} \citep{pepe2021}. In addition to the two planets Proxima Centauri b and d, the RV signal was modeled using a Gaussian process with an RV amplitude of \AS{1.7}{0.7}{0.5}~\si{\meter\per\second} for the contribution of stellar RV signals. In the analysis, additional jitter terms of \SIrange{0.07}{0.40}{\meter\per\second} were retained. In this work, any stellar or other signal is modeled as jitter. When we consider the RV amplitude of the Gaussian process, the jitter floor of this work is thus consistent with the results of \citet{faria2022}.

    \subsection{Jitter-rotation relation}
        The linear part of our relation of RV jitter with the rotation velocity at fast rotation agrees in general with previous findings \citep{talor2018,saar1998}. \citet{saar1998} were the first to fit a power law in $v \sin i$, finding a close-to-linear relation for F-, G-, and K-type stars. Subsequently, the RV jitter caused by the flux effect of cool star spots was studied many times also in numerical simulations \citep{saar1997,hatzes1999,hatzes2002,desort2007,boisse2012,dumusque2014,herrero2016}. These simulations demonstrated a linear (or almost linear) relation of a spot-induced RV amplitude with $v \sin i$, which agrees with our observations in fast-rotating stars. They also found a linear relation with the spot filling factor \citep[e.g.][]{saar1997}. As stated by \citet{boisse2012}, a quantitative comparison between simulated and observed jitter relations is not a desirable approach because the models are constrained by simplifications and assumptions regarding the model parameters.
        
        A linear scaling of the RV jitter with the projected rotation velocity is expected for the flux effect as long as the spot parameters stay constant as a function of rotation velocity. The observed trend (Fig.~\ref{fig:vrot_bfield}) thus implies that spot parameters do not vary strongly with the rotation velocity in this regime. The activity-rotation relations predict increasing activity with increasing rotation velocity and saturation in fast-rotating stars. Most of the stars in the part that is dominated by activity and rotation in our plot in Fig.~\ref{fig:jitter_vrot_all} belong to the stars in the saturated regime (see Fig.~F.3).

        Our fit of the linear part of the relation in Eq.~(\ref{eq:jitter_vrot}) is mainly based on stars of type M4 and M5 because the CARMENES survey has a bias to early- and mid-type M dwarfs and mostly slow rotation in the early-type M dwarfs \citep{marfil2021}. Furthermore, CARMENES contains only a few young stars \citep{contreras2024}. As a result of this bias, we cannot exclude deviations from the observed relation of jitter with rotation for early- and late-type M dwarfs and young M dwarfs. The jitter-rotation relation might change with spectral subtype and evolutionary state of a star.

        The jitter-rotation relation allows us to estimate the expected RV jitter of M dwarfs given their rotation velocity. It might therefore be helpful in estimating detection limits of RV surveys as well as in preparing RV observations, such as the follow-up observations for mass determinations of transiting planets. Furthermore, an excess jitter compared to the jitter-rotation relation can be used as a quick indication of a planetary companion, in particular, for series with only a few measurements.

        We show the predictive aspect of the relation for six stars from the CARMENES-TESS program (Table~\ref{tab:carmn_tess}). CARMENES-TESS operated under the CARMENES GTO and Legacy surveys and followed-up planet candidates from the Transiting Exoplanet Survey Satellite mission \citep[TESS;][]{2015JATIS...1a4003R}. The CARMENES-TESS stars were excluded from our jitter sample because they host planets. In Table~\ref{tab:carmn_tess}, we predicted the RV jitter $\sigma^{\mathrm{pred}}_{\mathrm{jitter}}$ based on the rotation period and stellar radius adopted in the respective CARMENES-TESS publication and using Eq.~(\ref{eq:jitter_vrot}). The raw TAC RVs of TOI-442 jitter with $\sigma^{\mathrm{meas}}_{\mathrm{jitter}}=\SI{12.7}{\meter\per\second}$, which significantly exceeds our prediction from stars without planets. TOI-1201 and TOI-1801 are close to the upper edge of the prediction interval. No quick conclusions about the presence of planets can be made for the remaining three stars. When the planetary signals are removed, the remaining jitter should also be consistent with the prediction.
        As a proxy for the remaining jitter $\sigma^{\mathrm{lit}}_{\mathrm{jitter+GP}}$, we used the jitter given in the CARMENES-TESS publications. Since the stellar activity is typically taken out mostly by Gaussian processes \citep[GP;][]{rajpaul2015}, we quadratically added the GP amplitude back. These jitters $\sigma^{\mathrm{lit}}_{\mathrm{jitter+GP}}$ agree well with our predictions. Thus, our prediction can serve as a simple planet indicator that is even applicable for a few data points.

        \begin{table}
            \caption{Comparison of predicted and observed jitter for six CARMENES-TESS stars.}
            \begin{tabularx}{\linewidth}{@{}r@{\ \ }S[table-format=1.2(2),separate-uncertainty] l S[table-format=2.2(2),separate-uncertainty] l X@{}}  
                \toprule
                \toprule
                TOI &  $v_{\mathrm{eq}}$ &  $\sigma^{\mathrm{pred}}_{\mathrm{jitter}}$ & $\sigma^{\mathrm{meas}}_{\mathrm{jitter}}$ &  $\sigma^{\mathrm{lit}}_{\mathrm{jitter+GP}}$ & Ref \\
                 &  [\si{\kilo\meter\per\second}] &  [\si{\meter\per\second}] & [\si{\meter\per\second}] & [\si{\meter\per\second}] &  \\
                \midrule
                442  & 0.84(8)  & \AS{3.4}{3.0}{1.6}  &  12.7(11)  &  \AS{5.20}{0.77}{0.58}                  &  Drei20 \\[0.05cm]
                562  & 0.22(1)  & \AS{2.0}{1.7}{0.9}  &  1.58(32)  &  \AS{2.8}{1.0}{0.8}                     &  Luq19  \\[0.05cm]
                1201 & 1.22(12) & \AS{4.6}{4.0}{2.1}  &  7.22(90)  &  \AS{3.7}{1.7}{1.1}                     &  Kos21  \\[0.05cm]
                1801 & 1.72(12) & \AS{6.3}{5.4}{2.9}  &  11.68(48) &  \AS{6.49}{0.94}{0.87}                  &  Mal23  \\[0.05cm]
                4438 & 0.28(3)  & \AS{2.0}{1.8}{0.9}  &  2.88(57)  &  \AS{1.33}{0.71}{0.80}                  &  Gof24  \\[0.05cm]
                4599 & 0.59(1)  & \AS{2.7}{2.4}{1.3}  &  3.00(28)  &  \AS{2.79}{0.57}{0.50}                  &  Luq22  \\[0.05cm]
                \bottomrule
            \end{tabularx}
            \label{tab:carmn_tess}
            \textbf{References.} Drei20: \citet{2020A&A...644A.127D}, Luq19: \citet{2019A&A...628A..39L}, Kos21: \citet{kossakowski2021}, Mal23: \citet{2023A&A...680A..76M},  Gof24: \citet{2024A&A...685A.147G}, Luq22: \citet{2022A&A...664A.199L}. 
        \end{table}

    \subsection{Role of the magnetic field}
     
        The jitter-rotation relation describes the trend of RV jitter with rotation. Aiming to reduce the scatter in this relation, we included the average magnetic field strength in the relation. The improved relation is a function of the product of the average magnetic field with the equatorial rotation velocity. 
        
        The increased jitter floor in stars with strong magnetic fields indicates the suppression of convective shifts by the presence of magnetic fields. We observed a difference of  \SI{0.5 \pm 0.4}{\meter\per\second} in the fitted jitter floor between the jitter-rotation relation (Eq.~(\ref{eq:jitter_vrot})) and the relation that additionally included the average magnetic field (Eq.~(\ref{eq:jitter_bvrot})). To further investigate this difference in the jitter floor, we compared the mean jitter of stars with and without strong magnetic fields in the slowest rotators of our sample ($v_\mathrm{eq}<\SI{0.5}{\kilo\meter\per\second}$). The mean jitter of stars with weak magnetic fields (only upper limits on the average magnetic field strength) is lower by \SI{0.5 \pm 0.3}{\meter\per\second}. This difference in the jitter floor could be caused by the suppression of convective shifts by strong magnetic fields. Variability in the magnetic field results in a variable suppression of convective shifts and could thus produce RV jitter regardless of rotational modulation and the flux effect.

        The average magnetic field saturates in fast-rotating stars \citet{reiners2022}. A saturation of the average magnetic field with the equatorial rotation velocity can also be seen (Fig.~F.3). The saturation in fast-rotating stars might correspond to saturation in the expression of the surface features, such as spots. In this case, we would expect that RV jitter increases with rotation velocity at a slower rate for stars in the saturated regime. A saturation would affect stars with a rotation velocity upward of approximately \SI{5}{\kilo\meter\per\second}. No saturation is clearly visible in our plot of the RV jitter against the equatorial rotation velocity (Fig.~\ref{fig:jitter_vrot_all}). However, the improved fit including the average magnetic field (Eq.~(\ref{eq:jitter_bvrot})) compared to the rotation-only fit (Eq.~(\ref{eq:jitter_vrot})) may result from a slight saturation of the RV jitter in fast-rotating stars. 

        A correlation between the magnetic field strength and RV jitter in M dwarfs was previously investigated by \citet{moutou2017}. \citet{moutou2017} proposed an activity merit function based on the total chromospheric emission, the absolute longitudinal magnetic field, $v \sin i$, and the average magnetic field. From the weights of the parameters of their merit function, the authors concluded that most of the jitter is explained by the longitudinal magnetic field and chromospheric emission. In contrast, we observe that jitter is primarily a function of rotation (Sect.~\ref{sec:jitter_and_rotation} and \ref{sec:jitter_and_bfield}). Their sample mostly consisted of slow rotators, many only with upper limits on $v \sin i$, such that the effect of rotation likely is less prevalent in their sample.
        
        A series of stars that showed excess jitter compared to the jitter-rotation relation also showed similar distributions of their magnetic field components, with a single dominating magnetic field component at \SIrange{2}{4}{\kilo\gauss}. The filling factors of the dominant component lie between \numrange{0.35}{0.55}. \citet{shulyak2019} previously noted the similarity of the magnetic filling factor distributions between YZ~CMi, EV~Lac, and J23548+385. It now appears that this type of distribution extends to the entire group of stars with excess jitter, and the excess jitter in these stars appears to be related to the shape of their magnetic filling factor distributions. We characterize the shape of magnetic filling factor distribution with the index $z$ defined in Eq.~(\ref{eq:z-index}). The stars of the excess-jitter group have an index value between $z=$ \numrange{7}{9}, which is significantly above the mean for all filling factor distributions of our sample at \num{1.6}.

        The high $z$ values in the excess-jitter stars is indicative of a correlation between the magnetic filling factor distribution and the RV jitter. The distribution does not reveal the spatial distribution of the magnetic field on the stellar surface. It is therefore unclear whether the dominant component is associated with specific features on the stellar surface, that is, spots or faculae, or is distributed rather homogeneously over the stellar surface.  
        
        Active stars exhibit a variety of magnetic filling factor distributions, and it has been suggested that different distributions are related to different dynamo states \citep{reiners2022,shulyak2019}. Although it is unclear why the stars show different distributions, the distributions appear to affect the RV jitter. RV jitter arises from corotating spatial inhomogeneities, that is, it requires a difference in the magnetic field or temperature. Therefore, magnetic field or temperature differences must exist in the stars with excess jitter, even though their magnetic field components are concentrated in the range of \SIrange{2}{4}{\kilo\gauss}. These excess-jitter stars also have low filling factors for magnetic field components below $B=\SI{2}{\kilo\gauss}$. This implies that the magnetic field or temperature difference that causes the RV jitter is between wide spread regions with strong magnetic fields, and not between small regions of spots and faculae against the backdrop of a quiet stellar surface.  
        
        Presumably, the RV signal of a spot on a star with an overall active surface changes compared to a spot on a star that is predominantly inactive because the stellar surface brightness strongly changes with magnetic field strength \citep{norris2023} and the spectrum is altered by the Zeeman effect. Simulations should consider the concentration of magnetic field components, such as observed in the excess-jitter stars, in order to better understand the influence of the magnetic field on the stellar RV signal. Another pathway is the detailed study of individual magnetically sensitive lines such as the strong KI lines in the near-infrared \citep{fuhrmeister2022,terrien2022}.
                
        We suspect that the increased RV jitter in the stars with concentrated magnetic field components is related to the line shape of magnetically sensitive lines. Each line in the spectrum carries RV information depending on the line shape, with sharp lines carrying more RV information than broad lines \citep{bouchy2001}. As a result of the Zeeman effect, magnetically sensitive lines split in the presence of magnetic fields, where the splitting is proportional to the magnetic field strength. The resulting Zeeman components of the line are therefore sharper for a narrow distribution of the magnetic field components and broader for a wide distribution of the magnetic field components. Consequently, the Zeeman components of a line carry more RV information in the case of a narrow distribution of the magnetic field components. Hence, we expect the magnetically sensitive lines to have greater RV information in stars with concentrated magnetic field components than in stars with a broad distribution of the magnetic field components. Greater RV information in the magnetically sensitive lines means that these lines contribute more to the total RV. We therefore expect magnetic field variability to cause larger RV jitter in stars with concentrated magnetic field components. 
        
\section{Summary}

    We measured the jitter in the RV time series of M~dwarfs from the CARMENES survey. The CARMENES VIS RV jitter correlates with the rotation velocity, which we computed from stellar rotation period and stellar radius. We fit a relation of the RV jitter as a function of rotation velocity,
    \begin{equation}
        {\sigma}_{\mathrm{jitter}}^2 = (1.8~\si{\meter\per\second})^2 + \left( 3.5 \cdot \num{e-3} \cdot v_{\mathrm{eq}} \right)^2.    
    \end{equation}
    The relation can be used to predict jitter, where the prediction has to be multiplied (divided) by \num{1.9} to obtain upper (lower) bounds.
    In stars rotating slower than \SI{1}{\kilo\meter\per\second}, the RV jitter is independent of stellar rotation. Thus, the remaining jitter must result from residual instrumental noise, hidden companions, and granulation. We conclude that the reduction of rotationally modulated variability is not sufficient to reach below the jitter floor of $\sim$\SI{2}{\meter\per\second} in our survey.

    The effect of the average magnetic field on the RV jitter was considered as an extension of the jitter-rotation relation. A relation of the RV jitter as function of the product of the average magnetic field and rotation velocity is preferred to the jitter-rotation relation without a magnetic field. The best-fitting relation with a magnetic field is given by

    \begin{multline}
        {\sigma}_{\mathrm{jitter}}^2 = (2.3~\si{\meter\per\second})^2 +
        \left(3.9~\si{\meter\per\second} \cdot \left(\frac{\langle B \rangle}{\SI{1}{\kilo\gauss}} \cdot \frac{v_{\mathrm{eq}}}{\SI{1}{\kilo\meter\per\second}}\right)^{0.67} \right)^2 .
    \end{multline}

    A series of stars with excess jitter shares a distinctive distribution of the magnetic filling factors, which is characterized by magnetic filling factors that are concentrated around \SIrange{2}{4}{\kilo\gauss}. We conclude that the concentration of magnetic filling factors affects the Zeeman-split line profiles and leads to the observed excess jitter. The identification of distinctive magnetic field components in stars with excess RV jitter suggests that activity-induced RV signals are associated with specific distributions of the magnetic field components.

\section*{Data availability}
    Table~\ref{table:sample} is available in electronic form at the CDS via anonymous ftp to cdsarc.u-strasbg.fr (130.79.128.5) or via \href{http://cdsweb.u-strasbg.fr/cgi-bin/qcat?J/A+A/}{http://cdsweb.u-strasbg.fr/cgi-bin/qcat?J/A+A/}.

\begin{acknowledgements}  
    This publication was based on observations collected under the CARMENES Legacy+ project.

    CARMENES is an instrument at the Centro Astron\'omico Hispano en Andaluc\'ia (CAHA) at Calar Alto (Almer\'{\i}a, Spain), operated jointly by the Junta de Andaluc\'ia and the Instituto de Astrof\'isica de Andaluc\'ia (CSIC).
    
    The authors wish to express their sincere thanks to all members of the Calar Alto staff for their expert support of the instrument and telescope operation.
    
    CARMENES was funded by the Max-Planck-Gesellschaft (MPG), 
    the Consejo Superior de Investigaciones Cient\'{\i}ficas (CSIC),
    the Ministerio de Econom\'ia y Competitividad (MINECO) and the European Regional Development Fund (ERDF) through projects FICTS-2011-02, ICTS-2017-07-CAHA-4, and CAHA16-CE-3978, 
    and the members of the CARMENES Consortium 
    (Max-Planck-Institut f\"ur Astronomie,
    Instituto de Astrof\'{\i}sica de Andaluc\'{\i}a,
    Landessternwarte K\"onigstuhl,
    Institut de Ci\`encies de l'Espai,
    Institut f\"ur Astrophysik G\"ottingen,
    Universidad Complutense de Madrid,
    Th\"uringer Landessternwarte Tautenburg,
    Instituto de Astrof\'{\i}sica de Canarias,
    Hamburger Sternwarte,
    Centro de Astrobiolog\'{\i}a and
    Centro Astron\'omico Hispano-Alem\'an), 
    with additional contributions by the MINECO, 
    the Deutsche Forschungsgemeinschaft (DFG) through the Major Research Instrumentation Programme and Research Unit FOR2544 ``Blue Planets around Red Stars'', 
    the Klaus Tschira Stiftung, 
    the states of Baden-W\"urttemberg and Niedersachsen, 
    and by the Junta de Andaluc\'{\i}a.
    
      We acknowledge financial support from the Agencia Estatal de Investigaci\'on (AEI/10.13039/501100011033) of the Ministerio de Ciencia e Innovaci\'on and the ERDF ``A way of making Europe'' through projects 
  PID2022-137241NB-C4[1:4],     
  PID2021-125627OB-C31,         
and the Centre of Excellence ``Severo Ochoa'' and ``Mar\'ia de Maeztu'' awards to the Instituto de Astrof\'isica de Canarias (CEX2019-000920-S), Instituto de Astrof\'isica de Andaluc\'ia (CEX2021-001131-S) and Institut de Ci\`encies de l'Espai (CEX2020-001058-M).
    
    This work was also funded by the Generalitat de Catalunya/CERCA programme, 
    the DFG priority program SPP 1992 ``Exploring the Diversity of Extrasolar Planets'' through grants JE 701/5-1 and RE 1664/20-1, and the Israel Science Foundation through grant 1404/22.

    We thank the anonymous reviewer for helpful comments.
    
\end{acknowledgements}

\bibliographystyle{aa} 
\bibliography{jitter} 

\begin{thebibliography}{188}
\expandafter\ifx\csname natexlab\endcsname\relax\def\natexlab#1{#1}\fi

\bibitem[{{Affer} {et~al.}(2019){Affer}, {Damasso}, {Micela}, {Poretti},
  {Scandariato}, {Maldonado}, {Lanza}, {Covino}, {Garrido Rubio}, {Gonz{\'a}lez
  Hern{\'a}ndez}, {Gratton}, {Leto}, {Maggio}, {Perger}, {Sozzetti},
  {Su{\'a}rez Mascare{\~n}o}, {Bonomo}, {Borsa}, {Claudi}, {Cosentino},
  {Desidera}, {Giacobbe}, {Molinari}, {Pedani}, {Pinamonti}, {Rebolo}, {Ribas},
  \& {Toledo-Padr{\'o}n}}]{2019A&A...622A.193A}
{Affer}, L., {Damasso}, M., {Micela}, G., {et~al.} 2019,
  \href{https://ui.adsabs.harvard.edu/abs/2019A&A...622A.193A}{\aap, 622, A193}

\bibitem[{{Afram} \& {Berdyugina}(2019)}]{afram2019}
{Afram}, N. \& {Berdyugina}, S.~V. 2019,
  \href{https://ui.adsabs.harvard.edu/abs/2019A&A...629A..83A}{\aap, 629, A83}

\bibitem[{{Akaike}(1974)}]{akaike1974}
{Akaike}, H. 1974,
  \href{https://ui.adsabs.harvard.edu/abs/1974ITAC...19..716A}{IEEE
  Transactions on Automatic Control, 19, 716}

\bibitem[{{Al Moulla} {et~al.}(2023){Al Moulla}, {Dumusque}, {Figueira}, {Lo
  Curto}, {Santos}, \& {Wildi}}]{almoulla2022b}
{Al Moulla}, K., {Dumusque}, X., {Figueira}, P., {et~al.} 2023,
  \href{https://ui.adsabs.harvard.edu/abs/2023A&A...669A..39A}{\aap, 669, A39}

\bibitem[{{Alonso-Floriano} {et~al.}(2015){Alonso-Floriano}, {Morales},
  {Caballero}, {Montes}, {Klutsch}, {Mundt}, {Cort{\'e}s-Contreras}, {Ribas},
  {Reiners}, {Amado}, {Quirrenbach}, \& {Jeffers}}]{alonsofloriano2015}
{Alonso-Floriano}, F.~J., {Morales}, J.~C., {Caballero}, J.~A., {et~al.} 2015,
  \href{https://ui.adsabs.harvard.edu/abs/2015A&A...577A.128A}{\aap, 577, A128}

\bibitem[{{Amado} {et~al.}(2021){Amado}, {Bauer}, {Rodr{\'\i}guez L{\'o}pez},
  {Rodr{\'\i}guez}, {Cardona Guill{\'e}n}, {Perger}, {Caballero},
  {L{\'o}pez-Gonz{\'a}lez}, {Mu{\~n}oz Rodr{\'\i}guez}, {Pozuelos},
  {S{\'a}nchez-Rivero}, {Schlecker}, {Quirrenbach}, {Ribas}, {Reiners},
  {Almenara}, {Astudillo-Defru}, {Azzaro}, {B{\'e}jar}, {Bohemann}, {Bonfils},
  {Bouchy}, {Cifuentes}, {Cort{\'e}s-Contreras}, {Delfosse}, {Dreizler},
  {Forveille}, {Hatzes}, {Henning}, {Jeffers}, {Kaminski}, {K{\"u}rster},
  {Lafarga}, {Lodieu}, {Lovis}, {Mayor}, {Montes}, {Morales}, {Morales},
  {Murgas}, {Ortiz}, {Pall{\'e}}, {Pepe}, {Perdelwitz}, {Pollaco}, {Santos},
  {Sch{\"o}fer}, {Schweitzer}, {S{\'e}gransan}, {Shan}, {Stock}, {Tal-Or},
  {Udry}, {Zapatero Osorio}, \& {Zechmeister}}]{2021A&A...650A.188A}
{Amado}, P.~J., {Bauer}, F.~F., {Rodr{\'\i}guez L{\'o}pez}, C., {et~al.} 2021,
  \href{https://ui.adsabs.harvard.edu/abs/2021A&A...650A.188A}{\aap, 650, A188}

\bibitem[{{Astudillo-Defru} {et~al.}(2017{\natexlab{a}}){Astudillo-Defru},
  {D{\'\i}az}, {Bonfils}, {Almenara}, {Delisle}, {Bouchy}, {Delfosse},
  {Forveille}, {Lovis}, {Mayor}, {Murgas}, {Pepe}, {Santos}, {S{\'e}gransan},
  {Udry}, \& {W{\"u}nsche}}]{astudillo2017a}
{Astudillo-Defru}, N., {D{\'\i}az}, R.~F., {Bonfils}, X., {et~al.}
  2017{\natexlab{a}},
  \href{https://ui.adsabs.harvard.edu/abs/2017A&A...605L..11A}{\aap, 605, L11}

\bibitem[{{Astudillo-Defru} {et~al.}(2017{\natexlab{b}}){Astudillo-Defru},
  {Forveille}, {Bonfils}, {S{\'e}gransan}, {Bouchy}, {Delfosse}, {Lovis},
  {Mayor}, {Murgas}, {Pepe}, {Santos}, {Udry}, \&
  {W{\"u}nsche}}]{astudillo2017b}
{Astudillo-Defru}, N., {Forveille}, T., {Bonfils}, X., {et~al.}
  2017{\natexlab{b}},
  \href{https://ui.adsabs.harvard.edu/abs/2017A&A...602A..88A}{\aap, 602, A88}

\bibitem[{{Baluev}(2009)}]{baluev2009}
{Baluev}, R.~V. 2009,
  \href{https://ui.adsabs.harvard.edu/abs/2009MNRAS.393..969B}{\mnras, 393,
  969}

\bibitem[{{Baran} {et~al.}(2011){Baran}, {Winiarski}, {Krzesi{\'n}ski},
  {Fox-Machado}, {Kawaler}, {Dr{\'o}{\.z}dz}, {Faltenbacher}, {Thompson}, \&
  {Reed}}]{baran2011}
{Baran}, A.~S., {Winiarski}, M., {Krzesi{\'n}ski}, J., {et~al.} 2011,
  \href{https://ui.adsabs.harvard.edu/abs/2011AcA....61...37B}{\actaa, 61, 37}

\bibitem[{{Baroch} {et~al.}(2021){Baroch}, {Morales}, {Ribas}, {B{\'e}jar},
  {Reffert}, {Cardona Guill{\'e}n}, {Reiners}, {Caballero}, {Quirrenbach},
  {Amado}, {Anglada-Escud{\'e}}, {Colom{\'e}}, {Cort{\'e}s-Contreras},
  {Dreizler}, {Galad{\'\i}-Enr{\'\i}quez}, {Hatzes}, {Jeffers}, {Henning},
  {Herrero}, {Kaminski}, {K{\"u}rster}, {Lafarga}, {Lodieu},
  {L{\'o}pez-Gonz{\'a}lez}, {Montes}, {Pall{\'e}}, {Perger}, {Pollacco},
  {Rodr{\'\i}guez-L{\'o}pez}, {Rodr{\'\i}guez}, {Rosich}, {Sch{\"o}fer},
  {Schweitzer}, {Shan}, {Tal-Or}, \& {Zechmeister}}]{2021A&A...653A..49B}
{Baroch}, D., {Morales}, J.~C., {Ribas}, I., {et~al.} 2021,
  \href{https://ui.adsabs.harvard.edu/abs/2021A&A...653A..49B}{\aap, 653, A49}

\bibitem[{{Baroch} {et~al.}(2020){Baroch}, {Morales}, {Ribas}, {Herrero},
  {Rosich}, {Perger}, {Anglada-Escud{\'e}}, {Reiners}, {Caballero},
  {Quirrenbach}, {Amado}, {Jeffers}, {Cifuentes}, {Passegger}, {Schweitzer},
  {Lafarga}, {Bauer}, {B{\'e}jar}, {Colom{\'e}}, {Cort{\'e}s-Contreras},
  {Dreizler}, {Galad{\'\i}-Enr{\'\i}quez}, {Hatzes}, {Henning}, {Kaminski},
  {K{\"u}rster}, {Montes}, {Rodr{\'\i}guez-L{\'o}pez}, \&
  {Zechmeister}}]{baroch2020}
{Baroch}, D., {Morales}, J.~C., {Ribas}, I., {et~al.} 2020,
  \href{https://ui.adsabs.harvard.edu/abs/2020A&A...641A..69B}{\aap, 641, A69}

\bibitem[{{Baroch} {et~al.}(2018){Baroch}, {Morales}, {Ribas}, {Tal-Or},
  {Zechmeister}, {Reiners}, {Caballero}, {Quirrenbach}, {Amado}, {Dreizler},
  {Lalitha}, {Jeffers}, {Lafarga}, {B{\'e}jar}, {Colom{\'e}},
  {Cort{\'e}s-Contreras}, {D{\'\i}ez-Alonso}, {Galad{\'\i}-Enr{\'\i}quez},
  {Guenther}, {Hagen}, {Henning}, {Herrero}, {K{\"u}rster}, {Montes}, {Nagel},
  {Passegger}, {Perger}, {Rosich}, {Schweitzer}, \& {Seifert}}]{baroch2018}
{Baroch}, D., {Morales}, J.~C., {Ribas}, I., {et~al.} 2018,
  \href{https://ui.adsabs.harvard.edu/abs/2018A&A...619A..32B}{\aap, 619, A32}

\bibitem[{{Bauer} {et~al.}(2020){Bauer}, {Zechmeister}, {Kaminski},
  {Rodr{\'\i}guez L{\'o}pez}, {Caballero}, {Azzaro}, {Stahl}, {Kossakowski},
  {Quirrenbach}, {Becerril Jarque}, {Rodr{\'\i}guez}, {Amado}, {Seifert},
  {Reiners}, {Sch{\"a}fer}, {Ribas}, {B{\'e}jar}, {Cort{\'e}s-Contreras},
  {Dreizler}, {Hatzes}, {Henning}, {Jeffers}, {K{\"u}rster}, {Lafarga},
  {Montes}, {Morales}, {Schmitt}, {Schweitzer}, \& {Solano}}]{bauer2020}
{Bauer}, F.~F., {Zechmeister}, M., {Kaminski}, A., {et~al.} 2020,
  \href{https://ui.adsabs.harvard.edu/abs/2020A&A...640A..50B}{\aap, 640, A50}

\bibitem[{{Beeck} {et~al.}(2013){Beeck}, {Cameron}, {Reiners}, \&
  {Sch{\"u}ssler}}]{beeck2013}
{Beeck}, B., {Cameron}, R.~H., {Reiners}, A., \& {Sch{\"u}ssler}, M. 2013,
  \href{https://ui.adsabs.harvard.edu/abs/2013A&A...558A..49B}{\aap, 558, A49}

\bibitem[{{Bicz} {et~al.}(2022){Bicz}, {Falewicz}, {Pietras}, {Siarkowski}, \&
  {Pre{\'s}}}]{biscz2022}
{Bicz}, K., {Falewicz}, R., {Pietras}, M., {Siarkowski}, M., \& {Pre{\'s}}, P.
  2022, \href{https://ui.adsabs.harvard.edu/abs/2022ApJ...935..102B}{\apj, 935,
  102}

\bibitem[{{Bluhm} {et~al.}(2020){Bluhm}, {Luque}, {Espinoza}, {Pall{\'e}},
  {Caballero}, {Dreizler}, {Livingston}, {Mathur}, {Quirrenbach}, {Stock}, {Van
  Eylen}, {Nowak}, {L{\'o}pez}, {Csizmadia}, {Zapatero Osorio}, {Sch{\"o}fer},
  {Lillo-Box}, {Oshagh}, {Gonz{\'a}lez-{\'A}lvarez}, {Amado}, {Barrado},
  {B{\'e}jar}, {Cale}, {Chaturvedi}, {Cifuentes}, {Cochran}, {Collins},
  {Collins}, {Cort{\'e}s-Contreras}, {D{\'\i}ez Alonso}, {El Mufti},
  {Ercolino}, {Fridlund}, {Gaidos}, {Garc{\'\i}a}, {Georgieva},
  {Gonz{\'a}lez-Cuesta}, {Guerra}, {Hatzes}, {Henning}, {Herrero}, {Hidalgo},
  {Isopi}, {Jeffers}, {Jenkins}, {Jensen}, {K{\'a}bath}, {Kaminski}, {Kemmer},
  {Korth}, {Kossakowski}, {K{\"u}rster}, {Lafarga}, {Mallia}, {Montes},
  {Morales}, {Morales-Calder{\'o}n}, {Murgas}, {Narita}, {Passegger}, {Pedraz},
  {Persson}, {Plavchan}, {Rauer}, {Redfield}, {Reffert}, {Reiners}, {Ribas},
  {Ricker}, {Rodr{\'\i}guez-L{\'o}pez}, {Santos}, {Seager}, {Schlecker},
  {Schweitzer}, {Shan}, {Soto}, {Subjak}, {Tal-Or}, {Trifonov}, {Vanaverbeke},
  {Vanderspek}, {Wittrock}, {Zechmeister}, \& {Zohrabi}}]{2020A&A...639A.132B}
{Bluhm}, P., {Luque}, R., {Espinoza}, N., {et~al.} 2020,
  \href{https://ui.adsabs.harvard.edu/abs/2020A&A...639A.132B}{\aap, 639, A132}

\bibitem[{{Boisse} {et~al.}(2012){Boisse}, {Bonfils}, \& {Santos}}]{boisse2012}
{Boisse}, I., {Bonfils}, X., \& {Santos}, N.~C. 2012,
  \href{https://ui.adsabs.harvard.edu/abs/2012A&A...545A.109B}{\aap, 545, A109}

\bibitem[{{Bonfils} {et~al.}(2018){Bonfils}, {Astudillo-Defru}, {D{\'\i}az},
  {Almenara}, {Forveille}, {Bouchy}, {Delfosse}, {Lovis}, {Mayor}, {Murgas},
  {Pepe}, {Santos}, {S{\'e}gransan}, {Udry}, \&
  {W{\"u}nsche}}]{2018A&A...613A..25B}
{Bonfils}, X., {Astudillo-Defru}, N., {D{\'\i}az}, R., {et~al.} 2018,
  \href{https://ui.adsabs.harvard.edu/abs/2018A&A...613A..25B}{\aap, 613, A25}

\bibitem[{{Bonfils} {et~al.}(2013){Bonfils}, {Delfosse}, {Udry}, {Forveille},
  {Mayor}, {Perrier}, {Bouchy}, {Gillon}, {Lovis}, {Pepe}, {Queloz}, {Santos},
  {S{\'e}gransan}, \& {Bertaux}}]{bonfils2013}
{Bonfils}, X., {Delfosse}, X., {Udry}, S., {et~al.} 2013,
  \href{https://ui.adsabs.harvard.edu/abs/2013A&A...549A.109B}{\aap, 549, A109}

\bibitem[{{Bonfils} {et~al.}(2005){Bonfils}, {Forveille}, {Delfosse}, {Udry},
  {Mayor}, {Perrier}, {Bouchy}, {Pepe}, {Queloz}, \&
  {Bertaux}}]{2005A&A...443L..15B}
{Bonfils}, X., {Forveille}, T., {Delfosse}, X., {et~al.} 2005,
  \href{https://ui.adsabs.harvard.edu/abs/2005A&A...443L..15B}{\aap, 443, L15}

\bibitem[{{Bouchy} {et~al.}(2001){Bouchy}, {Pepe}, \& {Queloz}}]{bouchy2001}
{Bouchy}, F., {Pepe}, F., \& {Queloz}, D. 2001,
  \href{https://ui.adsabs.harvard.edu/abs/2001A&A...374..733B}{\aap, 374, 733}

\bibitem[{{Brandenburg} \& {Subramanian}(2005)}]{brandenburg2005}
{Brandenburg}, A. \& {Subramanian}, K. 2005,
  \href{https://ui.adsabs.harvard.edu/abs/2005PhR...417....1B}{\physrep, 417,
  1}

\bibitem[{{Burt} {et~al.}(2014){Burt}, {Vogt}, {Butler}, {Hanson}, {Meschiari},
  {Rivera}, {Henry}, \& {Laughlin}}]{2014ApJ...789..114B}
{Burt}, J., {Vogt}, S.~S., {Butler}, R.~P., {et~al.} 2014,
  \href{https://ui.adsabs.harvard.edu/abs/2014ApJ...789..114B}{\apj, 789, 114}

\bibitem[{{Butler} {et~al.}(2006){Butler}, {Johnson}, {Marcy}, {Wright},
  {Vogt}, \& {Fischer}}]{2006PASP..118.1685B}
{Butler}, R.~P., {Johnson}, J.~A., {Marcy}, G.~W., {et~al.} 2006,
  \href{https://ui.adsabs.harvard.edu/abs/2006PASP..118.1685B}{\pasp, 118,
  1685}

\bibitem[{{Butler} {et~al.}(2017){Butler}, {Vogt}, {Laughlin}, {Burt},
  {Rivera}, {Tuomi}, {Teske}, {Arriagada}, {Diaz}, {Holden}, \&
  {Keiser}}]{2017AJ....153..208B}
{Butler}, R.~P., {Vogt}, S.~S., {Laughlin}, G., {et~al.} 2017,
  \href{https://ui.adsabs.harvard.edu/abs/2017AJ....153..208B}{\aj, 153, 208}

\bibitem[{{Butler} {et~al.}(2004){Butler}, {Vogt}, {Marcy}, {Fischer},
  {Wright}, {Henry}, {Laughlin}, \& {Lissauer}}]{2004ApJ...617..580B}
{Butler}, R.~P., {Vogt}, S.~S., {Marcy}, G.~W., {et~al.} 2004,
  \href{https://ui.adsabs.harvard.edu/abs/2004ApJ...617..580B}{\apj, 617, 580}

\bibitem[{{Caballero} {et~al.}(2016){Caballero}, {Cort{\'e}s-Contreras},
  {Alonso-Floriano}, {Montes}, {Quirrenbach}, {Amado}, {Ribas}, {Reiners},
  {Abellan}, {B{\'e}jar}, {Brinkm{\"o}ller}, {Czesla}, {Dorda}, {Gallardo},
  {Gonz{\'a}lez-{\'A}lvarez}, {Hidalgo}, {Holgado}, {Jeffers}, {Kim},
  {Klutsch}, {Lamert}, {Llamas}, {L{\'o}pez-Santiago},
  {Mart{\'\i}nez-Rodr{\'\i}guez}, {Morales}, {Mundt}, {Passegger},
  {Sch{\"o}fer}, {Seifert}, \& {Zechmeister}}]{caballero2016}
{Caballero}, J.~A., {Cort{\'e}s-Contreras}, M., {Alonso-Floriano}, F.~J.,
  {et~al.} 2016, in 19th Cambridge Workshop on Cool Stars, Stellar Systems, and
  the Sun (CS19), 148

\bibitem[{{Cadieux} {et~al.}(2022){Cadieux}, {Doyon}, {Plotnykov},
  {H{\'e}brard}, {Jahandar}, {Artigau}, {Valencia}, {Cook}, {Martioli},
  {Vandal}, {Donati}, {Cloutier}, {Narita}, {Fukui}, {Hirano}, {Bouchy},
  {Cowan}, {Gonzales}, {Ciardi}, {Stassun}, {Arnold}, {Benneke}, {Boisse},
  {Bonfils}, {Carmona}, {Cort{\'e}s-Zuleta}, {Delfosse}, {Forveille},
  {Fouqu{\'e}}, {Gomes da Silva}, {Jenkins}, {Kiefer}, {K{\'o}sp{\'a}l},
  {Lafreni{\`e}re}, {Martins}, {Moutou}, {do Nascimento}, {Ould-Elhkim},
  {Pelletier}, {Twicken}, {Bouma}, {Cartwright}, {Darveau-Bernier}, {Grankin},
  {Ikoma}, {Kagetani}, {Kawauchi}, {Kodama}, {Kotani}, {Latham}, {Menou},
  {Ricker}, {Seager}, {Tamura}, {Vanderspek}, \&
  {Watanabe}}]{2022AJ....164...96C}
{Cadieux}, C., {Doyon}, R., {Plotnykov}, M., {et~al.} 2022,
  \href{https://ui.adsabs.harvard.edu/abs/2022AJ....164...96C}{\aj, 164, 96}

\bibitem[{{Chaturvedi} {et~al.}(2022){Chaturvedi}, {Bluhm}, {Nagel}, {Hatzes},
  {Morello}, {Brady}, {Korth}, {Molaverdikhani}, {Kossakowski}, {Caballero},
  {Guenther}, {Pall{\'e}}, {Espinoza}, {Seifahrt}, {Lodieu}, {Cifuentes},
  {Furlan}, {Amado}, {Barclay}, {Bean}, {B{\'e}jar}, {Bergond}, {Boyle},
  {Ciardi}, {Collins}, {Collins}, {Esparza-Borges}, {Fukui}, {Gnilka}, {Goeke},
  {Guerra}, {Henning}, {Herrero}, {Howell}, {Jeffers}, {Jenkins}, {Jensen},
  {Kasper}, {Kodama}, {Latham}, {L{\'o}pez-Gonz{\'a}lez}, {Luque}, {Montes},
  {Morales}, {Mori}, {Murgas}, {Narita}, {Nowak}, {Parviainen}, {Passegger},
  {Quirrenbach}, {Reffert}, {Reiners}, {Ribas}, {Ricker}, {Rodriguez},
  {Rodr{\'\i}guez-L{\'o}pez}, {Schlecker}, {Schwarz}, {Schweitzer}, {Seager},
  {Stef{\'a}nsson}, {Stockdale}, {Tal-Or}, {Twicken}, {Vanaverbeke}, {Wang},
  {Watanabe}, {Winn}, \& {Zechmeister}}]{chaturvedi2022}
{Chaturvedi}, P., {Bluhm}, P., {Nagel}, E., {et~al.} 2022,
  \href{https://ui.adsabs.harvard.edu/abs/2022A&A...666A.155C}{\aap, 666, A155}

\bibitem[{{Cortés-Contreras} {et~al.}(2024){Cortés-Contreras}, {Caballero},
  {Montes}, {Cardona-Guillén}, {Béjar}, \& {Cifuentes}}]{contreras2024}
{Cortés-Contreras}, M., {Caballero}, J.~A., {Montes}, D., {et~al.} 2024,
  {\aap}, submitted

\bibitem[{{Cosentino} {et~al.}(2012){Cosentino}, {Lovis}, {Pepe}, {Collier
  Cameron}, {Latham}, {Molinari}, {Udry}, {Bezawada}, {Black}, {Born},
  {Buchschacher}, {Charbonneau}, {Figueira}, {Fleury}, {Galli}, {Gallie},
  {Gao}, {Ghedina}, {Gonzalez}, {Gonzalez}, {Guerra}, {Henry}, {Horne},
  {Hughes}, {Kelly}, {Lodi}, {Lunney}, {Maire}, {Mayor}, {Micela}, {Ordway},
  {Peacock}, {Phillips}, {Piotto}, {Pollacco}, {Queloz}, {Rice}, {Riverol},
  {Riverol}, {San Juan}, {Sasselov}, {Segransan}, {Sozzetti}, {Sosnowska},
  {Stobie}, {Szentgyorgyi}, {Vick}, \& {Weber}}]{2012SPIE.8446E..1VC}
{Cosentino}, R., {Lovis}, C., {Pepe}, F., {et~al.} 2012, in \procspie, Vol.
  8446, Ground-based and Airborne Instrumentation for Astronomy IV, ed. I.~S.
  {McLean}, S.~K. {Ramsay}, \& H.~{Takami}, 84461V

\bibitem[{{Crass} {et~al.}(2021){Crass}, {Gaudi}, {Leifer}, {Beichman},
  {Bender}, {Blackwood}, {Burt}, {Callas}, {Cegla}, {Diddams}, {Dumusque},
  {Eastman}, {Ford}, {Fulton}, {Gibson}, {Halverson}, {Haywood}, {Hearty},
  {Howard}, {Latham}, {L{\"o}hner-B{\"o}ttcher}, {Mamajek}, {Mortier},
  {Newman}, {Plavchan}, {Quirrenbach}, {Reiners}, {Robertson}, {Roy}, {Schwab},
  {Seifahrt}, {Szentgyorgyi}, {Terrien}, {Teske}, {Thompson}, \&
  {Vasisht}}]{crass2021}
{Crass}, J., {Gaudi}, B.~S., {Leifer}, S., {et~al.} 2021,
  \href{https://ui.adsabs.harvard.edu/abs/2021arXiv210714291C}{arXiv e-prints,
  arXiv:2107.14291}

\bibitem[{{Cristofari} {et~al.}(2023){Cristofari}, {Donati}, {Folsom},
  {Masseron}, {Fouqu{\'e}}, {Moutou}, {Artigau}, {Carmona}, {Petit},
  {Delfosse}, {Martioli}, \& {the SLS consortium}}]{cristofari2023}
{Cristofari}, P.~I., {Donati}, J.~F., {Folsom}, C.~P., {et~al.} 2023,
  \href{https://ui.adsabs.harvard.edu/abs/2023MNRAS.522.1342C}{\mnras, 522,
  1342}

\bibitem[{{Damasso} {et~al.}(2022){Damasso}, {Perger}, {Almenara}, {Nardiello},
  {P{\'e}rez-Torres}, {Sozzetti}, {Hara}, {Quirrenbach}, {Bonfils}, {Zapatero
  Osorio}, {Astudillo-Defru}, {Gonz{\'a}lez Hern{\'a}ndez}, {Su{\'a}rez
  Mascareno}, {Amado}, {Forveille}, {Lillo-Box}, {Alibert}, {Caballero},
  {Cifuentes}, {Delfosse}, {Figueira}, {Galad{\'\i}-Enr{\'\i}quez}, {Hatzes},
  {Henning}, {Kaminski}, {Mayor}, {Murgas}, {Montes}, {Pinamonti}, {Reiners},
  {Ribas}, {B{\'e}jar}, {Schweitzer}, \& {Zechmeister}}]{2022A&A...666A.187D}
{Damasso}, M., {Perger}, M., {Almenara}, J.~M., {et~al.} 2022,
  \href{https://ui.adsabs.harvard.edu/abs/2022A&A...666A.187D}{\aap, 666, A187}

\bibitem[{{Dekker} {et~al.}(2000){Dekker}, {D'Odorico}, {Kaufer}, {Delabre}, \&
  {Kotzlowski}}]{2000SPIE.4008..534D}
{Dekker}, H., {D'Odorico}, S., {Kaufer}, A., {Delabre}, B., \& {Kotzlowski}, H.
  2000, in \procspie, Vol. 4008, Optical and IR Telescope Instrumentation and
  Detectors, ed. M.~{Iye} \& A.~F. {Moorwood}, 534--545

\bibitem[{{Desort} {et~al.}(2007){Desort}, {Lagrange}, {Galland}, {Udry}, \&
  {Mayor}}]{desort2007}
{Desort}, M., {Lagrange}, A.~M., {Galland}, F., {Udry}, S., \& {Mayor}, M.
  2007, \href{https://ui.adsabs.harvard.edu/abs/2007A&A...473..983D}{\aap, 473,
  983}

\bibitem[{{D{\'\i}az} {et~al.}(2019){D{\'\i}az}, {Delfosse}, {Hobson},
  {Boisse}, {Astudillo-Defru}, {Bonfils}, {Henry}, {Arnold}, {Bouchy},
  {Bourrier}, {Brugger}, {Dalal}, {Deleuil}, {Demangeon}, {Dolon}, {Dumusque},
  {Forveille}, {Hara}, {H{\'e}brard}, {Kiefer}, {Lopez}, {Mignon}, {Moreau},
  {Mousis}, {Moutou}, {Pepe}, {Perruchot}, {Richaud}, {Santerne}, {Santos},
  {Sottile}, {Stalport}, {S{\'e}gransan}, {Udry}, {Unger}, \&
  {Wilson}}]{2019A&A...625A..17D}
{D{\'\i}az}, R.~F., {Delfosse}, X., {Hobson}, M.~J., {et~al.} 2019,
  \href{https://ui.adsabs.harvard.edu/abs/2019A&A...625A..17D}{\aap, 625, A17}

\bibitem[{{D{\'\i}ez Alonso} {et~al.}(2019){D{\'\i}ez Alonso}, {Caballero},
  {Montes}, {de Cos Juez}, {Dreizler}, {Dubois}, {Jeffers}, {Lalitha}, {Naves},
  {Reiners}, {Ribas}, {Vanaverbeke}, {Amado}, {B{\'e}jar},
  {Cort{\'e}s-Contreras}, {Herrero}, {Hidalgo}, {K{\"u}rster}, {Logie},
  {Quirrenbach}, {Rau}, {Seifert}, {Sch{\"o}fer}, \& {Tal-Or}}]{diezalonso2019}
{D{\'\i}ez Alonso}, E., {Caballero}, J.~A., {Montes}, D., {et~al.} 2019,
  \href{https://ui.adsabs.harvard.edu/abs/2019A&A...621A.126D}{\aap, 621, A126}

\bibitem[{{Donati} {et~al.}(2023{\natexlab{a}}){Donati}, {Cristofari},
  {Finociety}, {Klein}, {Moutou}, {Gaidos}, {Cadieux}, {Artigau}, {Correia},
  {Bou{\'e}}, {Cook}, {Carmona}, {Lehmann}, {Bouvier}, {Martioli}, {Morin},
  {Fouqu{\'e}}, {Delfosse}, {Doyon}, {H{\'e}brard}, {Alencar}, {Laskar},
  {Arnold}, {Petit}, {K{\'o}sp{\'a}l}, {Vidotto}, {Folsom}, \&
  {collaboration}}]{donati2023}
{Donati}, J.~F., {Cristofari}, P.~I., {Finociety}, B., {et~al.}
  2023{\natexlab{a}},
  \href{https://ui.adsabs.harvard.edu/abs/2023MNRAS.525..455D}{\mnras, 525,
  455}

\bibitem[{{Donati} {et~al.}(2023{\natexlab{b}}){Donati}, {Lehmann},
  {Cristofari}, {Fouqu{\'e}}, {Moutou}, {Charpentier}, {Ould-Elhkim},
  {Carmona}, {Delfosse}, {Artigau}, {Alencar}, {Cadieux}, {Arnold}, {Petit},
  {Morin}, {Forveille}, {Cloutier}, {Doyon}, {H{\'e}brard}, \& {SLS
  Collaboration}}]{2023MNRAS.525.2015D}
{Donati}, J.~F., {Lehmann}, L.~T., {Cristofari}, P.~I., {et~al.}
  2023{\natexlab{b}},
  \href{https://ui.adsabs.harvard.edu/abs/2023MNRAS.525.2015D}{\mnras, 525,
  2015}

\bibitem[{{Dorda}(2011)}]{dorda2011}
{Dorda}, R. 2011, {CARMENCITA: CARMENES Cool star Information and daTa Archive}

\bibitem[{{Dravins} \& {Ludwig}(2023)}]{dravins2023}
{Dravins}, D. \& {Ludwig}, H.-G. 2023,
  \href{https://ui.adsabs.harvard.edu/abs/2023A&A...679A...3D}{\aap, 679, A3}

\bibitem[{{Dreizler} {et~al.}(2020){Dreizler}, {Crossfield}, {Kossakowski},
  {Plavchan}, {Jeffers}, {Kemmer}, {Luque}, {Espinoza}, {Pall{\'e}}, {Stassun},
  {Matthews}, {Cale}, {Caballero}, {Schlecker}, {Lillo-Box}, {Zechmeister},
  {Lalitha}, {Reiners}, {Soubkiou}, {Bitsch}, {Zapatero Osorio}, {Chaturvedi},
  {Hatzes}, {Ricker}, {Vanderspek}, {Latham}, {Seager}, {Winn}, {Jenkins},
  {Aceituno}, {Amado}, {Barkaoui}, {Barbieri}, {Batalha}, {Bauer}, {Benneke},
  {Benkhaldoun}, {Beichman}, {Berberian}, {Burt}, {Butler}, {Caldwell},
  {Chintada}, {Chontos}, {Christiansen}, {Ciardi}, {Cifuentes}, {Collins},
  {Collins}, {Combs}, {Cort{\'e}s-Contreras}, {Crane}, {Daylan}, {Dragomir},
  {Esparza-Borges}, {Evans}, {Feng}, {Flowers}, {Fukui}, {Fulton}, {Furlan},
  {Gaidos}, {Geneser}, {Giacalone}, {Gillon}, {Gonzales}, {Gorjian}, {Hellier},
  {Hidalgo}, {Howard}, {Howell}, {Huber}, {Isaacson}, {Jehin}, {Jensen},
  {Kaminski}, {Kane}, {Kawauchi}, {Kielkopf}, {Klahr}, {Kosiarek}, {Kreidberg},
  {K{\"u}rster}, {Lafarga}, {Livingston}, {Louie}, {Mann}, {Madrigal-Aguado},
  {Matson}, {Mocnik}, {Morales}, {Muirhead}, {Murgas}, {Nandakumar}, {Narita},
  {Nowak}, {Oshagh}, {Parviainen}, {Passegger}, {Pollacco}, {Pozuelos},
  {Quirrenbach}, {Reefe}, {Ribas}, {Robertson}, {Rodr{\'\i}guez-L{\'o}pez},
  {Rose}, {Roy}, {Schweitzer}, {Schlieder}, {Shectman}, {Tanner},
  {{\c{S}}enavc{\i}}, {Teske}, {Twicken}, {Villasenor}, {Wang}, {Weiss},
  {Wittrock}, {Y{\i}lmaz}, \& {Zohrabi}}]{2020A&A...644A.127D}
{Dreizler}, S., {Crossfield}, I.~J.~M., {Kossakowski}, D., {et~al.} 2020,
  \href{https://ui.adsabs.harvard.edu/abs/2020A&A...644A.127D}{\aap, 644, A127}

\bibitem[{{Dumusque} {et~al.}(2014){Dumusque}, {Boisse}, \&
  {Santos}}]{dumusque2014}
{Dumusque}, X., {Boisse}, I., \& {Santos}, N.~C. 2014,
  \href{https://ui.adsabs.harvard.edu/abs/2014ApJ...796..132D}{\apj, 796, 132}

\bibitem[{{Efron}(1979)}]{efron1979}
{Efron}, B. 1979, \href{http://www.jstor.org/stable/2958830}{Ann. Stat., 7, 1}

\bibitem[{{Ellwarth} {et~al.}(2023){Ellwarth}, {Sch{\"a}fer}, {Reiners}, \&
  {Zechmeister}}]{ellwarth2023}
{Ellwarth}, M., {Sch{\"a}fer}, S., {Reiners}, A., \& {Zechmeister}, M. 2023,
  \href{https://ui.adsabs.harvard.edu/abs/2023A&A...673A..19E}{\aap, 673, A19}

\bibitem[{{Endl} {et~al.}(2008){Endl}, {Cochran}, {Wittenmyer}, \&
  {Boss}}]{2008ApJ...673.1165E}
{Endl}, M., {Cochran}, W.~D., {Wittenmyer}, R.~A., \& {Boss}, A.~P. 2008,
  \href{https://ui.adsabs.harvard.edu/abs/2008ApJ...673.1165E}{\apj, 673, 1165}

\bibitem[{{Endl} {et~al.}(2022){Endl}, {Robertson}, {Cochran}, {MacQueen},
  {Bowler}, {Franson}, {Holcomb}, {Beard}, {Isaacson}, {Howard}, \&
  {Lubin}}]{2022AJ....164..238E}
{Endl}, M., {Robertson}, P., {Cochran}, W.~D., {et~al.} 2022,
  \href{https://ui.adsabs.harvard.edu/abs/2022AJ....164..238E}{\aj, 164, 238}

\bibitem[{{Espinoza} {et~al.}(2022){Espinoza}, {Pall{\'e}}, {Kemmer}, {Luque},
  {Caballero}, {Cifuentes}, {Herrero}, {S{\'a}nchez B{\'e}jar}, {Stock},
  {Molaverdikhani}, {Morello}, {Kossakowski}, {Schlecker}, {Amado}, {Bluhm},
  {Cort{\'e}s-Contreras}, {Henning}, {Kreidberg}, {K{\"u}rster}, {Lafarga},
  {Lodieu}, {Morales}, {Oshagh}, {Passegger}, {Pavlov}, {Quirrenbach},
  {Reffert}, {Reiners}, {Ribas}, {Rodr{\'\i}guez}, {Rodr{\'\i}guez L{\'o}pez},
  {Schweitzer}, {Trifonov}, {Chaturvedi}, {Dreizler}, {Jeffers}, {Kaminski},
  {L{\'o}pez-Gonz{\'a}lez}, {Lillo-Box}, {Montes}, {Nowak}, {Pedraz},
  {Vanaverbeke}, {Zapatero Osorio}, {Zechmeister}, {Collins}, {Girardin},
  {Guerra}, {Naves}, {Crossfield}, {Matthews}, {Howell}, {Ciardi}, {Gonzales},
  {Matson}, {Beichman}, {Schlieder}, {Barclay}, {Vezie}, {Villase{\~n}or},
  {Daylan}, {Mireies}, {Dragomir}, {Twicken}, {Jenkins}, {Winn}, {Latham},
  {Ricker}, \& {Seager}}]{2022AJ....163..133E}
{Espinoza}, N., {Pall{\'e}}, E., {Kemmer}, J., {et~al.} 2022,
  \href{https://ui.adsabs.harvard.edu/abs/2022AJ....163..133E}{\aj, 163, 133}

\bibitem[{{Faria} {et~al.}(2022){Faria}, {Su{\'a}rez Mascare{\~n}o},
  {Figueira}, {Silva}, {Damasso}, {Demangeon}, {Pepe}, {Santos}, {Rebolo},
  {Cristiani}, {Adibekyan}, {Alibert}, {Allart}, {Barros}, {Cabral},
  {D'Odorico}, {Di Marcantonio}, {Dumusque}, {Ehrenreich}, {Gonz{\'a}lez
  Hern{\'a}ndez}, {Hara}, {Lillo-Box}, {Lo Curto}, {Lovis}, {Martins},
  {M{\'e}gevand}, {Mehner}, {Micela}, {Molaro}, {Nunes}, {Pall{\'e}},
  {Poretti}, {Sousa}, {Sozzetti}, {Tabernero}, {Udry}, \& {Zapatero
  Osorio}}]{faria2022}
{Faria}, J.~P., {Su{\'a}rez Mascare{\~n}o}, A., {Figueira}, P., {et~al.} 2022,
  \href{https://ui.adsabs.harvard.edu/abs/2022A&A...658A.115F}{\aap, 658, A115}

\bibitem[{{Feng} {et~al.}(2020){Feng}, {Shectman}, {Clement}, {Vogt}, {Tuomi},
  {Teske}, {Burt}, {Crane}, {Holden}, {Wang}, {Thompson}, {D{\'\i}az}, \&
  {Butler}}]{feng2020}
{Feng}, F., {Shectman}, S.~A., {Clement}, M.~S., {et~al.} 2020,
  \href{https://ui.adsabs.harvard.edu/abs/2020ApJS..250...29F}{\apjs, 250, 29}

\bibitem[{{Fischer} {et~al.}(2016){Fischer}, {Anglada-Escude}, {Arriagada},
  {Baluev}, {Bean}, {Bouchy}, {Buchhave}, {Carroll}, {Chakraborty}, {Crepp},
  {Dawson}, {Diddams}, {Dumusque}, {Eastman}, {Endl}, {Figueira}, {Ford},
  {Foreman-Mackey}, {Fournier}, {F{\H{u}}r{\'e}sz}, {Gaudi}, {Gregory},
  {Grundahl}, {Hatzes}, {H{\'e}brard}, {Herrero}, {Hogg}, {Howard}, {Johnson},
  {Jorden}, {Jurgenson}, {Latham}, {Laughlin}, {Loredo}, {Lovis}, {Mahadevan},
  {McCracken}, {Pepe}, {Perez}, {Phillips}, {Plavchan}, {Prato}, {Quirrenbach},
  {Reiners}, {Robertson}, {Santos}, {Sawyer}, {Segransan}, {Sozzetti},
  {Steinmetz}, {Szentgyorgyi}, {Udry}, {Valenti}, {Wang}, {Wittenmyer}, \&
  {Wright}}]{fischer2016}
{Fischer}, D.~A., {Anglada-Escude}, G., {Arriagada}, P., {et~al.} 2016,
  \href{https://ui.adsabs.harvard.edu/abs/2016PASP..128f6001F}{\pasp, 128,
  066001}

\bibitem[{{Foreman-Mackey} {et~al.}(2013){Foreman-Mackey}, {Hogg}, {Lang}, \&
  {Goodman}}]{emcee}
{Foreman-Mackey}, D., {Hogg}, D.~W., {Lang}, D., \& {Goodman}, J. 2013,
  \href{https://ui.adsabs.harvard.edu/abs/2013PASP..125..306F}{\pasp, 125, 306}

\bibitem[{{Forveille} {et~al.}(2008){Forveille}, {Bonfils}, {Delfosse},
  {Beuzit}, {Perrier}, {S{\'e}gransan}, {Udry}, {Mayor}, {Pepe}, {Queloz},
  {Bouchy}, \& {Bertaux}}]{2008psa..conf..191F}
{Forveille}, T., {Bonfils}, X., {Delfosse}, X., {et~al.} 2008, in Precision
  Spectroscopy in Astrophysics, ed. N.~C. {Santos}, L.~{Pasquini}, A.~C.~M.
  {Correia}, \& M.~{Romaniello}, 191--196

\bibitem[{{Fuhrmeister} {et~al.}(2022){Fuhrmeister}, {Czesla}, {Nagel},
  {Reiners}, {Schmitt}, {Jeffers}, {Caballero}, {Shulyak}, {Johnson},
  {Zechmeister}, {Montes}, {L{\'o}pez-Gallifa}, {Ribas}, {Quirrenbach},
  {Amado}, {Galad{\'\i}-Enr{\'\i}quez}, {Hatzes}, {K{\"u}rster}, {Danielski},
  {B{\'e}jar}, {Kaminski}, {Morales}, \& {Zapatero Osorio}}]{fuhrmeister2022}
{Fuhrmeister}, B., {Czesla}, S., {Nagel}, E., {et~al.} 2022,
  \href{https://ui.adsabs.harvard.edu/abs/2022A&A...657A.125F}{\aap, 657, A125}

\bibitem[{{Fuhrmeister} {et~al.}(2023){Fuhrmeister}, {Czesla}, {Schmitt},
  {Schneider}, {Caballero}, {Jeffers}, {Nagel}, {Montes}, {G{\'a}lvez Ortiz},
  {Reiners}, {Ribas}, {Quirrenbach}, {Amado}, {Henning}, {Lodieu},
  {Mart{\'\i}n-Fern{\'a}ndez}, {Morales}, {Sch{\"o}fer}, {Seifert}, \&
  {Zechmeister}}]{fuhrmeister2023}
{Fuhrmeister}, B., {Czesla}, S., {Schmitt}, J.~H.~M.~M., {et~al.} 2023,
  \href{https://ui.adsabs.harvard.edu/abs/2023A&A...678A...1F}{\aap, 678, A1}

\bibitem[{{Gaia Collaboration}(2020)}]{2020yCat.1350....0G}
{Gaia Collaboration}. 2020,
  \href{https://ui.adsabs.harvard.edu/abs/2020yCat.1350....0G}{VizieR Online
  Data Catalog, I/350}

\bibitem[{{Gillon} {et~al.}(2016){Gillon}, {Jehin}, {Lederer}, {Delrez}, {de
  Wit}, {Burdanov}, {Van Grootel}, {Burgasser}, {Triaud}, {Opitom}, {Demory},
  {Sahu}, {Bardalez Gagliuffi}, {Magain}, \& {Queloz}}]{2016Natur.533..221G}
{Gillon}, M., {Jehin}, E., {Lederer}, S.~M., {et~al.} 2016,
  \href{https://ui.adsabs.harvard.edu/abs/2016Natur.533..221G}{\nat, 533, 221}

\bibitem[{{Gillon} {et~al.}(2017){Gillon}, {Triaud}, {Demory}, {Jehin}, {Agol},
  {Deck}, {Lederer}, {de Wit}, {Burdanov}, {Ingalls}, {Bolmont}, {Leconte},
  {Raymond}, {Selsis}, {Turbet}, {Barkaoui}, {Burgasser}, {Burleigh}, {Carey},
  {Chaushev}, {Copperwheat}, {Delrez}, {Fernandes}, {Holdsworth}, {Kotze}, {Van
  Grootel}, {Almleaky}, {Benkhaldoun}, {Magain}, \&
  {Queloz}}]{2017Natur.542..456G}
{Gillon}, M., {Triaud}, A. H.~M.~J., {Demory}, B.-O., {et~al.} 2017,
  \href{https://ui.adsabs.harvard.edu/abs/2017Natur.542..456G}{\nat, 542, 456}

\bibitem[{{Goffo} {et~al.}(2024){Goffo}, {Chaturvedi}, {Murgas}, {Morello},
  {Orell-Miquel}, {Acu{\~n}a}, {Pe{\~n}a-Mo{\~n}ino}, {Pall{\'e}}, {Hatzes},
  {Gerald{\'\i}a-Gonz{\'a}lez}, {Pozuelos}, {Lanza}, {Gandolfi}, {Caballero},
  {Schlecker}, {P{\'e}rez-Torres}, {Lodieu}, {Schweitzer}, {Hellier},
  {Jeffers}, {Duque-Arribas}, {Cifuentes}, {B{\'e}jar}, {Daspute}, {Dubois},
  {Dufoer}, {Esparza-Borges}, {Fukui}, {Hayashi}, {Herrero}, {Mori}, {Narita},
  {Parviainen}, {Tal-Or}, {Vanaverbeke}, {Hermelo}, {Amado}, {Dreizler},
  {Henning}, {Lillo-Box}, {Luque}, {Mallorqu{\'\i}n}, {Nagel}, {Quirrenbach},
  {Reffert}, {Reiners}, {Ribas}, {Sch{\"o}fer}, {Tabernero}, \&
  {Zechmeister}}]{2024A&A...685A.147G}
{Goffo}, E., {Chaturvedi}, P., {Murgas}, F., {et~al.} 2024,
  \href{https://ui.adsabs.harvard.edu/abs/2024A&A...685A.147G}{\aap, 685, A147}

\bibitem[{{Gonz{\'a}lez-{\'A}lvarez}
  {et~al.}(2023{\natexlab{a}}){Gonz{\'a}lez-{\'A}lvarez}, {Kemmer},
  {Chaturvedi}, {Caballero}, {Quirrenbach}, {Amado}, {B{\'e}jar}, {Cifuentes},
  {Herrero}, {Kossakowski}, {Reiners}, {Ribas}, {Rodr{\'\i}guez},
  {Rodr{\'\i}guez-L{\'o}pez}, {Sanz-Forcada}, {Shan}, {Stock}, {Tabernero},
  {Tal-Or}, {Osorio}, {Hatzes}, {Henning}, {L{\'o}pez-Gonz{\'a}lez}, {Montes},
  {Morales}, {Pall{\'e}}, {Pedraz}, {Perger}, {Reffert}, {Sabotta},
  {Schweitzer}, \& {Zechmeister}}]{2023A&A...675A.141G}
{Gonz{\'a}lez-{\'A}lvarez}, E., {Kemmer}, J., {Chaturvedi}, P., {et~al.}
  2023{\natexlab{a}},
  \href{https://ui.adsabs.harvard.edu/abs/2023A&A...675A.141G}{\aap, 675, A141}

\bibitem[{{Gonz{\'a}lez-{\'A}lvarez} {et~al.}(2021){Gonz{\'a}lez-{\'A}lvarez},
  {Petralia}, {Micela}, {Maldonado}, {Affer}, {Maggio}, {Covino}, {Damasso},
  {Lanza}, {Perger}, {Pinamonti}, {Poretti}, {Scandariato}, {Sozzetti},
  {Bignamini}, {Giacobbe}, {Leto}, {Pagano}, {Zanmar S{\'a}nchez},
  {Gonz{\'a}lez Hern{\'a}ndez}, {Rebolo}, {Ribas}, {Su{\'a}rez Mascare{\~n}o},
  \& {Toledo-Padr{\'o}n}}]{2021A&A...649A.157G}
{Gonz{\'a}lez-{\'A}lvarez}, E., {Petralia}, A., {Micela}, G., {et~al.} 2021,
  \href{https://ui.adsabs.harvard.edu/abs/2021A&A...649A.157G}{\aap, 649, A157}

\bibitem[{{Gonz{\'a}lez-{\'A}lvarez}
  {et~al.}(2023{\natexlab{b}}){Gonz{\'a}lez-{\'A}lvarez}, {Zapatero Osorio},
  {Caballero}, {B{\'e}jar}, {Cifuentes}, {Fukui}, {Herrero}, {Kawauchi},
  {Livingston}, {L{\'o}pez-Gonz{\'a}lez}, {Morello}, {Murgas}, {Narita},
  {Pall{\'e}}, {Passegger}, {Rodr{\'\i}guez}, {Rodr{\'\i}guez-L{\'o}pez},
  {Sanz-Forcada}, {Schweitzer}, {Tabernero}, {Quirrenbach}, {Amado},
  {Charbonneau}, {Ciardi}, {Cikota}, {Collins}, {Conti}, {Fausnaugh}, {Hatzes},
  {Hedges}, {Henning}, {Jenkins}, {Latham}, {Massey}, {Moldovan}, {Montes},
  {Panahi}, {Reiners}, {Ribas}, {Ricker}, {Seager}, {Shporer}, {Srdoc},
  {Tenenbaum}, {Vanderspek}, {Winn}, {Fukuda}, {Ikoma}, {Isogai}, {Kawai},
  {Mori}, {Tamura}, \& {Watanabe}}]{2023A&A...675A.177G}
{Gonz{\'a}lez-{\'A}lvarez}, E., {Zapatero Osorio}, M.~R., {Caballero}, J.~A.,
  {et~al.} 2023{\natexlab{b}},
  \href{https://ui.adsabs.harvard.edu/abs/2023A&A...675A.177G}{\aap, 675, A177}

\bibitem[{{Gonz{\'a}lez-{\'A}lvarez} {et~al.}(2020){Gonz{\'a}lez-{\'A}lvarez},
  {Zapatero Osorio}, {Caballero}, {Sanz-Forcada}, {B{\'e}jar},
  {Gonz{\'a}lez-Cuesta}, {Dreizler}, {Bauer}, {Rodr{\'\i}guez}, {Tal-Or},
  {Zechmeister}, {Montes}, {L{\'o}pez-Gonz{\'a}lez}, {Ribas}, {Reiners},
  {Quirrenbach}, {Amado}, {Anglada-Escud{\'e}}, {Azzaro},
  {Cort{\'e}s-Contreras}, {Hatzes}, {Henning}, {Jeffers}, {Kaminski},
  {K{\"u}rster}, {Lafarga}, {Morales}, {Pall{\'e}}, {Perger}, \&
  {Schmitt}}]{2020A&A...637A..93G}
{Gonz{\'a}lez-{\'A}lvarez}, E., {Zapatero Osorio}, M.~R., {Caballero}, J.~A.,
  {et~al.} 2020,
  \href{https://ui.adsabs.harvard.edu/abs/2020A&A...637A..93G}{\aap, 637, A93}

\bibitem[{{Gonz{\'a}lez-{\'A}lvarez} {et~al.}(2022){Gonz{\'a}lez-{\'A}lvarez},
  {Zapatero Osorio}, {Sanz-Forcada}, {Caballero}, {Reffert}, {B{\'e}jar},
  {Hatzes}, {Herrero}, {Jeffers}, {Kemmer}, {L{\'o}pez-Gonz{\'a}lez}, {Luque},
  {Molaverdikhani}, {Morello}, {Nagel}, {Quirrenbach}, {Rodr{\'\i}guez},
  {Rodr{\'\i}guez-L{\'o}pez}, {Schlecker}, {Schweitzer}, {Stock}, {Passegger},
  {Trifonov}, {Amado}, {Baker}, {Boyd}, {Cadieux}, {Charbonneau}, {Collins},
  {Doyon}, {Dreizler}, {Espinoza}, {F{\H{u}}r{\'e}sz}, {Furlan}, {Hesse},
  {Howell}, {Jenkins}, {Kidwell}, {Latham}, {McLeod}, {Montes}, {Morales},
  {O'Dwyer}, {Pall{\'e}}, {Pedraz}, {Reiners}, {Ribas}, {Quinn}, {Schnaible},
  {Seager}, {Skinner}, {Smith}, {Schwarz}, {Shporer}, {Vanderspek}, \&
  {Winn}}]{2022A&A...658A.138G}
{Gonz{\'a}lez-{\'A}lvarez}, E., {Zapatero Osorio}, M.~R., {Sanz-Forcada}, J.,
  {et~al.} 2022,
  \href{https://ui.adsabs.harvard.edu/abs/2022A&A...658A.138G}{\aap, 658, A138}

\bibitem[{{Gorrini} {et~al.}(2023){Gorrini}, {Kemmer}, {Dreizler}, {Burn},
  {Hirano}, {Pozuelos}, {Kuzuhara}, {Caballero}, {Amado}, {Harakawa}, {Kudo},
  {Quirrenbach}, {Reiners}, {Ribas}, {B{\'e}jar}, {Chaturvedi}, {Cifuentes},
  {Galad{\'\i}-Enr{\'\i}quez}, {Hatzes}, {Kaminski}, {Kotani}, {K{\"u}rster},
  {Livingston}, {L{\'o}pez Gonz{\'a}lez}, {Montes}, {Morales}, {Murgas},
  {Omiya}, {Pall{\'e}}, {Rodr{\'\i}guez}, {Sato}, {Schweitzer}, {Shan},
  {Takarada}, {Tal-Or}, {Tamura}, {Vievard}, {Zapatero Osorio}, \&
  {Zechmeister}}]{2023A&A...680A..28G}
{Gorrini}, P., {Kemmer}, J., {Dreizler}, S., {et~al.} 2023,
  \href{https://ui.adsabs.harvard.edu/abs/2023A&A...680A..28G}{\aap, 680, A28}

\bibitem[{{Haghighipour} {et~al.}(2010){Haghighipour}, {Vogt}, {Butler},
  {Rivera}, {Laughlin}, {Meschiari}, \& {Henry}}]{2010ApJ...715..271H}
{Haghighipour}, N., {Vogt}, S.~S., {Butler}, R.~P., {et~al.} 2010,
  \href{https://ui.adsabs.harvard.edu/abs/2010ApJ...715..271H}{\apj, 715, 271}

\bibitem[{{Hatzes}(1999)}]{hatzes1999}
{Hatzes}, A.~P. 1999, in Astronomical Society of the Pacific Conference Series,
  Vol. 185, IAU Colloq. 170: Precise Stellar Radial Velocities, ed. J.~B.
  {Hearnshaw} \& C.~D. {Scarfe}, 259

\bibitem[{{Hatzes}(2002)}]{hatzes2002}
{Hatzes}, A.~P. 2002,
  \href{https://ui.adsabs.harvard.edu/abs/2002AN....323..392H}{Astronomische
  Nachrichten, 323, 392}

\bibitem[{{Haywood} {et~al.}(2022){Haywood}, {Milbourne}, {Saar}, {Mortier},
  {Phillips}, {Charbonneau}, {Cameron}, {Cegla}, {Meunier}, \&
  {}}]{haywood2022}
{Haywood}, R.~D., {Milbourne}, T.~W., {Saar}, S.~H., {et~al.} 2022,
  \href{https://ui.adsabs.harvard.edu/abs/2022ApJ...935....6H}{\apj, 935, 6}

\bibitem[{{Herrero} {et~al.}(2016){Herrero}, {Ribas}, {Jordi}, {Morales},
  {Perger}, \& {Rosich}}]{herrero2016}
{Herrero}, E., {Ribas}, I., {Jordi}, C., {et~al.} 2016,
  \href{https://ui.adsabs.harvard.edu/abs/2016A&A...586A.131H}{\aap, 586, A131}

\bibitem[{{Hobson} {et~al.}(2019){Hobson}, {Delfosse}, {Astudillo-Defru},
  {Boisse}, {D{\'\i}az}, {Bouchy}, {Bonfils}, {Forveille}, {Arnold},
  {Borgniet}, {Bourrier}, {Brugger}, {Cabrera Salazar}, {Courcol}, {Dalal},
  {Deleuil}, {Demangeon}, {Dumusque}, {Hara}, {H{\'e}brard}, {Kiefer}, {Lopez},
  {Mignon}, {Montagnier}, {Mousis}, {Moutou}, {Pepe}, {Rey}, {Santerne},
  {Santos}, {Stalport}, {S{\'e}gransan}, {Udry}, \&
  {Wilson}}]{2019A&A...625A..18H}
{Hobson}, M.~J., {Delfosse}, X., {Astudillo-Defru}, N., {et~al.} 2019,
  \href{https://ui.adsabs.harvard.edu/abs/2019A&A...625A..18H}{\aap, 625, A18}

\bibitem[{{Hobson} {et~al.}(2018){Hobson}, {D{\'\i}az}, {Delfosse},
  {Astudillo-Defru}, {Boisse}, {Bouchy}, {Bonfils}, {Forveille}, {Hara},
  {Arnold}, {Borgniet}, {Bourrier}, {Brugger}, {Cabrera}, {Courcol}, {Dalal},
  {Deleuil}, {Demangeon}, {Dumusque}, {Ehrenreich}, {H{\'e}brard}, {Kiefer},
  {Lopez}, {Mignon}, {Montagnier}, {Mousis}, {Moutou}, {Pepe}, {Rey},
  {Santerne}, {Santos}, {Stalport}, {S{\'e}gransan}, {Udry}, \&
  {Wilson}}]{hobson2018}
{Hobson}, M.~J., {D{\'\i}az}, R.~F., {Delfosse}, X., {et~al.} 2018,
  \href{https://ui.adsabs.harvard.edu/abs/2018A&A...618A.103H}{\aap, 618, A103}

\bibitem[{{Howard} {et~al.}(2010){Howard}, {Johnson}, {Marcy}, {Fischer},
  {Wright}, {Bernat}, {Henry}, {Peek}, {Isaacson}, {Apps}, {Endl}, {Cochran},
  {Valenti}, {Anderson}, \& {Piskunov}}]{2010ApJ...721.1467H}
{Howard}, A.~W., {Johnson}, J.~A., {Marcy}, G.~W., {et~al.} 2010,
  \href{https://ui.adsabs.harvard.edu/abs/2010ApJ...721.1467H}{\apj, 721, 1467}

\bibitem[{{Howard} {et~al.}(2014){Howard}, {Marcy}, {Fischer}, {Isaacson},
  {Muirhead}, {Henry}, {Boyajian}, {von Braun}, {Becker}, {Wright}, \&
  {Johnson}}]{howard2014}
{Howard}, A.~W., {Marcy}, G.~W., {Fischer}, D.~A., {et~al.} 2014,
  \href{https://ui.adsabs.harvard.edu/abs/2014ApJ...794...51H}{\apj, 794, 51}

\bibitem[{{Hudson}(1988)}]{hudson1988}
{Hudson}, H.~S. 1988,
  \href{https://ui.adsabs.harvard.edu/abs/1988ARA&A..26..473H}{\araa, 26, 473}

\bibitem[{{Ikuta} {et~al.}(2023){Ikuta}, {Namekata}, {Notsu}, {Maehara},
  {Okamoto}, {Honda}, {Nogami}, \& {Shibata}}]{ikuta2023}
{Ikuta}, K., {Namekata}, K., {Notsu}, Y., {et~al.} 2023,
  \href{https://ui.adsabs.harvard.edu/abs/2023ApJ...948...64I}{\apj, 948, 64}

\bibitem[{{Isaacson} \& {Fischer}(2010)}]{isaacson2010}
{Isaacson}, H. \& {Fischer}, D. 2010,
  \href{https://ui.adsabs.harvard.edu/abs/2010ApJ...725..875I}{\apj, 725, 875}

\bibitem[{{Jeffers} {et~al.}(2022){Jeffers}, {Barnes}, {Sch{\"o}fer},
  {Quirrenbach}, {Zechmeister}, {Amado}, {Caballero}, {Fern{\'a}ndez},
  {Rodr{\'\i}guez}, {Ribas}, {Reiners}, {Cardona Guill{\'e}n}, {Cifuentes},
  {Czesla}, {Hatzes}, {K{\"u}rster}, {Montes}, {Morales}, {Pedraz}, \&
  {Sadegi}}]{jeffers2022}
{Jeffers}, S.~V., {Barnes}, J.~R., {Sch{\"o}fer}, P., {et~al.} 2022,
  \href{https://ui.adsabs.harvard.edu/abs/2022A&A...663A..27J}{\aap, 663, A27}

\bibitem[{{Jeffers} {et~al.}(2018){Jeffers}, {Sch{\"o}fer}, {Lamert},
  {Reiners}, {Montes}, {Caballero}, {Cort{\'e}s-Contreras}, {Marvin},
  {Passegger}, {Zechmeister}, {Quirrenbach}, {Alonso-Floriano}, {Amado},
  {Bauer}, {Casal}, {Diez Alonso}, {Herrero}, {Morales}, {Mundt}, {Ribas}, \&
  {Sarmiento}}]{2018A&A...614A..76J}
{Jeffers}, S.~V., {Sch{\"o}fer}, P., {Lamert}, A., {et~al.} 2018,
  \href{https://ui.adsabs.harvard.edu/abs/2018A&A...614A..76J}{\aap, 614, A76}

\bibitem[{{Jenkins} {et~al.}(2009){Jenkins}, {Ramsey}, {Jones}, {Pavlenko},
  {Gallardo}, {Barnes}, \& {Pinfield}}]{2009ApJ...704..975J}
{Jenkins}, J.~S., {Ramsey}, L.~W., {Jones}, H.~R.~A., {et~al.} 2009,
  \href{https://ui.adsabs.harvard.edu/abs/2009ApJ...704..975J}{\apj, 704, 975}

\bibitem[{{Johnson} {et~al.}(2007){Johnson}, {Butler}, {Marcy}, {Fischer},
  {Vogt}, {Wright}, \& {Peek}}]{2007ApJ...670..833J}
{Johnson}, J.~A., {Butler}, R.~P., {Marcy}, G.~W., {et~al.} 2007,
  \href{https://ui.adsabs.harvard.edu/abs/2007ApJ...670..833J}{\apj, 670, 833}

\bibitem[{{Johnson} {et~al.}(2010){Johnson}, {Howard}, {Marcy}, {Bowler},
  {Henry}, {Fischer}, {Apps}, {Isaacson}, \& {Wright}}]{2010PASP..122..149J}
{Johnson}, J.~A., {Howard}, A.~W., {Marcy}, G.~W., {et~al.} 2010,
  \href{https://ui.adsabs.harvard.edu/abs/2010PASP..122..149J}{\pasp, 122, 149}

\bibitem[{{Kaminski} {et~al.}(2018){Kaminski}, {Trifonov}, {Caballero},
  {Quirrenbach}, {Ribas}, {Reiners}, {Amado}, {Zechmeister}, {Dreizler},
  {Perger}, {Tal-Or}, {Bonfils}, {Mayor}, {Astudillo-Defru}, {Bauer},
  {B{\'e}jar}, {Cifuentes}, {Colom{\'e}}, {Cort{\'e}s-Contreras}, {Delfosse},
  {D{\'\i}ez-Alonso}, {Forveille}, {Guenther}, {Hatzes}, {Henning}, {Jeffers},
  {K{\"u}rster}, {Lafarga}, {Luque}, {Mandel}, {Montes}, {Morales},
  {Passegger}, {Pedraz}, {Reffert}, {Sadegi}, {Schweitzer}, {Seifert}, {Stahl},
  \& {Udry}}]{2018A&A...618A.115K}
{Kaminski}, A., {Trifonov}, T., {Caballero}, J.~A., {et~al.} 2018,
  \href{https://ui.adsabs.harvard.edu/abs/2018A&A...618A.115K}{\aap, 618, A115}

\bibitem[{Kass \& Raftery(1995)}]{kass1995}
Kass, R.~E. \& Raftery, A.~E. 1995,
  \href{http://www.jstor.org/stable/2291327}{J. Am. Stat. Assoc., 90, 773}

\bibitem[{{Kemmer} {et~al.}(2022){Kemmer}, {Dreizler}, {Kossakowski}, {Stock},
  {Quirrenbach}, {Caballero}, {Amado}, {Collins}, {Espinoza}, {Herrero},
  {Jenkins}, {Latham}, {Lillo-Box}, {Narita}, {Pall{\'e}}, {Reiners}, {Ribas},
  {Ricker}, {Rodr{\'\i}guez}, {Seager}, {Vanderspek}, {Wells}, {Winn},
  {Aceituno}, {B{\'e}jar}, {Barclay}, {Bluhm}, {Chaturvedi}, {Cifuentes},
  {Collins}, {Cort{\'e}s-Contreras}, {Demory}, {Fausnaugh}, {Fukui}, {G{\'o}mez
  Maqueo Chew}, {Galad{\'\i}-Enr{\'\i}quez}, {Gan}, {Gillon}, {Golovin},
  {Hatzes}, {Henning}, {Huang}, {Jeffers}, {Kaminski}, {Kunimoto},
  {K{\"u}rster}, {L{\'o}pez-Gonz{\'a}lez}, {Lafarga}, {Luque}, {McCormac},
  {Molaverdikhani}, {Montes}, {Morales}, {Passegger}, {Reffert}, {Sabin},
  {Sch{\"o}fer}, {Schanche}, {Schlecker}, {Schroffenegger}, {Schwarz},
  {Schweitzer}, {Sota}, {Tenenbaum}, {Trifonov}, {Vanaverbeke}, \&
  {Zechmeister}}]{2022A&A...659A..17K}
{Kemmer}, J., {Dreizler}, S., {Kossakowski}, D., {et~al.} 2022,
  \href{https://ui.adsabs.harvard.edu/abs/2022A&A...659A..17K}{\aap, 659, A17}

\bibitem[{{Kemmer} {et~al.}(2020){Kemmer}, {Stock}, {Kossakowski}, {Kaminski},
  {Molaverdikhani}, {Schlecker}, {Caballero}, {Amado}, {Astudillo-Defru},
  {Bonfils}, {Ciardi}, {Collins}, {Espinoza}, {Fukui}, {Hirano}, {Jenkins},
  {Latham}, {Matthews}, {Narita}, {Pall{\'e}}, {Parviainen}, {Quirrenbach},
  {Reiners}, {Ribas}, {Ricker}, {Schlieder}, {Seager}, {Vanderspek}, {Winn},
  {Almenara}, {B{\'e}jar}, {Bluhm}, {Bouchy}, {Boyd}, {Christiansen},
  {Cifuentes}, {Cloutier}, {Collins}, {Cort{\'e}s-Contreras}, {Crossfield},
  {Crouzet}, {de Leon}, {Della-Rose}, {Delfosse}, {Dreizler}, {Esparza-Borges},
  {Essack}, {Forveille}, {Figueira}, {Galad{\'\i}-Enr{\'\i}quez}, {Gan},
  {Glidden}, {Gonzales}, {Guerra}, {Harakawa}, {Hatzes}, {Henning}, {Herrero},
  {Hodapp}, {Hori}, {Howell}, {Ikoma}, {Isogai}, {Jeffers}, {K{\"u}rster},
  {Kawauchi}, {Kimura}, {Klagyivik}, {Kotani}, {Kurokawa}, {Kusakabe},
  {Kuzuhara}, {Lafarga}, {Livingston}, {Luque}, {Matson}, {Morales}, {Mori},
  {Muirhead}, {Murgas}, {Nishikawa}, {Nishiumi}, {Omiya}, {Reffert},
  {Rodr{\'\i}guez L{\'o}pez}, {Santos}, {Sch{\"o}fer}, {Schwarz}, {Shiao},
  {Tamura}, {Terada}, {Twicken}, {Ueda}, {Vievard}, {Watanabe}, \&
  {Zechmeister}}]{2020A&A...642A.236K}
{Kemmer}, J., {Stock}, S., {Kossakowski}, D., {et~al.} 2020,
  \href{https://ui.adsabs.harvard.edu/abs/2020A&A...642A.236K}{\aap, 642, A236}

\bibitem[{{Kiraga}(2012)}]{2012AcA....62...67K}
{Kiraga}, M. 2012,
  \href{https://ui.adsabs.harvard.edu/abs/2012AcA....62...67K}{\actaa, 62, 67}

\bibitem[{{Kochukhov}(2021)}]{2021A&ARv..29....1K}
{Kochukhov}, O. 2021,
  \href{https://ui.adsabs.harvard.edu/abs/2021A&ARv..29....1K}{\aapr, 29, 1}

\bibitem[{{Kossakowski} {et~al.}(2021){Kossakowski}, {Kemmer}, {Bluhm},
  {Stock}, {Caballero}, {B{\'e}jar}, {Guill{\'e}n}, {Lodieu}, {Collins},
  {Oshagh}, {Schlecker}, {Espinoza}, {Pall{\'e}}, {Henning}, {Kreidberg},
  {K{\"u}rster}, {Amado}, {Anderson}, {Morales}, {Cartwright}, {Charbonneau},
  {Chaturvedi}, {Cifuentes}, {Conti}, {Cort{\'e}s-Contreras}, {Dreizler},
  {Galad{\'\i}-Enr{\'\i}quez}, {Guerra}, {Hart}, {Hellier}, {Henze}, {Herrero},
  {Jeffers}, {Jenkins}, {Jensen}, {Kaminski}, {Kielkopf}, {Kunimoto},
  {Lafarga}, {Latham}, {Lillo-Box}, {Luque}, {Molaverdikhani}, {Montes},
  {Morello}, {Morgan}, {Nowak}, {Pavlov}, {Perger}, {Quintana}, {Quirrenbach},
  {Reffert}, {Reiners}, {Ricker}, {Ribas}, {Rodr{\'\i}guez L{\'o}pez},
  {Zapatero Osorio}, {Seager}, {Sch{\"o}fer}, {Schweitzer}, {Trifonov},
  {Vanaverbeke}, {Vanderspek}, {West}, {Winn}, \&
  {Zechmeister}}]{kossakowski2021}
{Kossakowski}, D., {Kemmer}, J., {Bluhm}, P., {et~al.} 2021,
  \href{https://ui.adsabs.harvard.edu/abs/2021A&A...656A.124K}{\aap, 656, A124}

\bibitem[{{Kossakowski} {et~al.}(2022){Kossakowski}, {K{\"u}rster}, {Henning},
  {Trifonov}, {Caballero}, {Lafarga}, {Bauer}, {Stock}, {Kemmer}, {Jeffers},
  {Amado}, {P{\'e}rez-Torres}, {B{\'e}jar}, {Cort{\'e}s-Contreras}, {Ribas},
  {Reiners}, {Quirrenbach}, {Aceituno}, {Baroch}, {Cifuentes}, {Dreizler},
  {Hatzes}, {Kaminski}, {Montes}, {Morales}, {Pavlov}, {Pena}, {Perdelwitz},
  {Reffert}, {Revilla}, {Rodr{\'\i}guez Lopez}, {Rosich}, {Sadegi},
  {Sanz-Forcada}, {Sch{\"o}fer}, {Schweitzer}, \&
  {Zechmeister}}]{kossakowski2022}
{Kossakowski}, D., {K{\"u}rster}, M., {Henning}, T., {et~al.} 2022,
  \href{https://ui.adsabs.harvard.edu/abs/2022A&A...666A.143K}{\aap, 666, A143}

\bibitem[{{Kossakowski} {et~al.}(2023){Kossakowski}, {K{\"u}rster}, {Trifonov},
  {Henning}, {Kemmer}, {Caballero}, {Burn}, {Sabotta}, {Crouse}, {Fauchez},
  {Nagel}, {Kaminski}, {Herrero}, {Rodr{\'\i}guez}, {Gonz{\'a}lez-{\'A}lvarez},
  {Quirrenbach}, {Amado}, {Ribas}, {Reiners}, {Aceituno}, {B{\'e}jar},
  {Baroch}, {Bastelberger}, {Chaturvedi}, {Cifuentes}, {Dreizler}, {Jeffers},
  {Kopparapu}, {Lafarga}, {L{\'o}pez-Gonz{\'a}lez}, {Mart{\'\i}n-Ruiz},
  {Montes}, {Morales}, {Pall{\'e}}, {Pavlov}, {Pedraz}, {Perdelwitz},
  {P{\'e}rez-Torres}, {Perger}, {Reffert}, {Rodr{\'\i}guez L{\'o}pez},
  {Schlecker}, {Sch{\"o}fer}, {Schweitzer}, {Shan}, {Shields}, {Stock}, {Wolf},
  {Zapatero Osorio}, \& {Zechmeister}}]{2023A&A...670A..84K}
{Kossakowski}, D., {K{\"u}rster}, M., {Trifonov}, T., {et~al.} 2023,
  \href{https://ui.adsabs.harvard.edu/abs/2023A&A...670A..84K}{\aap, 670, A84}

\bibitem[{{K{\"u}rster} {et~al.}(2003){K{\"u}rster}, {Endl}, {Rouesnel}, {Els},
  {Kaufer}, {Brillant}, {Hatzes}, {Saar}, \& {Cochran}}]{kuester2003}
{K{\"u}rster}, M., {Endl}, M., {Rouesnel}, F., {et~al.} 2003,
  \href{https://ui.adsabs.harvard.edu/abs/2003A&A...403.1077K}{\aap, 403, 1077}

\bibitem[{{Lafarga} {et~al.}(2021){Lafarga}, {Ribas}, {Reiners}, {Quirrenbach},
  {Amado}, {Caballero}, {Azzaro}, {B{\'e}jar}, {Cort{\'e}s-Contreras},
  {Dreizler}, {Hatzes}, {Henning}, {Jeffers}, {Kaminski}, {K{\"u}rster},
  {Montes}, {Morales}, {Oshagh}, {Rodr{\'\i}guez-L{\'o}pez}, {Sch{\"o}fer},
  {Schweitzer}, \& {Zechmeister}}]{2021A&A...652A..28L}
{Lafarga}, M., {Ribas}, I., {Reiners}, A., {et~al.} 2021,
  \href{https://ui.adsabs.harvard.edu/abs/2021A&A...652A..28L}{\aap, 652, A28}

\bibitem[{{Lalitha} {et~al.}(2019){Lalitha}, {Baroch}, {Morales}, {Passegger},
  {Bauer}, {Cardona Guill{\'e}n}, {Dreizler}, {Oshagh}, {Reiners}, {Ribas},
  {Caballero}, {Quirrenbach}, {Amado}, {B{\'e}jar}, {Colom{\'e}},
  {Cort{\'e}s-Contreras}, {Galad{\'\i}-Enr{\'\i}quez}, {Gonz{\'a}lez-Cuesta},
  {Guenther}, {Hagen}, {Henning}, {Herrero}, {Husser}, {Jeffers}, {Kaminski},
  {K{\"u}rster}, {Lafarga}, {Lodieu}, {L{\'o}pez-Gonz{\'a}lez}, {Montes},
  {Perger}, {Rosich}, {Rodr{\'\i}guez}, {Rodr{\'\i}guez-L{\'o}pez}, {Schmitt},
  {Tal-Or}, \& {Zechmeister}}]{2019A&A...627A.116L}
{Lalitha}, S., {Baroch}, D., {Morales}, J.~C., {et~al.} 2019,
  \href{https://ui.adsabs.harvard.edu/abs/2019A&A...627A.116L}{\aap, 627, A116}

\bibitem[{{Liebing} {et~al.}(2021){Liebing}, {Jeffers}, {Reiners}, \&
  {Zechmeister}}]{liebing2021}
{Liebing}, F., {Jeffers}, S.~V., {Reiners}, A., \& {Zechmeister}, M. 2021,
  \href{https://ui.adsabs.harvard.edu/abs/2021A&A...654A.168L}{\aap, 654, A168}

\bibitem[{{Lienhard} {et~al.}(2023){Lienhard}, {Mortier}, {Cegla}, {Cameron},
  {Klein}, \& {Watson}}]{lienhard2023}
{Lienhard}, F., {Mortier}, A., {Cegla}, H.~M., {et~al.} 2023,
  \href{https://ui.adsabs.harvard.edu/abs/2023MNRAS.522.5862L}{\mnras, 522,
  5862}

\bibitem[{{Ludwig} {et~al.}(2002){Ludwig}, {Allard}, \&
  {Hauschildt}}]{ludwig2002}
{Ludwig}, H.~G., {Allard}, F., \& {Hauschildt}, P.~H. 2002,
  \href{https://ui.adsabs.harvard.edu/abs/2002A&A...395...99L}{\aap, 395, 99}

\bibitem[{{Luhn} {et~al.}(2020{\natexlab{a}}){Luhn}, {Wright}, {Howard}, \&
  {Isaacson}}]{luhn2020a}
{Luhn}, J.~K., {Wright}, J.~T., {Howard}, A.~W., \& {Isaacson}, H.
  2020{\natexlab{a}},
  \href{https://ui.adsabs.harvard.edu/abs/2020AJ....159..235L}{\aj, 159, 235}

\bibitem[{{Luhn} {et~al.}(2020{\natexlab{b}}){Luhn}, {Wright}, \&
  {Isaacson}}]{luhn2020b}
{Luhn}, J.~K., {Wright}, J.~T., \& {Isaacson}, H. 2020{\natexlab{b}},
  \href{https://ui.adsabs.harvard.edu/abs/2020AJ....159..236L}{\aj, 159, 236}

\bibitem[{{Luque} {et~al.}(2022){Luque}, {Fulton}, {Kunimoto}, {Amado},
  {Gorrini}, {Dreizler}, {Hellier}, {Henry}, {Molaverdikhani}, {Morello},
  {Pe{\~n}a-Mo{\~n}ino}, {P{\'e}rez-Torres}, {Pozuelos}, {Shan},
  {Anglada-Escud{\'e}}, {B{\'e}jar}, {Bergond}, {Boyle}, {Caballero},
  {Charbonneau}, {Ciardi}, {Dufoer}, {Espinoza}, {Everett}, {Fischer},
  {Hatzes}, {Henning}, {Hesse}, {Howard}, {Howell}, {Isaacson}, {Jeffers},
  {Jenkins}, {Kane}, {Kemmer}, {Khalafinejad}, {Kidwell}, {Kossakowski},
  {Latham}, {Lillo-Box}, {Lissauer}, {Montes}, {Orell-Miquel}, {Pall{\'e}},
  {Pollacco}, {Quirrenbach}, {Reffert}, {Reiners}, {Ribas}, {Ricker}, {Rogers},
  {Sanz-Forcada}, {Schlecker}, {Schweitzer}, {Seager}, {Shporer}, {Stassun},
  {Stock}, {Tal-Or}, {Ting}, {Trifonov}, {Vanaverbeke}, {Vanderspek},
  {Villase{\~n}or}, {Winn}, {Winters}, \& {Zapatero
  Osorio}}]{2022A&A...664A.199L}
{Luque}, R., {Fulton}, B.~J., {Kunimoto}, M., {et~al.} 2022,
  \href{https://ui.adsabs.harvard.edu/abs/2022A&A...664A.199L}{\aap, 664, A199}

\bibitem[{{Luque} {et~al.}(2018){Luque}, {Nowak}, {Pall{\'e}}, {Kossakowski},
  {Trifonov}, {Zechmeister}, {B{\'e}jar}, {Cardona Guill{\'e}n}, {Tal-Or},
  {Hidalgo}, {Ribas}, {Reiners}, {Caballero}, {Amado}, {Quirrenbach},
  {Aceituno}, {Cort{\'e}s-Contreras}, {D{\'\i}ez-Alonso}, {Dreizler},
  {Guenther}, {Henning}, {Jeffers}, {Kaminski}, {K{\"u}rster}, {Lafarga},
  {Montes}, {Morales}, {Passegger}, {Schmitt}, \&
  {Schweitzer}}]{2018A&A...620A.171L}
{Luque}, R., {Nowak}, G., {Pall{\'e}}, E., {et~al.} 2018,
  \href{https://ui.adsabs.harvard.edu/abs/2018A&A...620A.171L}{\aap, 620, A171}

\bibitem[{{Luque} {et~al.}(2019){Luque}, {Pall{\'e}}, {Kossakowski},
  {Dreizler}, {Kemmer}, {Espinoza}, {Burt}, {Anglada-Escud{\'e}}, {B{\'e}jar},
  {Caballero}, {Collins}, {Collins}, {Cort{\'e}s-Contreras},
  {D{\'\i}ez-Alonso}, {Feng}, {Hatzes}, {Hellier}, {Henning}, {Jeffers},
  {Kaltenegger}, {K{\"u}rster}, {Madden}, {Molaverdikhani}, {Montes}, {Narita},
  {Nowak}, {Ofir}, {Oshagh}, {Parviainen}, {Quirrenbach}, {Reffert}, {Reiners},
  {Rodr{\'\i}guez-L{\'o}pez}, {Schlecker}, {Stock}, {Trifonov}, {Winn},
  {Zapatero Osorio}, {Zechmeister}, {Amado}, {Anderson}, {Batalha}, {Bauer},
  {Bluhm}, {Burke}, {Butler}, {Caldwell}, {Chen}, {Crane}, {Dragomir},
  {Dressing}, {Dynes}, {Jenkins}, {Kaminski}, {Klahr}, {Kotani}, {Lafarga},
  {Latham}, {Lewin}, {McDermott}, {Monta{\~n}{\'e}s-Rodr{\'\i}guez}, {Morales},
  {Murgas}, {Nagel}, {Pedraz}, {Ribas}, {Ricker}, {Rowden}, {Seager},
  {Shectman}, {Tamura}, {Teske}, {Twicken}, {Vanderspeck}, {Wang}, \&
  {Wohler}}]{2019A&A...628A..39L}
{Luque}, R., {Pall{\'e}}, E., {Kossakowski}, D., {et~al.} 2019,
  \href{https://ui.adsabs.harvard.edu/abs/2019A&A...628A..39L}{\aap, 628, A39}

\bibitem[{{Mahadevan} {et~al.}(2021){Mahadevan}, {Stef{\'a}nsson}, {Robertson},
  {Terrien}, {Ninan}, {Holcomb}, {Halverson}, {Cochran}, {Kanodia}, {Ramsey},
  {Wolszczan}, {Endl}, {Bender}, {Diddams}, {Fredrick}, {Hearty}, {Monson},
  {Metcalf}, {Roy}, \& {Schwab}}]{2021ApJ...919L...9M}
{Mahadevan}, S., {Stef{\'a}nsson}, G., {Robertson}, P., {et~al.} 2021,
  \href{https://ui.adsabs.harvard.edu/abs/2021ApJ...919L...9M}{\apjl, 919, L9}

\bibitem[{{Mallorqu{\'\i}n} {et~al.}(2023){Mallorqu{\'\i}n}, {Goffo},
  {Pall{\'e}}, {Lodieu}, {B{\'e}jar}, {Isaacson}, {Zapatero Osorio},
  {Dreizler}, {Stock}, {Luque}, {Murgas}, {Pe{\~n}a}, {Sanz-Forcada},
  {Morello}, {Ciardi}, {Furlan}, {Collins}, {Herrero}, {Vanaverbeke},
  {Plavchan}, {Narita}, {Schweitzer}, {P{\'e}rez-Torres}, {Quirrenbach},
  {Kemmer}, {Hatzes}, {Howard}, {Schlecker}, {Reffert}, {Nagel}, {Morales},
  {Orell-Miquel}, {Duque-Arribas}, {Carleo}, {Cifuentes}, {Nowak}, {Ribas},
  {Reiners}, {Amado}, {Caballero}, {Henning}, {Pinter}, {Murphy}, {Beard},
  {Blunt}, {Brinkman}, {Cale}, {Chontos}, {Collins}, {Crossfield}, {Dai},
  {Dalba}, {Dufoer}, {El Mufti}, {Espinoza}, {Fetherolf}, {Fukui}, {Giacalone},
  {Gnilka}, {Gonzales}, {Grunblatt}, {Howell}, {Huber}, {Kane}, {de Le{\'o}n},
  {Lubin}, {MacDougall}, {Massey}, {Montes}, {Mori}, {Parviainen}, {Passegger},
  {Polanski}, {Robertson}, {Schwarz}, {Srdoc}, {Tabernero}, {Tanner},
  {Turtelboom}, {Van Zandt}, {Weiss}, \& {Zechmeister}}]{2023A&A...680A..76M}
{Mallorqu{\'\i}n}, M., {Goffo}, E., {Pall{\'e}}, E., {et~al.} 2023,
  \href{https://ui.adsabs.harvard.edu/abs/2023A&A...680A..76M}{\aap, 680, A76}

\bibitem[{{Malo} {et~al.}(2013){Malo}, {Doyon}, {Lafreni{\`e}re}, {Artigau},
  {Gagn{\'e}}, {Baron}, \& {Riedel}}]{malo2013}
{Malo}, L., {Doyon}, R., {Lafreni{\`e}re}, D., {et~al.} 2013,
  \href{https://ui.adsabs.harvard.edu/abs/2013ApJ...762...88M}{\apj, 762, 88}

\bibitem[{{Marcy} {et~al.}(2001){Marcy}, {Butler}, {Fischer}, {Vogt},
  {Lissauer}, \& {Rivera}}]{2001ApJ...556..296M}
{Marcy}, G.~W., {Butler}, R.~P., {Fischer}, D., {et~al.} 2001,
  \href{https://ui.adsabs.harvard.edu/abs/2001ApJ...556..296M}{\apj, 556, 296}

\bibitem[{{Marcy} {et~al.}(1998){Marcy}, {Butler}, {Vogt}, {Fischer}, \&
  {Lissauer}}]{1998ApJ...505L.147M}
{Marcy}, G.~W., {Butler}, R.~P., {Vogt}, S.~S., {Fischer}, D., \& {Lissauer},
  J.~J. 1998,
  \href{https://ui.adsabs.harvard.edu/abs/1998ApJ...505L.147M}{\apjl, 505,
  L147}

\bibitem[{{Marfil} {et~al.}(2021){Marfil}, {Tabernero}, {Montes}, {Caballero},
  {L{\'a}zaro}, {Gonz{\'a}lez Hern{\'a}ndez}, {Nagel}, {Passegger},
  {Schweitzer}, {Ribas}, {Reiners}, {Quirrenbach}, {Amado}, {Cifuentes},
  {Cort{\'e}s-Contreras}, {Dreizler}, {Duque-Arribas},
  {Galad{\'\i}-Enr{\'\i}quez}, {Henning}, {Jeffers}, {Kaminski}, {K{\"u}rster},
  {Lafarga}, {L{\'o}pez-Gallifa}, {Morales}, {Shan}, \&
  {Zechmeister}}]{marfil2021}
{Marfil}, E., {Tabernero}, H.~M., {Montes}, D., {et~al.} 2021,
  \href{https://ui.adsabs.harvard.edu/abs/2021A&A...656A.162M}{\aap, 656, A162}

\bibitem[{{Martioli} {et~al.}(2021){Martioli}, {H{\'e}brard}, {Correia},
  {Laskar}, \& {Lecavelier des Etangs}}]{2021A&A...649A.177M}
{Martioli}, E., {H{\'e}brard}, G., {Correia}, A.~C.~M., {Laskar}, J., \&
  {Lecavelier des Etangs}, A. 2021,
  \href{https://ui.adsabs.harvard.edu/abs/2021A&A...649A.177M}{\aap, 649, A177}

\bibitem[{{Mayor} {et~al.}(2009){Mayor}, {Bonfils}, {Forveille}, {Delfosse},
  {Udry}, {Bertaux}, {Beust}, {Bouchy}, {Lovis}, {Pepe}, {Perrier}, {Queloz},
  \& {Santos}}]{2009A&A...507..487M}
{Mayor}, M., {Bonfils}, X., {Forveille}, T., {et~al.} 2009,
  \href{https://ui.adsabs.harvard.edu/abs/2009A&A...507..487M}{\aap, 507, 487}

\bibitem[{{Mayor} {et~al.}(2003){Mayor}, {Pepe}, {Queloz}, {Bouchy},
  {Rupprecht}, {Lo Curto}, {Avila}, {Benz}, {Bertaux}, {Bonfils}, {Dall},
  {Dekker}, {Delabre}, {Eckert}, {Fleury}, {Gilliotte}, {Gojak}, {Guzman},
  {Kohler}, {Lizon}, {Longinotti}, {Lovis}, {Megevand}, {Pasquini}, {Reyes},
  {Sivan}, {Sosnowska}, {Soto}, {Udry}, {van Kesteren}, {Weber}, \&
  {Weilenmann}}]{2003Msngr.114...20M}
{Mayor}, M., {Pepe}, F., {Queloz}, D., {et~al.} 2003,
  \href{https://ui.adsabs.harvard.edu/abs/2003Msngr.114...20M}{The Messenger,
  114, 20}

\bibitem[{{Montet} {et~al.}(2014){Montet}, {Crepp}, {Johnson}, {Howard}, \&
  {Marcy}}]{2014ApJ...781...28M}
{Montet}, B.~T., {Crepp}, J.~R., {Johnson}, J.~A., {Howard}, A.~W., \& {Marcy},
  G.~W. 2014,
  \href{https://ui.adsabs.harvard.edu/abs/2014ApJ...781...28M}{\apj, 781, 28}

\bibitem[{{Morales} {et~al.}(2019){Morales}, {Mustill}, {Ribas}, {Davies},
  {Reiners}, {Bauer}, {Kossakowski}, {Herrero}, {Rodr{\'\i}guez},
  {L{\'o}pez-Gonz{\'a}lez}, {Rodr{\'\i}guez-L{\'o}pez}, {B{\'e}jar},
  {Gonz{\'a}lez-Cuesta}, {Luque}, {Pall{\'e}}, {Perger}, {Baroch}, {Johansen},
  {Klahr}, {Mordasini}, {Anglada-Escud{\'e}}, {Caballero},
  {Cort{\'e}s-Contreras}, {Dreizler}, {Lafarga}, {Nagel}, {Passegger},
  {Reffert}, {Rosich}, {Schweitzer}, {Tal-Or}, {Trifonov}, {Zechmeister},
  {Quirrenbach}, {Amado}, {Guenther}, {Hagen}, {Henning}, {Jeffers},
  {Kaminski}, {K{\"u}rster}, {Montes}, {Seifert}, {Abell{\'a}n}, {Abril},
  {Aceituno}, {Aceituno}, {Alonso-Floriano}, {Ammler-von Eiff}, {Antona},
  {Arroyo-Torres}, {Azzaro}, {Barrado}, {Becerril-Jarque}, {Ben{\'\i}tez},
  {Berdi{\~n}as}, {Bergond}, {Brinkm{\"o}ller}, {del Burgo}, {Burn},
  {Calvo-Ortega}, {Cano}, {C{\'a}rdenas}, {Cardona Guill{\'e}n}, {Carro},
  {Casal}, {Casanova}, {Casasayas-Barris}, {Chaturvedi}, {Cifuentes}, {Claret},
  {Colom{\'e}}, {Czesla}, {D{\'\i}ez-Alonso}, {Dorda}, {Emsenhuber},
  {Fern{\'a}ndez}, {Fern{\'a}ndez-Mart{\'\i}n}, {Ferro}, {Fuhrmeister},
  {Galad{\'\i}-Enr{\'\i}quez}, {Gallardo Cava}, {Garc{\'\i}a Vargas},
  {Garcia-Piquer}, {Gesa}, {Gonz{\'a}lez-{\'A}lvarez}, {Gonz{\'a}lez
  Hern{\'a}ndez}, {Gonz{\'a}lez-Peinado}, {Gu{\`a}rdia}, {Guijarro}, {de
  Guindos}, {Hatzes}, {Hauschildt}, {Hedrosa}, {Hermelo}, {Hern{\'a}ndez
  Arabi}, {Hern{\'a}ndez Otero}, {Hintz}, {Holgado}, {Huber}, {Huke},
  {Johnson}, {de Juan}, {Kehr}, {Kemmer}, {Kim}, {Kl{\"u}ter}, {Klutsch},
  {Labarga}, {Labiche}, {Lalitha}, {Lamp{\'o}n}, {Lara}, {Launhardt},
  {L{\'a}zaro}, {Lizon}, {Llamas}, {Lodieu}, {L{\'o}pez del Fresno}, {L{\'o}pez
  Salas}, {L{\'o}pez-Santiago}, {Mag{\'a}n Madinabeitia}, {Mall}, {Mancini},
  {Mandel}, {Marfil}, {Mar{\'\i}n Molina}, {Mart{\'\i}n},
  {Mart{\'\i}n-Fern{\'a}ndez}, {Mart{\'\i}n-Ruiz},
  {Mart{\'\i}nez-Rodr{\'\i}guez}, {Marvin}, {Mirabet}, {Moya}, {Naranjo},
  {Nelson}, {Nortmann}, {Nowak}, {Ofir}, {Pascual}, {Pavlov}, {Pedraz},
  {P{\'e}rez Medialdea}, {P{\'e}rez-Calpena}, {Perryman}, {Rabaza}, {Ram{\'o}n
  Ballesta}, {Rebolo}, {Redondo}, {Rix}, {Rodler}, {Rodr{\'\i}guez Trinidad},
  {Sabotta}, {Sadegi}, {Salz}, {S{\'a}nchez-Blanco}, {S{\'a}nchez Carrasco},
  {S{\'a}nchez-L{\'o}pez}, {Sanz-Forcada}, {Sarkis}, {Sarmiento},
  {Sch{\"a}fer}, {Schlecker}, {Schmitt}, {Sch{\"o}fer}, {Solano}, {Sota},
  {Stahl}, {Stock}, {Stuber}, {St{\"u}rmer}, {Su{\'a}rez}, {Tabernero},
  {Tulloch}, {Veredas}, {Vico-Linares}, {Vilardell}, {Wagner}, {Winkler},
  {Wolthoff}, {Yan}, \& {Zapatero Osorio}}]{2019Sci...365.1441M}
{Morales}, J.~C., {Mustill}, A.~J., {Ribas}, I., {et~al.} 2019,
  \href{https://ui.adsabs.harvard.edu/abs/2019Sci...365.1441M}{Science, 365,
  1441}

\bibitem[{{Morin} {et~al.}(2008){Morin}, {Donati}, {Petit}, {Delfosse},
  {Forveille}, {Albert}, {Auri{\`e}re}, {Cabanac}, {Dintrans}, {Fares},
  {Gastine}, {Jardine}, {Ligni{\`e}res}, {Paletou}, {Ramirez Velez}, \&
  {Th{\'e}ado}}]{morin2008}
{Morin}, J., {Donati}, J.~F., {Petit}, P., {et~al.} 2008,
  \href{https://ui.adsabs.harvard.edu/abs/2008MNRAS.390..567M}{\mnras, 390,
  567}

\bibitem[{{Morin} {et~al.}(2010){Morin}, {Donati}, {Petit}, {Delfosse},
  {Forveille}, \& {Jardine}}]{2010MNRAS.407.2269M}
{Morin}, J., {Donati}, J.~F., {Petit}, P., {et~al.} 2010,
  \href{https://ui.adsabs.harvard.edu/abs/2010MNRAS.407.2269M}{\mnras, 407,
  2269}

\bibitem[{{Moutou} {et~al.}(2017){Moutou}, {H{\'e}brard}, {Morin}, {Malo},
  {Fouqu{\'e}}, {Torres-Rivas}, {Martioli}, {Delfosse}, {Artigau}, \&
  {Doyon}}]{moutou2017}
{Moutou}, C., {H{\'e}brard}, E.~M., {Morin}, J., {et~al.} 2017,
  \href{https://ui.adsabs.harvard.edu/abs/2017MNRAS.472.4563M}{\mnras, 472,
  4563}

\bibitem[{{Nagel} {et~al.}(2023){Nagel}, {Czesla}, {Kaminski}, {Zechmeister},
  {Tal-Or}, {Schmitt}, {Reiners}, {Quirrenbach}, {Garc{\'\i}a L{\'o}pez},
  {Caballero}, {Ribas}, {Amado}, {B{\'e}jar}, {Cort{\'e}s-Contreras},
  {Dreizler}, {Hatzes}, {Henning}, {Jeffers}, {K{\"u}rster}, {Lafarga},
  {L{\'o}pez-Puertas}, {Montes}, {Morales}, {Pedraz}, \&
  {Schweitzer}}]{nagel2023}
{Nagel}, E., {Czesla}, S., {Kaminski}, A., {et~al.} 2023,
  \href{https://ui.adsabs.harvard.edu/abs/2023A&A...680A..73N}{\aap, 680, A73}

\bibitem[{{Nagel} {et~al.}(2019){Nagel}, {Czesla}, {Schmitt}, {Dreizler},
  {Anglada-Escud{\'e}}, {Rodr{\'\i}guez}, {Ribas}, {Reiners}, {Quirrenbach},
  {Amado}, {Caballero}, {Aceituno}, {B{\'e}jar}, {Cort{\'e}s-Contreras},
  {Gonz{\'a}lez-Cuesta}, {Guenther}, {Henning}, {Jeffers}, {Kaminski},
  {K{\"u}rster}, {Lafarga}, {L{\'o}pez-Gonz{\'a}lez}, {Montes}, {Morales},
  {Passegger}, {Rodr{\'\i}guez-L{\'o}pez}, {Schweitzer}, \&
  {Zechmeister}}]{2019A&A...622A.153N}
{Nagel}, E., {Czesla}, S., {Schmitt}, J.~H.~M.~M., {et~al.} 2019,
  \href{https://ui.adsabs.harvard.edu/abs/2019A&A...622A.153N}{\aap, 622, A153}

\bibitem[{{Newton} {et~al.}(2016){Newton}, {Irwin}, {Charbonneau},
  {Berta-Thompson}, {Dittmann}, \& {West}}]{2016ApJ...821...93N}
{Newton}, E.~R., {Irwin}, J., {Charbonneau}, D., {et~al.} 2016,
  \href{https://ui.adsabs.harvard.edu/abs/2016ApJ...821...93N}{\apj, 821, 93}

\bibitem[{{Newton} {et~al.}(2018){Newton}, {Mondrik}, {Irwin}, {Winters}, \&
  {Charbonneau}}]{2018AJ....156..217N}
{Newton}, E.~R., {Mondrik}, N., {Irwin}, J., {Winters}, J.~G., \&
  {Charbonneau}, D. 2018,
  \href{https://ui.adsabs.harvard.edu/abs/2018AJ....156..217N}{\aj, 156, 217}

\bibitem[{{Norris} {et~al.}(2023){Norris}, {Unruh}, {Witzke}, {Solanki},
  {Krivova}, {Shapiro}, {Yeo}, {Cameron}, \& {Beeck}}]{norris2023}
{Norris}, C.~M., {Unruh}, Y.~C., {Witzke}, V., {et~al.} 2023,
  \href{https://ui.adsabs.harvard.edu/abs/2023MNRAS.524.1139N}{\mnras, 524,
  1139}

\bibitem[{{Nowak} {et~al.}(2020){Nowak}, {Luque}, {Parviainen}, {Pall{\'e}},
  {Molaverdikhani}, {B{\'e}jar}, {Lillo-Box}, {Rodr{\'\i}guez-L{\'o}pez},
  {Caballero}, {Zechmeister}, {Passegger}, {Cifuentes}, {Schweitzer}, {Narita},
  {Cale}, {Espinoza}, {Murgas}, {Hidalgo}, {Zapatero Osorio}, {Pozuelos},
  {Aceituno}, {Amado}, {Barkaoui}, {Barrado}, {Bauer}, {Benkhaldoun},
  {Caldwell}, {Casasayas Barris}, {Chaturvedi}, {Chen}, {Collins}, {Collins},
  {Cort{\'e}s-Contreras}, {Crossfield}, {de Le{\'o}n}, {D{\'\i}ez Alonso},
  {Dreizler}, {El Mufti}, {Esparza-Borges}, {Essack}, {Fukui}, {Gaidos},
  {Gillon}, {Gonzales}, {Guerra}, {Hatzes}, {Henning}, {Herrero}, {Hesse},
  {Hirano}, {Howell}, {Jeffers}, {Jehin}, {Jenkins}, {Kaminski}, {Kemmer},
  {Kielkopf}, {Kossakowski}, {Kotani}, {K{\"u}rster}, {Lafarga}, {Latham},
  {Law}, {Lissauer}, {Lodieu}, {Madrigal-Aguado}, {Mann}, {Massey}, {Matson},
  {Matthews}, {Monta{\~n}{\'e}s-Rodr{\'\i}guez}, {Montes}, {Morales}, {Mori},
  {Nagel}, {Oshagh}, {Pedraz}, {Plavchan}, {Pollacco}, {Quirrenbach},
  {Reffert}, {Reiners}, {Ribas}, {Ricker}, {Rose}, {Schlecker}, {Schlieder},
  {Seager}, {Stangret}, {Stock}, {Tamura}, {Tanner}, {Teske}, {Trifonov},
  {Twicken}, {Vanderspek}, {Watanabe}, {Wittrock}, {Ziegler}, \&
  {Zohrabi}}]{2020A&A...642A.173N}
{Nowak}, G., {Luque}, R., {Parviainen}, H., {et~al.} 2020,
  \href{https://ui.adsabs.harvard.edu/abs/2020A&A...642A.173N}{\aap, 642, A173}

\bibitem[{{Palle} {et~al.}(2023){Palle}, {Orell-Miquel}, {Brady}, {Bean},
  {Hatzes}, {Morello}, {Morales}, {Murgas}, {Molaverdikhani}, {Parviainen},
  {Sanz-Forcada}, {B{\'e}jar}, {Caballero}, {Sreenivas}, {Schlecker}, {Ribas},
  {Perdelwitz}, {Tal-Or}, {P{\'e}rez-Torres}, {Luque}, {Dreizler},
  {Fuhrmeister}, {Aceituno}, {Amado}, {Anglada-Escud{\'e}}, {Caldwell},
  {Charbonneau}, {Cifuentes}, {de Leon}, {Collins}, {Dufoer}, {Espinoza},
  {Essack}, {Fukui}, {Chew}, {G{\'o}mez-Mu{\~n}oz}, {Henning}, {Herrero},
  {Jeffers}, {Jenkins}, {Kaminski}, {Kasper}, {Kunimoto}, {Latham},
  {Lillo-Box}, {L{\'o}pez-Gonz{\'a}lez}, {Montes}, {Mori}, {Narita},
  {Quirrenbach}, {Pedraz}, {Reiners}, {Rodr{\'\i}guez},
  {Rodr{\'\i}guez-L{\'o}pez}, {Sabin}, {Schanche}, {Schwarz}, {Schweitzer},
  {Seifahrt}, {Stefansson}, {Sturmer}, {Trifonov}, {Vanaverbeke}, {Wells},
  {Zapatero-Osorio}, \& {Zechmeister}}]{2023A&A...678A..80P}
{Palle}, E., {Orell-Miquel}, J., {Brady}, M., {et~al.} 2023,
  \href{https://ui.adsabs.harvard.edu/abs/2023A&A...678A..80P}{\aap, 678, A80}

\bibitem[{{Pass} {et~al.}(2023{\natexlab{a}}){Pass}, {Winters}, {Charbonneau},
  {Irwin}, {Latham}, {Berlind}, {Calkins}, {Esquerdo}, \& {Mink}}]{pass2023a}
{Pass}, E.~K., {Winters}, J.~G., {Charbonneau}, D., {et~al.}
  2023{\natexlab{a}},
  \href{https://ui.adsabs.harvard.edu/abs/2023AJ....166...11P}{\aj, 166, 11}

\bibitem[{{Pass} {et~al.}(2023{\natexlab{b}}){Pass}, {Winters}, {Charbonneau},
  {Irwin}, \& {Medina}}]{2023AJ....166...16P}
{Pass}, E.~K., {Winters}, J.~G., {Charbonneau}, D., {Irwin}, J.~M., \&
  {Medina}, A.~A. 2023{\natexlab{b}},
  \href{https://ui.adsabs.harvard.edu/abs/2023AJ....166...16P}{\aj, 166, 16}

\bibitem[{{Pepe} {et~al.}(2021){Pepe}, {Cristiani}, {Rebolo}, {Santos},
  {Dekker}, {Cabral}, {Di Marcantonio}, {Figueira}, {Lo Curto}, {Lovis},
  {Mayor}, {M{\'e}gevand}, {Molaro}, {Riva}, {Zapatero Osorio}, {Amate},
  {Manescau}, {Pasquini}, {Zerbi}, {Adibekyan}, {Abreu}, {Affolter}, {Alibert},
  {Aliverti}, {Allart}, {Allende Prieto}, {{\'A}lvarez}, {Alves}, {Avila},
  {Baldini}, {Bandy}, {Barros}, {Benz}, {Bianco}, {Borsa}, {Bourrier},
  {Bouchy}, {Broeg}, {Calderone}, {Cirami}, {Coelho}, {Conconi}, {Coretti},
  {Cumani}, {Cupani}, {D'Odorico}, {Damasso}, {Deiries}, {Delabre},
  {Demangeon}, {Dumusque}, {Ehrenreich}, {Faria}, {Fragoso}, {Genolet},
  {Genoni}, {G{\'e}nova Santos}, {Gonz{\'a}lez Hern{\'a}ndez}, {Hughes},
  {Iwert}, {Kerber}, {Knudstrup}, {Landoni}, {Lavie}, {Lillo-Box}, {Lizon},
  {Maire}, {Martins}, {Mehner}, {Micela}, {Modigliani}, {Monteiro}, {Monteiro},
  {Moschetti}, {Murphy}, {Nunes}, {Oggioni}, {Oliveira}, {Oshagh}, {Pall{\'e}},
  {Pariani}, {Poretti}, {Rasilla}, {Rebord{\~a}o}, {Redaelli}, {Santana
  Tschudi}, {Santin}, {Santos}, {S{\'e}gransan}, {Schmidt}, {Segovia},
  {Sosnowska}, {Sozzetti}, {Sousa}, {Span{\`o}}, {Su{\'a}rez Mascare{\~n}o},
  {Tabernero}, {Tenegi}, {Udry}, \& {Zanutta}}]{pepe2021}
{Pepe}, F., {Cristiani}, S., {Rebolo}, R., {et~al.} 2021,
  \href{https://ui.adsabs.harvard.edu/abs/2021A&A...645A..96P}{\aap, 645, A96}

\bibitem[{{Perger} {et~al.}(2017){Perger}, {Garc{\'\i}a-Piquer}, {Ribas},
  {Morales}, {Affer}, {Micela}, {Damasso}, {Su{\'a}rez-Mascare{\~n}o},
  {Gonz{\'a}lez-Hern{\'a}ndez}, {Rebolo}, {Herrero}, {Rosich}, {Lafarga},
  {Bignamini}, {Sozzetti}, {Claudi}, {Cosentino}, {Molinari}, {Maldonado},
  {Maggio}, {Lanza}, {Poretti}, {Pagano}, {Desidera}, {Gratton}, {Piotto},
  {Bonomo}, {Martinez Fiorenzano}, {Giacobbe}, {Malavolta}, {Nascimbeni},
  {Rainer}, \& {Scandariato}}]{perger2017}
{Perger}, M., {Garc{\'\i}a-Piquer}, A., {Ribas}, I., {et~al.} 2017,
  \href{https://ui.adsabs.harvard.edu/abs/2017A&A...598A..26P}{\aap, 598, A26}

\bibitem[{{Perger} {et~al.}(2019){Perger}, {Scandariato}, {Ribas}, {Morales},
  {Affer}, {Azzaro}, {Amado}, {Anglada-Escud{\'e}}, {Baroch}, {Barrado},
  {Bauer}, {B{\'e}jar}, {Caballero}, {Cort{\'e}s-Contreras}, {Damasso},
  {Dreizler}, {Gonz{\'a}lez-Cuesta}, {Gonz{\'a}lez Hern{\'a}ndez}, {Guenther},
  {Henning}, {Herrero}, {Jeffers}, {Kaminski}, {K{\"u}rster}, {Lafarga},
  {Leto}, {L{\'o}pez-Gonz{\'a}lez}, {Maldonado}, {Micela}, {Montes},
  {Pinamonti}, {Quirrenbach}, {Rebolo}, {Reiners}, {Rodr{\'\i}guez},
  {Rodr{\'\i}guez-L{\'o}pez}, {Schmitt}, {Sozzetti}, {Su{\'a}rez
  Mascare{\~n}o}, {Toledo-Padr{\'o}n}, {Zanmar S{\'a}nchez}, {Zapatero Osorio},
  \& {Zechmeister}}]{perger2019}
{Perger}, M., {Scandariato}, G., {Ribas}, I., {et~al.} 2019,
  \href{https://ui.adsabs.harvard.edu/abs/2019A&A...624A.123P}{\aap, 624, A123}

\bibitem[{{Pinamonti} {et~al.}(2019){Pinamonti}, {Sozzetti}, {Giacobbe},
  {Damasso}, {Scandariato}, {Perger}, {Gonz{\'a}lez Hern{\'a}ndez}, {Lanza},
  {Maldonado}, {Micela}, {Su{\'a}rez Mascare{\~n}o}, {Toledo-Padr{\'o}n},
  {Affer}, {Benatti}, {Bignamini}, {Bonomo}, {Claudi}, {Cosentino}, {Desidera},
  {Maggio}, {Martinez Fiorenzano}, {Pagano}, {Piotto}, {Rainer}, {Rebolo}, \&
  {Ribas}}]{2019A&A...625A.126P}
{Pinamonti}, M., {Sozzetti}, A., {Giacobbe}, P., {et~al.} 2019,
  \href{https://ui.adsabs.harvard.edu/abs/2019A&A...625A.126P}{\aap, 625, A126}

\bibitem[{{Plavchan} {et~al.}(2020){Plavchan}, {Barclay}, {Gagn{\'e}}, {Gao},
  {Cale}, {Matzko}, {Dragomir}, {Quinn}, {Feliz}, {Stassun}, {Crossfield},
  {Berardo}, {Latham}, {Tieu}, {Anglada-Escud{\'e}}, {Ricker}, {Vanderspek},
  {Seager}, {Winn}, {Jenkins}, {Rinehart}, {Krishnamurthy}, {Dynes}, {Doty},
  {Adams}, {Afanasev}, {Beichman}, {Bottom}, {Bowler}, {Brinkworth}, {Brown},
  {Cancino}, {Ciardi}, {Clampin}, {Clark}, {Collins}, {Davison},
  {Foreman-Mackey}, {Furlan}, {Gaidos}, {Geneser}, {Giddens}, {Gilbert},
  {Hall}, {Hellier}, {Henry}, {Horner}, {Howard}, {Huang}, {Huber}, {Kane},
  {Kenworthy}, {Kielkopf}, {Kipping}, {Klenke}, {Kruse}, {Latouf}, {Lowrance},
  {Mennesson}, {Mengel}, {Mills}, {Morton}, {Narita}, {Newton}, {Nishimoto},
  {Okumura}, {Palle}, {Pepper}, {Quintana}, {Roberge}, {Roccatagliata},
  {Schlieder}, {Tanner}, {Teske}, {Tinney}, {Vanderburg}, {von Braun}, {Walp},
  {Wang}, {Wang}, {Weigand}, {White}, {Wittenmyer}, {Wright}, {Youngblood},
  {Zhang}, \& {Zilberman}}]{2020Natur.582..497P}
{Plavchan}, P., {Barclay}, T., {Gagn{\'e}}, J., {et~al.} 2020,
  \href{https://ui.adsabs.harvard.edu/abs/2020Natur.582..497P}{\nat, 582, 497}

\bibitem[{{Quirrenbach} {et~al.}(2016){Quirrenbach}, {Amado}, {Caballero},
  {Mundt}, {Reiners}, {Ribas}, {Seifert}, {Abril}, {Aceituno},
  {Alonso-Floriano}, {Anwand-Heerwart}, {Azzaro}, {Bauer}, {Barrado},
  {Becerril}, {Bejar}, {Benitez}, {Berdinas}, {Brinkm{\"o}ller}, {Cardenas},
  {Casal}, {Claret}, {Colom{\'e}}, {Cortes-Contreras}, {Czesla}, {Doellinger},
  {Dreizler}, {Feiz}, {Fernandez}, {Ferro}, {Fuhrmeister}, {Galadi},
  {Gallardo}, {G{\'a}lvez-Ortiz}, {Garcia-Piquer}, {Garrido}, {Gesa},
  {G{\'o}mez Galera}, {Gonz{\'a}lez Hern{\'a}ndez}, {Gonzalez Peinado},
  {Gr{\"o}zinger}, {Gu{\`a}rdia}, {Guenther}, {de Guindos}, {Hagen}, {Hatzes},
  {Hauschildt}, {Helmling}, {Henning}, {Hermann}, {Hern{\'a}ndez Arabi},
  {Hern{\'a}ndez Casta{\~n}o}, {Hern{\'a}ndez Hernando}, {Herrero}, {Huber},
  {Huber}, {Huke}, {Jeffers}, {de Juan}, {Kaminski}, {Kehr}, {Kim}, {Klein},
  {Kl{\"u}ter}, {K{\"u}rster}, {Lafarga}, {Lara}, {Lamert}, {Laun},
  {Launhardt}, {Lemke}, {Lenzen}, {Llamas}, {Lopez del Fresno},
  {L{\'o}pez-Puertas}, {L{\'o}pez-Santiago}, {Lopez Salas}, {Magan
  Madinabeitia}, {Mall}, {Mandel}, {Mancini}, {Marin Molina}, {Maroto
  Fern{\'a}ndez}, {Mart{\'\i}n}, {Mart{\'\i}n-Ruiz}, {Marvin}, {Mathar},
  {Mirabet}, {Montes}, {Morales}, {Morales Mu{\~n}oz}, {Nagel}, {Naranjo},
  {Nowak}, {Palle}, {Panduro}, {Passegger}, {Pavlov}, {Pedraz}, {Perez},
  {P{\'e}rez-Medialdea}, {Perger}, {Pluto}, {Ram{\'o}n}, {Rebolo}, {Redondo},
  {Reffert}, {Reinhart}, {Rhode}, {Rix}, {Rodler}, {Rodr{\'\i}guez},
  {Rodr{\'\i}guez L{\'o}pez}, {Rohloff}, {Rosich}, {Sanchez Carrasco},
  {Sanz-Forcada}, {Sarkis}, {Sarmiento}, {Sch{\"a}fer}, {Schiller}, {Schmidt},
  {Schmitt}, {Sch{\"o}fer}, {Schweitzer}, {Shulyak}, {Solano}, {Stahl},
  {Storz}, {Tabernero}, {Tala}, {Tal-Or}, {Ulbrich}, {Veredas}, {Vico Linares},
  {Vilardell}, {Wagner}, {Winkler}, {Zapatero Osorio}, {Zechmeister},
  {Ammler-von Eiff}, {Anglada-Escud{\'e}}, {del Burgo}, {Garcia-Vargas},
  {Klutsch}, {Lizon}, {Lopez-Morales}, {Ofir}, {P{\'e}rez-Calpena}, {Perryman},
  {S{\'a}nchez-Blanco}, {Strachan}, {St{\"u}rmer}, {Su{\'a}rez}, {Trifonov},
  {Tulloch}, \& {Xu}}]{quirrenbach2016}
{Quirrenbach}, A., {Amado}, P.~J., {Caballero}, J.~A., {et~al.} 2016, in
  \procspie, Vol. 9908, Ground-based and Airborne Instrumentation for Astronomy
  VI, ed. C.~J. {Evans}, L.~{Simard}, \& H.~{Takami}, 990812

\bibitem[{{Quirrenbach} {et~al.}(2022){Quirrenbach}, {Passegger}, {Trifonov},
  {Amado}, {Caballero}, {Reiners}, {Ribas}, {Aceituno}, {B{\'e}jar},
  {Chaturvedi}, {Gonz{\'a}lez-Cuesta}, {Henning}, {Herrero}, {Kaminski},
  {K{\"u}rster}, {Lalitha}, {Lodieu}, {L{\'o}pez-Gonz{\'a}lez}, {Montes},
  {Pall{\'e}}, {Perger}, {Pollacco}, {Reffert}, {Rodr{\'\i}guez}, {L{\'o}pez},
  {Shan}, {Tal-Or}, {Osorio}, \& {Zechmeister}}]{2022A&A...663A..48Q}
{Quirrenbach}, A., {Passegger}, V.~M., {Trifonov}, T., {et~al.} 2022,
  \href{https://ui.adsabs.harvard.edu/abs/2022A&A...663A..48Q}{\aap, 663, A48}

\bibitem[{{Raetz} {et~al.}(2020){Raetz}, {Stelzer}, {Damasso}, \&
  {Scholz}}]{2020A&A...637A..22R}
{Raetz}, S., {Stelzer}, B., {Damasso}, M., \& {Scholz}, A. 2020,
  \href{https://ui.adsabs.harvard.edu/abs/2020A&A...637A..22R}{\aap, 637, A22}

\bibitem[{{Rajpaul} {et~al.}(2015){Rajpaul}, {Aigrain}, {Osborne}, {Reece}, \&
  {Roberts}}]{rajpaul2015}
{Rajpaul}, V., {Aigrain}, S., {Osborne}, M.~A., {Reece}, S., \& {Roberts}, S.
  2015, \href{https://ui.adsabs.harvard.edu/abs/2015MNRAS.452.2269R}{\mnras,
  452, 2269}

\bibitem[{{Reiners}(2012)}]{2012LRSP....9....1R}
{Reiners}, A. 2012,
  \href{https://ui.adsabs.harvard.edu/abs/2012LRSP....9....1R}{Living Reviews
  in Solar Physics, 9, 1}

\bibitem[{{Reiners} \& {Basri}(2007)}]{reiners2007}
{Reiners}, A. \& {Basri}, G. 2007,
  \href{https://ui.adsabs.harvard.edu/abs/2007ApJ...656.1121R}{\apj, 656, 1121}

\bibitem[{{Reiners} {et~al.}(2009){Reiners}, {Basri}, \&
  {Browning}}]{reiners2009}
{Reiners}, A., {Basri}, G., \& {Browning}, M. 2009,
  \href{https://ui.adsabs.harvard.edu/abs/2009ApJ...692..538R}{\apj, 692, 538}

\bibitem[{{Reiners} {et~al.}(2010){Reiners}, {Bean}, {Huber}, {Dreizler},
  {Seifahrt}, \& {Czesla}}]{reiners2010}
{Reiners}, A., {Bean}, J.~L., {Huber}, K.~F., {et~al.} 2010,
  \href{https://ui.adsabs.harvard.edu/abs/2010ApJ...710..432R}{\apj, 710, 432}

\bibitem[{{Reiners} {et~al.}(2012){Reiners}, {Joshi}, \&
  {Goldman}}]{2012AJ....143...93R}
{Reiners}, A., {Joshi}, N., \& {Goldman}, B. 2012,
  \href{https://ui.adsabs.harvard.edu/abs/2012AJ....143...93R}{\aj, 143, 93}

\bibitem[{{Reiners} {et~al.}(2018{\natexlab{a}}){Reiners}, {Ribas},
  {Zechmeister}, {Caballero}, {Trifonov}, {Dreizler}, {Morales}, {Tal-Or},
  {Lafarga}, {Quirrenbach}, {Amado}, {Kaminski}, {Jeffers}, {Aceituno},
  {B{\'e}jar}, {Gu{\`a}rdia}, {Guenther}, {Hagen}, {Montes}, {Passegger},
  {Seifert}, {Schweitzer}, {Cort{\'e}s-Contreras}, {Abril}, {Alonso-Floriano},
  {Ammler-von Eiff}, {Antona}, {Anglada-Escud{\'e}}, {Anwand-Heerwart},
  {Arroyo-Torres}, {Azzaro}, {Baroch}, {Barrado}, {Bauer}, {Becerril},
  {Ben{\'\i}tez}, {Berdi{\~n}as}, {Bergond}, {Bl{\"u}mcke}, {Brinkm{\"o}ller},
  {del Burgo}, {Cano}, {C{\'a}rdenas V{\'a}zquez}, {Casal}, {Cifuentes},
  {Claret}, {Colom{\'e}}, {Czesla}, {D{\'\i}ez-Alonso}, {Feiz},
  {Fern{\'a}ndez}, {Ferro}, {Fuhrmeister}, {Galad{\'\i}-Enr{\'\i}quez},
  {Garcia-Piquer}, {Garc{\'\i}a Vargas}, {Gesa}, {G{\'o}mez Galera},
  {Gonz{\'a}lez Hern{\'a}ndez}, {Gonz{\'a}lez-Peinado}, {Gr{\"o}zinger},
  {Grohnert}, {Guijarro}, {de Guindos}, {Guti{\'e}rrez-Soto}, {Hatzes},
  {Hauschildt}, {Hedrosa}, {Helmling}, {Henning}, {Hermelo}, {Hern{\'a}ndez
  Arab{\'\i}}, {Hern{\'a}ndez Casta{\~n}o}, {Hern{\'a}ndez Hernando},
  {Herrero}, {Huber}, {Huke}, {Johnson}, {de Juan}, {Kim}, {Klein},
  {Kl{\"u}ter}, {Klutsch}, {K{\"u}rster}, {Labarga}, {Lamert}, {Lamp{\'o}n},
  {Lara}, {Laun}, {Lemke}, {Lenzen}, {Launhardt}, {L{\'o}pez del Fresno},
  {L{\'o}pez-Gonz{\'a}lez}, {L{\'o}pez-Puertas}, {L{\'o}pez Salas},
  {L{\'o}pez-Santiago}, {Luque}, {Mag{\'a}n Madinabeitia}, {Mall}, {Mancini},
  {Mandel}, {Marfil}, {Mar{\'\i}n Molina}, {Maroto Fern{\'a}ndez},
  {Mart{\'\i}n}, {Mart{\'\i}n-Ruiz}, {Marvin}, {Mathar}, {Mirabet},
  {Moreno-Raya}, {Moya}, {Mundt}, {Nagel}, {Naranjo}, {Nortmann}, {Nowak},
  {Ofir}, {Oreiro}, {Pall{\'e}}, {Panduro}, {Pascual}, {Pavlov}, {Pedraz},
  {P{\'e}rez-Calpena}, {P{\'e}rez Medialdea}, {Perger}, {Perryman}, {Pluto},
  {Rabaza}, {Ram{\'o}n}, {Rebolo}, {Redondo}, {Reffert}, {Reinhart}, {Rhode},
  {Rix}, {Rodler}, {Rodr{\'\i}guez}, {Rodr{\'\i}guez-L{\'o}pez},
  {Rodr{\'\i}guez Trinidad}, {Rohloff}, {Rosich}, {Sadegi},
  {S{\'a}nchez-Blanco}, {S{\'a}nchez Carrasco}, {S{\'a}nchez-L{\'o}pez},
  {Sanz-Forcada}, {Sarkis}, {Sarmiento}, {Sch{\"a}fer}, {Schmitt}, {Schiller},
  {Sch{\"o}fer}, {Solano}, {Stahl}, {Strachan}, {St{\"u}rmer}, {Su{\'a}rez},
  {Tabernero}, {Tala}, {Tulloch}, {Ulbrich}, {Veredas}, {Vico Linares},
  {Vilardell}, {Wagner}, {Winkler}, {Wolthoff}, {Xu}, {Yan}, \& {Zapatero
  Osorio}}]{2018A&A...609L...5R}
{Reiners}, A., {Ribas}, I., {Zechmeister}, M., {et~al.} 2018{\natexlab{a}},
  \href{https://ui.adsabs.harvard.edu/abs/2018A&A...609L...5R}{\aap, 609, L5}

\bibitem[{{Reiners} {et~al.}(2022){Reiners}, {Shulyak}, {K{\"a}pyl{\"a}},
  {Ribas}, {Nagel}, {Zechmeister}, {Caballero}, {Shan}, {Fuhrmeister},
  {Quirrenbach}, {Amado}, {Montes}, {Jeffers}, {Azzaro}, {B{\'e}jar},
  {Chaturvedi}, {Henning}, {K{\"u}rster}, \& {Pall{\'e}}}]{reiners2022}
{Reiners}, A., {Shulyak}, D., {K{\"a}pyl{\"a}}, P.~J., {et~al.} 2022,
  \href{https://ui.adsabs.harvard.edu/abs/2022A&A...662A..41R}{\aap, 662, A41}

\bibitem[{{Reiners} {et~al.}(2018{\natexlab{b}}){Reiners}, {Zechmeister},
  {Caballero}, {Ribas}, {Morales}, {Jeffers}, {Sch{\"o}fer}, {Tal-Or},
  {Quirrenbach}, {Amado}, {Kaminski}, {Seifert}, {Abril}, {Aceituno},
  {Alonso-Floriano}, {Ammler-von Eiff}, {Antona}, {Anglada-Escud{\'e}},
  {Anwand-Heerwart}, {Arroyo-Torres}, {Azzaro}, {Baroch}, {Barrado}, {Bauer},
  {Becerril}, {B{\'e}jar}, {Ben{\'\i}tez}, {Berdinas}, {Bergond},
  {Bl{\"u}mcke}, {Brinkm{\"o}ller}, {del Burgo}, {Cano}, {C{\'a}rdenas
  V{\'a}zquez}, {Casal}, {Cifuentes}, {Claret}, {Colom{\'e}},
  {Cort{\'e}s-Contreras}, {Czesla}, {D{\'\i}ez-Alonso}, {Dreizler}, {Feiz},
  {Fern{\'a}ndez}, {Ferro}, {Fuhrmeister}, {Galad{\'\i}-Enr{\'\i}quez},
  {Garcia-Piquer}, {Garc{\'\i}a Vargas}, {Gesa}, {G{\'o}mez Galera},
  {Gonz{\'a}lez Hern{\'a}ndez}, {Gonz{\'a}lez-Peinado}, {Gr{\"o}zinger},
  {Grohnert}, {Gu{\`a}rdia}, {Guenther}, {Guijarro}, {de Guindos},
  {Guti{\'e}rrez-Soto}, {Hagen}, {Hatzes}, {Hauschildt}, {Hedrosa}, {Helmling},
  {Henning}, {Hermelo}, {Hern{\'a}ndez Arab{\'\i}}, {Hern{\'a}ndez
  Casta{\~n}o}, {Hern{\'a}ndez Hernando}, {Herrero}, {Huber}, {Huke},
  {Johnson}, {de Juan}, {Kim}, {Klein}, {Kl{\"u}ter}, {Klutsch}, {K{\"u}rster},
  {Lafarga}, {Lamert}, {Lamp{\'o}n}, {Lara}, {Laun}, {Lemke}, {Lenzen},
  {Launhardt}, {L{\'o}pez del Fresno}, {L{\'o}pez-Gonz{\'a}lez},
  {L{\'o}pez-Puertas}, {L{\'o}pez Salas}, {L{\'o}pez-Santiago}, {Luque},
  {Mag{\'a}n Madinabeitia}, {Mall}, {Mancini}, {Mandel}, {Marfil}, {Mar{\'\i}n
  Molina}, {Maroto Fern{\'a}ndez}, {Mart{\'\i}n}, {Mart{\'\i}n-Ruiz}, {Marvin},
  {Mathar}, {Mirabet}, {Montes}, {Moreno-Raya}, {Moya}, {Mundt}, {Nagel},
  {Naranjo}, {Nortmann}, {Nowak}, {Ofir}, {Oreiro}, {Pall{\'e}}, {Panduro},
  {Pascual}, {Passegger}, {Pavlov}, {Pedraz}, {P{\'e}rez-Calpena}, {P{\'e}rez
  Medialdea}, {Perger}, {Perryman}, {Pluto}, {Rabaza}, {Ram{\'o}n}, {Rebolo},
  {Redondo}, {Reffert}, {Reinhart}, {Rhode}, {Rix}, {Rodler}, {Rodr{\'\i}guez},
  {Rodr{\'\i}guez-L{\'o}pez}, {Rodr{\'\i}guez Trinidad}, {Rohloff}, {Rosich},
  {Sadegi}, {S{\'a}nchez-Blanco}, {S{\'a}nchez Carrasco},
  {S{\'a}nchez-L{\'o}pez}, {Sanz-Forcada}, {Sarkis}, {Sarmiento},
  {Sch{\"a}fer}, {Schmitt}, {Schiller}, {Schweitzer}, {Solano}, {Stahl},
  {Strachan}, {St{\"u}rmer}, {Su{\'a}rez}, {Tabernero}, {Tala}, {Trifonov},
  {Tulloch}, {Ulbrich}, {Veredas}, {Vico Linares}, {Vilardell}, {Wagner},
  {Winkler}, {Wolthoff}, {Xu}, {Yan}, \& {Zapatero Osorio}}]{reiners2018}
{Reiners}, A., {Zechmeister}, M., {Caballero}, J.~A., {et~al.}
  2018{\natexlab{b}},
  \href{https://ui.adsabs.harvard.edu/abs/2018A&A...612A..49R}{\aap, 612, A49}

\bibitem[{{Ribas} {et~al.}(2023){Ribas}, {Reiners}, {Zechmeister}, {Caballero},
  {Morales}, {Sabotta}, {Baroch}, {Amado}, {Quirrenbach}, {Abril}, {Aceituno},
  {Anglada-Escud{\'e}}, {Azzaro}, {Barrado}, {B{\'e}jar}, {Ben{\'\i}tez de
  Haro}, {Bergond}, {Bluhm}, {Calvo Ortega}, {Cardona Guill{\'e}n},
  {Chaturvedi}, {Cifuentes}, {Colom{\'e}}, {Cont}, {Cort{\'e}s-Contreras},
  {Czesla}, {D{\'\i}ez-Alonso}, {Dreizler}, {Duque-Arribas}, {Espinoza},
  {Fern{\'a}ndez}, {Fuhrmeister}, {Galad{\'\i}-Enr{\'\i}quez},
  {Garc{\'\i}a-L{\'o}pez}, {Gonz{\'a}lez-{\'A}lvarez}, {Gonz{\'a}lez
  Hern{\'a}ndez}, {Guenther}, {de Guindos}, {Hatzes}, {Henning}, {Herrero},
  {Hintz}, {Huelmo}, {Jeffers}, {Johnson}, {de Juan}, {Kaminski}, {Kemmer},
  {Khaimova}, {Khalafinejad}, {Kossakowski}, {K{\"u}rster}, {Labarga},
  {Lafarga}, {Lalitha}, {Lamp{\'o}n}, {Lillo-Box}, {Lodieu}, {L{\'o}pez
  Gonz{\'a}lez}, {L{\'o}pez-Puertas}, {Luque}, {Mag{\'a}n}, {Mancini},
  {Marfil}, {Mart{\'\i}n}, {Mart{\'\i}n-Ruiz}, {Molaverdikhani}, {Montes},
  {Nagel}, {Nortmann}, {Nowak}, {Pall{\'e}}, {Passegger}, {Pavlov}, {Pedraz},
  {Perdelwitz}, {Perger}, {Ram{\'o}n-Ballesta}, {Reffert}, {Revilla},
  {Rodr{\'\i}guez}, {Rodr{\'\i}guez-L{\'o}pez}, {Sadegi}, {S{\'a}nchez
  Carrasco}, {S{\'a}nchez-L{\'o}pez}, {Sanz-Forcada}, {Sch{\"a}fer},
  {Schlecker}, {Schmitt}, {Sch{\"o}fer}, {Schweitzer}, {Seifert}, {Shan},
  {Skrzypinski}, {Solano}, {Stahl}, {Stangret}, {Stock}, {St{\"u}rmer},
  {Tabernero}, {Tal-Or}, {Trifonov}, {Vanaverbeke}, {Yan}, \& {Zapatero
  Osorio}}]{ribas2023}
{Ribas}, I., {Reiners}, A., {Zechmeister}, M., {et~al.} 2023,
  \href{https://ui.adsabs.harvard.edu/abs/2023A&A...670A.139R}{\aap, 670, A139}

\bibitem[{{Ribas} {et~al.}(2018){Ribas}, {Tuomi}, {Reiners}, {Butler},
  {Morales}, {Perger}, {Dreizler}, {Rodr{\'\i}guez-L{\'o}pez}, {Gonz{\'a}lez
  Hern{\'a}ndez}, {Rosich}, {Feng}, {Trifonov}, {Vogt}, {Caballero}, {Hatzes},
  {Herrero}, {Jeffers}, {Lafarga}, {Murgas}, {Nelson}, {Rodr{\'\i}guez},
  {Strachan}, {Tal-Or}, {Teske}, {Toledo-Padr{\'o}n}, {Zechmeister},
  {Quirrenbach}, {Amado}, {Azzaro}, {B{\'e}jar}, {Barnes}, {Berdi{\~n}as},
  {Burt}, {Coleman}, {Cort{\'e}s-Contreras}, {Crane}, {Engle}, {Guinan},
  {Haswell}, {Henning}, {Holden}, {Jenkins}, {Jones}, {Kaminski}, {Kiraga},
  {K{\"u}rster}, {Lee}, {L{\'o}pez-Gonz{\'a}lez}, {Montes}, {Morin}, {Ofir},
  {Pall{\'e}}, {Rebolo}, {Reffert}, {Schweitzer}, {Seifert}, {Shectman},
  {Staab}, {Street}, {Su{\'a}rez Mascare{\~n}o}, {Tsapras}, {Wang}, \&
  {Anglada-Escud{\'e}}}]{2018Natur.563..365R}
{Ribas}, I., {Tuomi}, M., {Reiners}, A., {et~al.} 2018,
  \href{https://ui.adsabs.harvard.edu/abs/2018Natur.563..365R}{\nat, 563, 365}

\bibitem[{{Ricker} {et~al.}(2015){Ricker}, {Winn}, {Vanderspek}, {Latham},
  {Bakos}, {Bean}, {Berta-Thompson}, {Brown}, {Buchhave}, {Butler}, {Butler},
  {Chaplin}, {Charbonneau}, {Christensen-Dalsgaard}, {Clampin}, {Deming},
  {Doty}, {De Lee}, {Dressing}, {Dunham}, {Endl}, {Fressin}, {Ge}, {Henning},
  {Holman}, {Howard}, {Ida}, {Jenkins}, {Jernigan}, {Johnson}, {Kaltenegger},
  {Kawai}, {Kjeldsen}, {Laughlin}, {Levine}, {Lin}, {Lissauer}, {MacQueen},
  {Marcy}, {McCullough}, {Morton}, {Narita}, {Paegert}, {Palle}, {Pepe},
  {Pepper}, {Quirrenbach}, {Rinehart}, {Sasselov}, {Sato}, {Seager},
  {Sozzetti}, {Stassun}, {Sullivan}, {Szentgyorgyi}, {Torres}, {Udry}, \&
  {Villasenor}}]{2015JATIS...1a4003R}
{Ricker}, G.~R., {Winn}, J.~N., {Vanderspek}, R., {et~al.} 2015,
  \href{https://ui.adsabs.harvard.edu/abs/2015JATIS...1a4003R}{JATIS, 1,
  014003}

\bibitem[{{Rivera} {et~al.}(2010){Rivera}, {Laughlin}, {Butler}, {Vogt},
  {Haghighipour}, \& {Meschiari}}]{2010ApJ...719..890R}
{Rivera}, E.~J., {Laughlin}, G., {Butler}, R.~P., {et~al.} 2010,
  \href{https://ui.adsabs.harvard.edu/abs/2010ApJ...719..890R}{\apj, 719, 890}

\bibitem[{{Rivera} {et~al.}(2005){Rivera}, {Lissauer}, {Butler}, {Marcy},
  {Vogt}, {Fischer}, {Brown}, {Laughlin}, \& {Henry}}]{2005ApJ...634..625R}
{Rivera}, E.~J., {Lissauer}, J.~J., {Butler}, R.~P., {et~al.} 2005,
  \href{https://ui.adsabs.harvard.edu/abs/2005ApJ...634..625R}{\apj, 634, 625}

\bibitem[{{Saar} {et~al.}(1998){Saar}, {Butler}, \& {Marcy}}]{saar1998}
{Saar}, S.~H., {Butler}, R.~P., \& {Marcy}, G.~W. 1998,
  \href{https://ui.adsabs.harvard.edu/abs/1998ApJ...498L.153S}{\apjl, 498,
  L153}

\bibitem[{{Saar} \& {Donahue}(1997)}]{saar1997}
{Saar}, S.~H. \& {Donahue}, R.~A. 1997,
  \href{https://ui.adsabs.harvard.edu/abs/1997ApJ...485..319S}{\apj, 485, 319}

\bibitem[{{Sch{\"a}fer} {et~al.}(2018){Sch{\"a}fer}, {Guenther}, {Reiners},
  {Winkler}, {Pluto}, \& {Schiller}}]{2018schaefer}
{Sch{\"a}fer}, S., {Guenther}, E.~W., {Reiners}, A., {et~al.} 2018, in
  \procspie, Vol. 10702, Ground-based and Airborne Instrumentation for
  Astronomy VII, ed. C.~J. {Evans}, L.~{Simard}, \& H.~{Takami}, 1070276

\bibitem[{{Sch{\"o}fer} {et~al.}(2022){Sch{\"o}fer}, {Jeffers}, {Reiners},
  {Zechmeister}, {Fuhrmeister}, {Lafarga}, {Ribas}, {Quirrenbach}, {Amado},
  {Caballero}, {Anglada-Escud{\'e}}, {Bauer}, {B{\'e}jar},
  {Cort{\'e}s-Contreras}, {Alonso}, {Dreizler}, {Guenther}, {Herbort},
  {Johnson}, {Kaminski}, {K{\"u}rster}, {Montes}, {Morales}, {Pedraz}, \&
  {Tal-Or}}]{schoefer2022}
{Sch{\"o}fer}, P., {Jeffers}, S.~V., {Reiners}, A., {et~al.} 2022,
  \href{https://ui.adsabs.harvard.edu/abs/2022A&A...663A..68S}{\aap, 663, A68}

\bibitem[{{Schwarz}(1978)}]{schwarz1978}
{Schwarz}, G. 1978,
  \href{https://ui.adsabs.harvard.edu/abs/1978AnSta...6..461S}{Ann. Stat., 6,
  461}

\bibitem[{{Schweitzer} {et~al.}(2019){Schweitzer}, {Passegger}, {Cifuentes},
  {B{\'e}jar}, {Cort{\'e}s-Contreras}, {Caballero}, {del Burgo}, {Czesla},
  {K{\"u}rster}, {Montes}, {Zapatero Osorio}, {Ribas}, {Reiners},
  {Quirrenbach}, {Amado}, {Aceituno}, {Anglada-Escud{\'e}}, {Bauer},
  {Dreizler}, {Jeffers}, {Guenther}, {Henning}, {Kaminski}, {Lafarga},
  {Marfil}, {Morales}, {Schmitt}, {Seifert}, {Solano}, {Tabernero}, \&
  {Zechmeister}}]{schweitzer2019}
{Schweitzer}, A., {Passegger}, V.~M., {Cifuentes}, C., {et~al.} 2019,
  \href{https://ui.adsabs.harvard.edu/abs/2019A&A...625A..68S}{\aap, 625, A68}

\bibitem[{{See} {et~al.}(2019){See}, {Matt}, {Folsom}, {Boro Saikia}, {Donati},
  {Fares}, {Finley}, {H{\'e}brard}, {Jardine}, {Jeffers}, {Lehmann}, {Marsden},
  {Mengel}, {Morin}, {Petit}, {Vidotto}, {Waite}, \& {BCool
  Collaboration}}]{see2019}
{See}, V., {Matt}, S.~P., {Folsom}, C.~P., {et~al.} 2019,
  \href{https://ui.adsabs.harvard.edu/abs/2019ApJ...876..118S}{\apj, 876, 118}

\bibitem[{{Seifahrt} {et~al.}(2022){Seifahrt}, {Bean}, {Kasper}, {St{\"u}rmer},
  {Brady}, {Liu}, {Zechmeister}, {Stef{\'a}nsson}, {Montet}, {White}, {Tapia},
  {Mocnik}, {Xu}, \& {Schwab}}]{seifahrt2022}
{Seifahrt}, A., {Bean}, J.~L., {Kasper}, D., {et~al.} 2022, in \procspie, Vol.
  12184, Ground-based and Airborne Instrumentation for Astronomy IX, ed. C.~J.
  {Evans}, J.~J. {Bryant}, \& K.~{Motohara}, 121841G

\bibitem[{{Shan} {et~al.}(2024){Shan}, {Revilla}, {Skrzypinski}, {Dreizler},
  {B{\'e}jar}, {Caballero}, {Cardona Guill{\'e}n}, {Cifuentes}, {Fuhrmeister},
  {Reiners}, {Vanaverbeke}, {Ribas}, {Quirrenbach}, {Amado}, {Aceituno},
  {Casanova}, {Cort{\'e}s-Contreras}, {Dubois}, {Gorrini}, {Henning},
  {Herrero}, {Jeffers}, {Kemmer}, {Lalitha}, {Lodieu}, {Logie}, {L{\'o}pez
  Gonz{\'a}lez}, {Mart{\'\i}n-Ruiz}, {Montes}, {Morales}, {Nagel}, {Pall{\'e}},
  {Perdelwitz}, {P{\'e}rez-Torres}, {Pollacco}, {Rau},
  {Rodr{\'\i}guez-L{\'o}pez}, {Rodr{\'\i}guez}, {Sch{\"o}fer}, {Seifert},
  {Sota}, {Zapatero Osorio}, \& {Zechmeister}}]{2024A&A...684A...9S}
{Shan}, Y., {Revilla}, D., {Skrzypinski}, S.~L., {et~al.} 2024,
  \href{https://ui.adsabs.harvard.edu/abs/2024A&A...684A...9S}{\aap, 684, A9}

\bibitem[{{Shulyak} {et~al.}(2019){Shulyak}, {Reiners}, {Nagel}, {Tal-Or},
  {Caballero}, {Zechmeister}, {B{\'e}jar}, {Cort{\'e}s-Contreras}, {Martin},
  {Kaminski}, {Ribas}, {Quirrenbach}, {Amado}, {Anglada-Escud{\'e}}, {Bauer},
  {Dreizler}, {Guenther}, {Henning}, {Jeffers}, {K{\"u}rster}, {Lafarga},
  {Montes}, {Morales}, \& {Pedraz}}]{shulyak2019}
{Shulyak}, D., {Reiners}, A., {Nagel}, E., {et~al.} 2019,
  \href{https://ui.adsabs.harvard.edu/abs/2019A&A...626A..86S}{\aap, 626, A86}

\bibitem[{{Skumanich}(1972)}]{skumanich1972}
{Skumanich}, A. 1972,
  \href{https://ui.adsabs.harvard.edu/abs/1972ApJ...171..565S}{\apj, 171, 565}

\bibitem[{{Soto} {et~al.}(2021){Soto}, {Anglada-Escud{\'e}}, {Dreizler},
  {Molaverdikhani}, {Kemmer}, {Rodr{\'\i}guez-L{\'o}pez}, {Lillo-Box},
  {Pall{\'e}}, {Espinoza}, {Caballero}, {Quirrenbach}, {Ribas}, {Reiners},
  {Narita}, {Hirano}, {Amado}, {B{\'e}jar}, {Bluhm}, {Burke}, {Caldwell},
  {Charbonneau}, {Cloutier}, {Collins}, {Cort{\'e}s-Contreras}, {Girardin},
  {Guerra}, {Harakawa}, {Hatzes}, {Irwin}, {Jenkins}, {Jensen}, {Kawauchi},
  {Kotani}, {Kudo}, {Kunimoto}, {Kuzuhara}, {Latham}, {Montes}, {Morales},
  {Mori}, {Nelson}, {Omiya}, {Pedraz}, {Passegger}, {Rackham}, {Rudat},
  {Schlieder}, {Sch{\"o}fer}, {Schweitzer}, {Selezneva}, {Stockdale}, {Tamura},
  {Trifonov}, {Vanderspek}, \& {Watanabe}}]{2021A&A...649A.144S}
{Soto}, M.~G., {Anglada-Escud{\'e}}, G., {Dreizler}, S., {et~al.} 2021,
  \href{https://ui.adsabs.harvard.edu/abs/2021A&A...649A.144S}{\aap, 649, A144}

\bibitem[{{Stock} {et~al.}(2020){Stock}, {Nagel}, {Kemmer}, {Passegger},
  {Reffert}, {Quirrenbach}, {Caballero}, {Czesla}, {B{\'e}jar}, {Cardona},
  {D{\'\i}ez-Alonso}, {Herrero}, {Lalitha}, {Schlecker}, {Tal-Or},
  {Rodr{\'\i}guez}, {Rodr{\'\i}guez-L{\'o}pez}, {Ribas}, {Reiners}, {Amado},
  {Bauer}, {Bluhm}, {Cort{\'e}s-Contreras}, {Gonz{\'a}lez-Cuesta}, {Dreizler},
  {Hatzes}, {Henning}, {Jeffers}, {Kaminski}, {K{\"u}rster}, {Lafarga},
  {L{\'o}pez-Gonz{\'a}lez}, {Montes}, {Morales}, {Pedraz}, {Sch{\"o}fer},
  {Schweitzer}, {Trifonov}, {Zapatero Osorio}, \&
  {Zechmeister}}]{2020A&A...643A.112S}
{Stock}, S., {Nagel}, E., {Kemmer}, J., {et~al.} 2020,
  \href{https://ui.adsabs.harvard.edu/abs/2020A&A...643A.112S}{\aap, 643, A112}

\bibitem[{{Su{\'a}rez Mascare{\~n}o} {et~al.}(2023){Su{\'a}rez Mascare{\~n}o},
  {Gonz{\'a}lez-{\'A}lvarez}, {Zapatero Osorio}, {Lillo-Box}, {Faria},
  {Passegger}, {Gonz{\'a}lez Hern{\'a}ndez}, {Figueira}, {Sozzetti}, {Rebolo},
  {Pepe}, {Santos}, {Cristiani}, {Lovis}, {Silva}, {Ribas}, {Amado},
  {Caballero}, {Quirrenbach}, {Reiners}, {Zechmeister}, {Adibekyan}, {Alibert},
  {B{\'e}jar}, {Benatti}, {D'Odorico}, {Damasso}, {Delisle}, {Di Marcantonio},
  {Dreizler}, {Ehrenreich}, {Hatzes}, {Hara}, {Henning}, {Kaminski},
  {L{\'o}pez-Gonz{\'a}lez}, {Martins}, {Micela}, {Montes}, {Pall{\'e}},
  {Pedraz}, {Rodr{\'\i}guez}, {Rodr{\'\i}guez-L{\'o}pez}, {Tal-Or}, {Sousa}, \&
  {Udry}}]{mascareno2023}
{Su{\'a}rez Mascare{\~n}o}, A., {Gonz{\'a}lez-{\'A}lvarez}, E., {Zapatero
  Osorio}, M.~R., {et~al.} 2023,
  \href{https://ui.adsabs.harvard.edu/abs/2023A&A...670A...5S}{\aap, 670, A5}

\bibitem[{{Su{\'a}rez Mascare{\~n}o} {et~al.}(2017{\natexlab{a}}){Su{\'a}rez
  Mascare{\~n}o}, {Gonz{\'a}lez Hern{\'a}ndez}, {Rebolo}, {Astudillo-Defru},
  {Bonfils}, {Bouchy}, {Delfosse}, {Forveille}, {Lovis}, {Mayor}, {Murgas},
  {Pepe}, {Santos}, {Udry}, {W{\"u}nsche}, \& {Velasco}}]{2017A&A...597A.108S}
{Su{\'a}rez Mascare{\~n}o}, A., {Gonz{\'a}lez Hern{\'a}ndez}, J.~I., {Rebolo},
  R., {et~al.} 2017{\natexlab{a}},
  \href{https://ui.adsabs.harvard.edu/abs/2017A&A...597A.108S}{\aap, 597, A108}

\bibitem[{{Su{\'a}rez Mascare{\~n}o} {et~al.}(2017{\natexlab{b}}){Su{\'a}rez
  Mascare{\~n}o}, {Gonz{\'a}lez Hern{\'a}ndez}, {Rebolo}, {Velasco},
  {Toledo-Padr{\'o}n}, {Affer}, {Perger}, {Micela}, {Ribas}, {Maldonado},
  {Leto}, {Zanmar Sanchez}, {Scandariato}, {Damasso}, {Sozzetti}, {Esposito},
  {Covino}, {Maggio}, {Lanza}, {Desidera}, {Rosich}, {Bignamini}, {Claudi},
  {Benatti}, {Borsa}, {Pedani}, {Molinari}, {Morales}, {Herrero}, \&
  {Lafarga}}]{2017A&A...605A..92S}
{Su{\'a}rez Mascare{\~n}o}, A., {Gonz{\'a}lez Hern{\'a}ndez}, J.~I., {Rebolo},
  R., {et~al.} 2017{\natexlab{b}},
  \href{https://ui.adsabs.harvard.edu/abs/2017A&A...605A..92S}{\aap, 605, A92}

\bibitem[{{Su{\'a}rez Mascare{\~n}o} {et~al.}(2015){Su{\'a}rez Mascare{\~n}o},
  {Rebolo}, {Gonz{\'a}lez Hern{\'a}ndez}, \& {Esposito}}]{2015MNRAS.452.2745S}
{Su{\'a}rez Mascare{\~n}o}, A., {Rebolo}, R., {Gonz{\'a}lez Hern{\'a}ndez},
  J.~I., \& {Esposito}, M. 2015,
  \href{https://ui.adsabs.harvard.edu/abs/2015MNRAS.452.2745S}{\mnras, 452,
  2745}

\bibitem[{{Su{\'a}rez Mascare{\~n}o} {et~al.}(2017{\natexlab{c}}){Su{\'a}rez
  Mascare{\~n}o}, {Rebolo}, {Gonz{\'a}lez Hern{\'a}ndez}, \&
  {Esposito}}]{suarez2017}
{Su{\'a}rez Mascare{\~n}o}, A., {Rebolo}, R., {Gonz{\'a}lez Hern{\'a}ndez},
  J.~I., \& {Esposito}, M. 2017{\natexlab{c}},
  \href{https://ui.adsabs.harvard.edu/abs/2017MNRAS.468.4772S}{\mnras, 468,
  4772}

\bibitem[{{Su{\'a}rez Mascare{\~n}o} {et~al.}(2018){Su{\'a}rez Mascare{\~n}o},
  {Rebolo}, {Gonz{\'a}lez Hern{\'a}ndez}, {Toledo-Padr{\'o}n}, {Perger},
  {Ribas}, {Affer}, {Micela}, {Damasso}, {Maldonado}, {Gonz{\'a}lez-Alvarez},
  {Leto}, {Pagano}, {Scandariato}, {Sozzetti}, {Lanza}, {Malavolta}, {Claudi},
  {Cosentino}, {Desidera}, {Giacobbe}, {Maggio}, {Rainer}, {Esposito},
  {Benatti}, {Pedani}, {Morales}, {Herrero}, {Lafarga}, {Rosich}, \&
  {Pinamonti}}]{2018A&A...612A..89S}
{Su{\'a}rez Mascare{\~n}o}, A., {Rebolo}, R., {Gonz{\'a}lez Hern{\'a}ndez},
  J.~I., {et~al.} 2018,
  \href{https://ui.adsabs.harvard.edu/abs/2018A&A...612A..89S}{\aap, 612, A89}

\bibitem[{{Szentgyorgyi} \& {Fur{\'e}sz}(2007)}]{2007RMxAC..28..129S}
{Szentgyorgyi}, A.~H. \& {Fur{\'e}sz}, G. 2007,
  \href{https://ui.adsabs.harvard.edu/abs/2007RMxAC..28..129S}{\rmxaa, 28, 129}

\bibitem[{{Tal-Or} {et~al.}(2019){Tal-Or}, {Trifonov}, {Zucker}, {Mazeh}, \&
  {Zechmeister}}]{talor2019}
{Tal-Or}, L., {Trifonov}, T., {Zucker}, S., {Mazeh}, T., \& {Zechmeister}, M.
  2019, \href{https://ui.adsabs.harvard.edu/abs/2019MNRAS.484L...8T}{\mnras,
  484, L8}

\bibitem[{{Tal-Or} {et~al.}(2018){Tal-Or}, {Zechmeister}, {Reiners}, {Jeffers},
  {Sch{\"o}fer}, {Quirrenbach}, {Amado}, {Ribas}, {Caballero}, {Aceituno},
  {Bauer}, {B{\'e}jar}, {Czesla}, {Dreizler}, {Fuhrmeister}, {Hatzes},
  {Johnson}, {K{\"u}rster}, {Lafarga}, {Montes}, {Morales}, {Reffert},
  {Sadegi}, {Seifert}, \& {Shulyak}}]{talor2018}
{Tal-Or}, L., {Zechmeister}, M., {Reiners}, A., {et~al.} 2018,
  \href{https://ui.adsabs.harvard.edu/abs/2018A&A...614A.122T}{\aap, 614, A122}

\bibitem[{{Tamura} {et~al.}(2016){Tamura}, {Takato}, {Shimono}, {Moritani},
  {Yabe}, {Ishizuka}, {Ueda}, {Kamata}, {Aghazarian}, {Arnouts}, {Barban},
  {Barkhouser}, {Borges}, {Braun}, {Carr}, {Chabaud}, {Chang}, {Chen}, {Chiba},
  {Chou}, {Chu}, {Cohen}, {de Almeida}, {de Oliveira}, {de Oliveira}, {Dekany},
  {Dohlen}, {dos Santos}, {dos Santos}, {Ellis}, {Fabricius}, {Ferrand},
  {Ferreira}, {Golebiowski}, {Greene}, {Gross}, {Gunn}, {Hammond}, {Harding},
  {Hart}, {Heckman}, {Hirata}, {Ho}, {Hope}, {Hovland}, {Hsu}, {Hu}, {Huang},
  {Jaquet}, {Jing}, {Karr}, {Kimura}, {King}, {Komatsu}, {Le Brun}, {Le
  F{\`e}vre}, {Le Fur}, {Le Mignant}, {Ling}, {Loomis}, {Lupton}, {Madec},
  {Mao}, {Marrara}, {Mendes de Oliveira}, {Minowa}, {Morantz}, {Murayama},
  {Murray}, {Ohyama}, {Orndorff}, {Pascal}, {Pereira}, {Reiley}, {Reinecke},
  {Ritter}, {Roberts}, {Schwochert}, {Seiffert}, {Smee}, {Sodre}, {Spergel},
  {Steinkraus}, {Strauss}, {Surace}, {Suto}, {Suzuki}, {Swinbank}, {Tait},
  {Takada}, {Tamura}, {Tanaka}, {Tresse}, {Verducci}, {Vibert}, {Vidal},
  {Wang}, {Wen}, {Yan}, \& {Yasuda}}]{2016SPIE.9908E..1MT}
{Tamura}, N., {Takato}, N., {Shimono}, A., {et~al.} 2016, in \procspie, Vol.
  9908, Ground-based and Airborne Instrumentation for Astronomy VI, ed. C.~J.
  {Evans}, L.~{Simard}, \& H.~{Takami}, 99081M

\bibitem[{{Terrien} {et~al.}(2022){Terrien}, {Keen}, {Oda}, {Parts(they/them)},
  {Stef{\'a}nsson}, {Mahadevan}, {Robertson}, {Ninan}, {Beard}, {Bender},
  {Cochran}, {Cunha}, {Diddams}, {Fredrick}, {Halverson}, {Hearty}, {Ickler},
  {Kanodia}, {Libby-Roberts}, {Lubin}, {Metcalf}, {Olsen}, {Ramsey}, {Roy},
  {Schwab}, {Smith}, \& {Turner}}]{terrien2022}
{Terrien}, R.~C., {Keen}, A., {Oda}, K., {et~al.} 2022,
  \href{https://ui.adsabs.harvard.edu/abs/2022ApJ...927L..11T}{\apjl, 927, L11}

\bibitem[{{Tokovinin} {et~al.}(2013){Tokovinin}, {Fischer}, {Bonati},
  {Giguere}, {Moore}, {Schwab}, {Spronck}, \&
  {Szymkowiak}}]{2013PASP..125.1336T}
{Tokovinin}, A., {Fischer}, D.~A., {Bonati}, M., {et~al.} 2013,
  \href{https://ui.adsabs.harvard.edu/abs/2013PASP..125.1336T}{\pasp, 125,
  1336}

\bibitem[{{Toledo-Padr{\'o}n} {et~al.}(2021){Toledo-Padr{\'o}n}, {Su{\'a}rez
  Mascare{\~n}o}, {Gonz{\'a}lez Hern{\'a}ndez}, {Rebolo}, {Pinamonti},
  {Perger}, {Scandariato}, {Damasso}, {Sozzetti}, {Maldonado}, {Desidera},
  {Ribas}, {Micela}, {Affer}, {Gonz{\'a}lez-Alvarez}, {Leto}, {Pagano}, {Zanmar
  S{\'a}nchez}, {Giacobbe}, {Herrero}, {Morales}, {Amado}, {Caballero},
  {Quirrenbach}, {Reiners}, \& {Zechmeister}}]{2021A&A...648A..20T}
{Toledo-Padr{\'o}n}, B., {Su{\'a}rez Mascare{\~n}o}, A., {Gonz{\'a}lez
  Hern{\'a}ndez}, J.~I., {et~al.} 2021,
  \href{https://ui.adsabs.harvard.edu/abs/2021A&A...648A..20T}{\aap, 648, A20}

\bibitem[{{Trifonov} {et~al.}(2021){Trifonov}, {Caballero}, {Morales},
  {Seifahrt}, {Ribas}, {Reiners}, {Bean}, {Luque}, {Parviainen}, {Pall{\'e}},
  {Stock}, {Zechmeister}, {Amado}, {Anglada-Escud{\'e}}, {Azzaro}, {Barclay},
  {B{\'e}jar}, {Bluhm}, {Casasayas-Barris}, {Cifuentes}, {Collins}, {Collins},
  {Cort{\'e}s-Contreras}, {de Leon}, {Dreizler}, {Dressing}, {Esparza-Borges},
  {Espinoza}, {Fausnaugh}, {Fukui}, {Hatzes}, {Hellier}, {Henning}, {Henze},
  {Herrero}, {Jeffers}, {Jenkins}, {Jensen}, {Kaminski}, {Kasper},
  {Kossakowski}, {K{\"u}rster}, {Lafarga}, {Latham}, {Mann}, {Molaverdikhani},
  {Montes}, {Montet}, {Murgas}, {Narita}, {Oshagh}, {Passegger}, {Pollacco},
  {Quinn}, {Quirrenbach}, {Ricker}, {Rodr{\'\i}guez L{\'o}pez}, {Sanz-Forcada},
  {Schwarz}, {Schweitzer}, {Seager}, {Shporer}, {Stangret}, {St{\"u}rmer},
  {Tan}, {Tenenbaum}, {Twicken}, {Vanderspek}, \& {Winn}}]{2021Sci...371.1038T}
{Trifonov}, T., {Caballero}, J.~A., {Morales}, J.~C., {et~al.} 2021,
  \href{https://ui.adsabs.harvard.edu/abs/2021Sci...371.1038T}{Science, 371,
  1038}

\bibitem[{{Trifonov} {et~al.}(2018){Trifonov}, {K{\"u}rster}, {Zechmeister},
  {Tal-Or}, {Caballero}, {Quirrenbach}, {Amado}, {Ribas}, {Reiners}, {Reffert},
  {Dreizler}, {Hatzes}, {Kaminski}, {Launhardt}, {Henning}, {Montes},
  {B{\'e}jar}, {Mundt}, {Pavlov}, {Schmitt}, {Seifert}, {Morales}, {Nowak},
  {Jeffers}, {Rodr{\'\i}guez-L{\'o}pez}, {del Burgo}, {Anglada-Escud{\'e}},
  {L{\'o}pez-Santiago}, {Mathar}, {Ammler-von Eiff}, {Guenther}, {Barrado},
  {Gonz{\'a}lez Hern{\'a}ndez}, {Mancini}, {St{\"u}rmer}, {Abril}, {Aceituno},
  {Alonso-Floriano}, {Antona}, {Anwand-Heerwart}, {Arroyo-Torres}, {Azzaro},
  {Baroch}, {Bauer}, {Becerril}, {Ben{\'\i}tez}, {Berdi{\~n}as}, {Bergond},
  {Bl{\"u}mcke}, {Brinkm{\"o}ller}, {Cano}, {C{\'a}rdenas V{\'a}zquez},
  {Casal}, {Cifuentes}, {Claret}, {Colom{\'e}}, {Cort{\'e}s-Contreras},
  {Czesla}, {D{\'\i}ez-Alonso}, {Feiz}, {Fern{\'a}ndez}, {Ferro},
  {Fuhrmeister}, {Galad{\'\i}-Enr{\'\i}quez}, {Garcia-Piquer}, {Garc{\'\i}a
  Vargas}, {Gesa}, {G{\'o}mez Galera}, {Gonz{\'a}lez-Peinado}, {Gr{\"o}zinger},
  {Grohnert}, {Gu{\`a}rdia}, {Guijarro}, {de Guindos}, {Guti{\'e}rrez-Soto},
  {Hagen}, {Hauschildt}, {Hedrosa}, {Helmling}, {Hermelo}, {Hern{\'a}ndez
  Arab{\'\i}}, {Hern{\'a}ndez Casta{\~n}o}, {Hern{\'a}ndez Hernando},
  {Herrero}, {Huber}, {Huke}, {Johnson}, {de Juan}, {Kim}, {Klein},
  {Kl{\"u}ter}, {Klutsch}, {Lafarga}, {Lamp{\'o}n}, {Lara}, {Laun}, {Lemke},
  {Lenzen}, {L{\'o}pez del Fresno}, {L{\'o}pez-Gonz{\'a}lez},
  {L{\'o}pez-Puertas}, {L{\'o}pez Salas}, {Luque}, {Mag{\'a}n Madinabeitia},
  {Mall}, {Mandel}, {Marfil}, {Mar{\'\i}n Molina}, {Maroto Fern{\'a}ndez},
  {Mart{\'\i}n}, {Mart{\'\i}n-Ruiz}, {Marvin}, {Mirabet}, {Moya},
  {Moreno-Raya}, {Nagel}, {Naranjo}, {Nortmann}, {Ofir}, {Oreiro}, {Pall{\'e}},
  {Panduro}, {Pascual}, {Passegger}, {Pedraz}, {P{\'e}rez-Calpena}, {P{\'e}rez
  Medialdea}, {Perger}, {Perryman}, {Pluto}, {Rabaza}, {Ram{\'o}n}, {Rebolo},
  {Redondo}, {Reinhardt}, {Rhode}, {Rix}, {Rodler}, {Rodr{\'\i}guez},
  {Rodr{\'\i}guez Trinidad}, {Rohloff}, {Rosich}, {Sadegi},
  {S{\'a}nchez-Blanco}, {S{\'a}nchez Carrasco}, {S{\'a}nchez-L{\'o}pez},
  {Sanz-Forcada}, {Sarkis}, {Sarmiento}, {Sch{\"a}fer}, {Schiller},
  {Sch{\"o}fer}, {Schweitzer}, {Solano}, {Stahl}, {Strachan}, {Su{\'a}rez},
  {Tabernero}, {Tala}, {Tulloch}, {Veredas}, {Vico Linares}, {Vilardell},
  {Wagner}, {Winkler}, {Wolthoff}, {Xu}, {Yan}, \& {Zapatero
  Osorio}}]{trifonov2018}
{Trifonov}, T., {K{\"u}rster}, M., {Zechmeister}, M., {et~al.} 2018,
  \href{https://ui.adsabs.harvard.edu/abs/2018A&A...609A.117T}{\aap, 609, A117}

\bibitem[{{Trifonov} {et~al.}(2020){Trifonov}, {Lee}, {K{\"u}rster}, {Henning},
  {Grishin}, {Stock}, {Tjoa}, {Caballero}, {Wong}, {Bauer}, {Quirrenbach},
  {Zechmeister}, {Ribas}, {Reffert}, {Reiners}, {Amado}, {Kossakowski},
  {Azzaro}, {B{\'e}jar}, {Cort{\'e}s-Contreras}, {Dreizler}, {Hatzes},
  {Jeffers}, {Kaminski}, {Lafarga}, {Montes}, {Morales}, {Pavlov},
  {Rodr{\'\i}guez-L{\'o}pez}, {Schmitt}, {Solano}, \&
  {Barnes}}]{2020A&A...638A..16T}
{Trifonov}, T., {Lee}, M.~H., {K{\"u}rster}, M., {et~al.} 2020,
  \href{https://ui.adsabs.harvard.edu/abs/2020A&A...638A..16T}{\aap, 638, A16}

\bibitem[{{Tuomi} {et~al.}(2018){Tuomi}, {Jones}, {Barnes},
  {Anglada-Escud{\'e}}, {Butler}, {Kiraga}, \& {Vogt}}]{2018AJ....155..192T}
{Tuomi}, M., {Jones}, H. R.~A., {Barnes}, J.~R., {et~al.} 2018,
  \href{https://ui.adsabs.harvard.edu/abs/2018AJ....155..192T}{\aj, 155, 192}

\bibitem[{{Tuomi} {et~al.}(2014){Tuomi}, {Jones}, {Barnes},
  {Anglada-Escud{\'e}}, \& {Jenkins}}]{2014MNRAS.441.1545T}
{Tuomi}, M., {Jones}, H. R.~A., {Barnes}, J.~R., {Anglada-Escud{\'e}}, G., \&
  {Jenkins}, J.~S. 2014,
  \href{https://ui.adsabs.harvard.edu/abs/2014MNRAS.441.1545T}{\mnras, 441,
  1545}

\bibitem[{{Tuomi} {et~al.}(2019){Tuomi}, {Jones}, {Butler}, {Arriagada},
  {Vogt}, {Burt}, {Laughlin}, {Holden}, {Shectman}, {Crane}, {Thompson},
  {Keiser}, {Jenkins}, {Berdi{\~n}as}, {Diaz}, {Kiraga}, \&
  {Barnes}}]{tuomi2019}
{Tuomi}, M., {Jones}, H.~R.~A., {Butler}, R.~P., {et~al.} 2019,
  \href{https://ui.adsabs.harvard.edu/abs/2019arXiv190604644T}{arXiv e-prints,
  arXiv:1906.04644}

\bibitem[{{Udry} {et~al.}(2007){Udry}, {Bonfils}, {Delfosse}, {Forveille},
  {Mayor}, {Perrier}, {Bouchy}, {Lovis}, {Pepe}, {Queloz}, \&
  {Bertaux}}]{2007A&A...469L..43U}
{Udry}, S., {Bonfils}, X., {Delfosse}, X., {et~al.} 2007,
  \href{https://ui.adsabs.harvard.edu/abs/2007A&A...469L..43U}{\aap, 469, L43}

\bibitem[{{Vogt} {et~al.}(1994){Vogt}, {Allen}, {Bigelow}, {Bresee}, {Brown},
  {Cantrall}, {Conrad}, {Couture}, {Delaney}, {Epps}, {Hilyard}, {Hilyard},
  {Horn}, {Jern}, {Kanto}, {Keane}, {Kibrick}, {Lewis}, {Osborne},
  {Pardeilhan}, {Pfister}, {Ricketts}, {Robinson}, {Stover}, {Tucker}, {Ward},
  \& {Wei}}]{1994SPIE.2198..362V}
{Vogt}, S.~S., {Allen}, S.~L., {Bigelow}, B.~C., {et~al.} 1994, in \procspie,
  Vol. 2198, Instrumentation in Astronomy VIII, ed. D.~L. {Crawford} \& E.~R.
  {Craine}, 362

\bibitem[{{Winters} {et~al.}(2020){Winters}, {Irwin}, {Charbonneau}, {Latham},
  {Medina}, {Mink}, {Esquerdo}, {Berlind}, {Calkins}, \&
  {Berta-Thompson}}]{winters2020}
{Winters}, J.~G., {Irwin}, J.~M., {Charbonneau}, D., {et~al.} 2020,
  \href{https://ui.adsabs.harvard.edu/abs/2020AJ....159..290W}{\aj, 159, 290}

\bibitem[{{Wright} {et~al.}(2016){Wright}, {Wittenmyer}, {Tinney}, {Bentley},
  \& {Zhao}}]{2016ApJ...817L..20W}
{Wright}, D.~J., {Wittenmyer}, R.~A., {Tinney}, C.~G., {Bentley}, J.~S., \&
  {Zhao}, J. 2016,
  \href{https://ui.adsabs.harvard.edu/abs/2016ApJ...817L..20W}{\apjl, 817, L20}

\bibitem[{{Wright}(2005)}]{wright2005}
{Wright}, J.~T. 2005,
  \href{https://ui.adsabs.harvard.edu/abs/2005PASP..117..657W}{\pasp, 117, 657}

\bibitem[{{Zechmeister} {et~al.}(2019){Zechmeister}, {Dreizler}, {Ribas},
  {Reiners}, {Caballero}, {Bauer}, {B{\'e}jar}, {Gonz{\'a}lez-Cuesta},
  {Herrero}, {Lalitha}, {L{\'o}pez-Gonz{\'a}lez}, {Luque}, {Morales},
  {Pall{\'e}}, {Rodr{\'\i}guez}, {Rodr{\'\i}guez L{\'o}pez}, {Tal-Or},
  {Anglada-Escud{\'e}}, {Quirrenbach}, {Amado}, {Abril}, {Aceituno},
  {Aceituno}, {Alonso-Floriano}, {Ammler-von Eiff}, {Antona Jim{\'e}nez},
  {Anwand-Heerwart}, {Arroyo-Torres}, {Azzaro}, {Baroch}, {Barrado},
  {Becerril}, {Ben{\'\i}tez}, {Berdi{\~n}as}, {Bergond}, {Bluhm},
  {Brinkm{\"o}ller}, {del Burgo}, {Calvo Ortega}, {Cano}, {Cardona
  Guill{\'e}n}, {Carro}, {C{\'a}rdenas V{\'a}zquez}, {Casal},
  {Casasayas-Barris}, {Casanova}, {Chaturvedi}, {Cifuentes}, {Claret},
  {Colom{\'e}}, {Cort{\'e}s-Contreras}, {Czesla}, {D{\'\i}ez-Alonso}, {Dorda},
  {Fern{\'a}ndez}, {Fern{\'a}ndez-Mart{\'\i}n}, {Fuhrmeister}, {Fukui},
  {Galad{\'\i}-Enr{\'\i}quez}, {Gallardo Cava}, {Garcia de la Fuente},
  {Garcia-Piquer}, {Garc{\'\i}a Vargas}, {Gesa}, {G{\'o}ngora Rueda},
  {Gonz{\'a}lez-{\'A}lvarez}, {Gonz{\'a}lez Hern{\'a}ndez},
  {Gonz{\'a}lez-Peinado}, {Gr{\"o}zinger}, {Gu{\`a}rdia}, {Guijarro}, {de
  Guindos}, {Hatzes}, {Hauschildt}, {Hedrosa}, {Helmling}, {Henning},
  {Hermelo}, {Hern{\'a}ndez Arabi}, {Hern{\'a}ndez Casta{\~n}o}, {Hern{\'a}ndez
  Otero}, {Hintz}, {Huke}, {Huber}, {Jeffers}, {Johnson}, {de Juan},
  {Kaminski}, {Kemmer}, {Kim}, {Klahr}, {Klein}, {Kl{\"u}ter}, {Klutsch},
  {Kossakowski}, {K{\"u}rster}, {Labarga}, {Lafarga}, {Llamas}, {Lamp{\'o}n},
  {Lara}, {Launhardt}, {L{\'a}zaro}, {Lodieu}, {L{\'o}pez del Fresno},
  {L{\'o}pez-Puertas}, {L{\'o}pez Salas}, {L{\'o}pez-Santiago}, {Mag{\'a}n
  Madinabeitia}, {Mall}, {Mancini}, {Mandel}, {Marfil}, {Mar{\'\i}n Molina},
  {Maroto Fern{\'a}ndez}, {Mart{\'\i}n}, {Mart{\'\i}n-Fern{\'a}ndez},
  {Mart{\'\i}n-Ruiz}, {Marvin}, {Mirabet}, {Monta{\~n}{\'e}s-Rodr{\'\i}guez},
  {Montes}, {Moreno-Raya}, {Nagel}, {Naranjo}, {Narita}, {Nortmann}, {Nowak},
  {Ofir}, {Oshagh}, {Panduro}, {Parviainen}, {Pascual}, {Passegger}, {Pavlov},
  {Pedraz}, {P{\'e}rez-Calpena}, {P{\'e}rez Medialdea}, {Perger}, {Perryman},
  {Rabaza}, {Ram{\'o}n Ballesta}, {Rebolo}, {Redondo}, {Reffert}, {Reinhardt},
  {Rhode}, {Rix}, {Rodler}, {Rodr{\'\i}guez Trinidad}, {Rosich}, {Sadegi},
  {S{\'a}nchez-Blanco}, {S{\'a}nchez Carrasco}, {S{\'a}nchez-L{\'o}pez},
  {Sanz-Forcada}, {Sarkis}, {Sarmiento}, {Sch{\"a}fer}, {Schmitt},
  {Sch{\"o}fer}, {Schweitzer}, {Seifert}, {Shulyak}, {Solano}, {Sota}, {Stahl},
  {Stock}, {Strachan}, {Stuber}, {St{\"u}rmer}, {Su{\'a}rez}, {Tabernero},
  {Tala Pinto}, {Trifonov}, {Veredas}, {Vico Linares}, {Vilardell}, {Wagner},
  {Wolthoff}, {Xu}, {Yan}, \& {Zapatero Osorio}}]{zechmeister2019}
{Zechmeister}, M., {Dreizler}, S., {Ribas}, I., {et~al.} 2019,
  \href{https://ui.adsabs.harvard.edu/abs/2019A&A...627A..49Z}{\aap, 627, A49}

\bibitem[{{Zechmeister} {et~al.}(2018){Zechmeister}, {Reiners}, {Amado},
  {Azzaro}, {Bauer}, {B{\'e}jar}, {Caballero}, {Guenther}, {Hagen}, {Jeffers},
  {Kaminski}, {K{\"u}rster}, {Launhardt}, {Montes}, {Morales}, {Quirrenbach},
  {Reffert}, {Ribas}, {Seifert}, {Tal-Or}, \& {Wolthoff}}]{zechmeister2018}
{Zechmeister}, M., {Reiners}, A., {Amado}, P.~J., {et~al.} 2018,
  \href{https://ui.adsabs.harvard.edu/abs/2018A&A...609A..12Z}{\aap, 609, A12}

\end{thebibliography}

\begin{appendix}

\section{Log-likelihood function jitter-rotation relation}
\label{sec:fitting}

We fit the jitter-rotation relation in logarithmic space. In logarithmic space, Eq.~(\ref{eq:jitter_vrot}) takes the shape of 
\begin{equation}
        \log {\sigma}_{\mathrm{jitter}}(\alpha,\beta,v_{\mathrm{eq}}) = \log \left( \sqrt{ \alpha^2 + \left( \beta  \cdot v_{\mathrm{eq} } \right)^2 } \right).
\end{equation}

In order to perform an unweighted fit, we assume the variance $\Delta^2$ is constant in all data points. We treat $\Delta$ as a nuisance parameter, hence its value is obtained from the fit itself. Performing the fit in logarithmic space effectively weights down data points with large values, since  residuals in logarithmic space correspond to relative rather than absolute deviations in the original space.

Best-fit values are obtained from MLE. The log-likelihood $\ln \mathcal{L}$ including the nuisance parameter $\Delta^2$ for the variance is given by
\begin{equation}
        \ln \mathcal{L}  = -\frac{1}{2} \sum_i \left[ \frac{(\log {\sigma}_{\mathrm{jitter},i} - \log {\sigma}_{\mathrm{jitter}}(\alpha,\beta,v_{\mathrm{eq},i}))^2}{\Delta^2} + \ln \left ( 2\pi\,\Delta^2 \right ) \right].
\end{equation}

\section{Expected value of the sine of the inclination}
\label{sec:sin_expectancy}

The inclination $i$ is in $[0,\pi/2]$. The probability density function (pdf) of $\cos i$ is uniform for isotropic distributed stellar spin axes
\begin{equation}
    f(\cos i) = 1 \mathrm{~for~} 0 \leq \cos i \leq 1.
\end{equation}
Then, the pdf $g(i)$ of the inclination $i$ can be computed as
\begin{equation}
    \begin{split}
            f(\cos i) \mathrm{~d}\cos i & = f(\cos i) \left| \frac{\mathrm{d}\cos i}{\mathrm{~d}i} \right| \mathrm{~d}i  \\
        & = \sin i \mathrm{~d}i = g(i) \mathrm{~d}i \\
        \implies g(i) & = \sin i.
    \end{split}
\end{equation}
The expected value of $\sin i$ is then given by
\begin{equation}
    \begin{split}
            \langle \sin i \rangle & = \int_0^{\pi/2} g(i) \sin i  \mathrm{~d}i
            & =  \int_0^{\pi/2} (\sin i)^2  \mathrm{~d}i
            & = \frac{\pi}{4}.
    \end{split}
\end{equation}
The expected value of $\frac{1}{\sin i}$ is then given by
\begin{equation}
    \begin{split}
            \langle \frac{1}{\sin i} \rangle & = \int_0^{\pi/2} g(i) \frac{1}{\sin i} \mathrm{~d}i
            & =  \int_0^{\pi/2} \mathrm{~d}i
            & = \frac{\pi}{2}.
    \end{split}
\end{equation}

\section{Hidden companions}
    \label{sec:hidden_planets}
    The RV signals of hidden companions can contribute to the observed RV jitter. Thus, we compare the RV semi-amplitude of planets from a model distribution to the observed RV jitter. We study the signals of potential low-mass planets, as they lie at the sensitivity threshold of the CARMENES survey.
    
    \citet{ribas2023} find that a fraction of \AS{0.89}{0.08}{0.11} of M~stars host planets when considering the planets in a mass bin of $\SI{1}{\mearth}\leq M_{\mathrm{pl}} \sin i \leq\SI{10}{\mearth}$ and periods of \SIrange{1}{1000}{\day}. The occurrence rate is computed applying the assumption of a log-uniform distribution of orbital periods. They fit a power-law dependence of the number of planets $N_{\mathrm{pl}}$ as function of their minimum mass $M_{\mathrm{pl}} \sin i$, where $N_{\mathrm{pl}} \propto \left( M_{\mathrm{pl}} \sin i\right)^{\alpha}$. For the coefficient $\alpha$, they find a value of \num{-1.05 \pm 0.01}. 
    
\begin{figure}
      \center
      \includegraphics[width=1\linewidth]{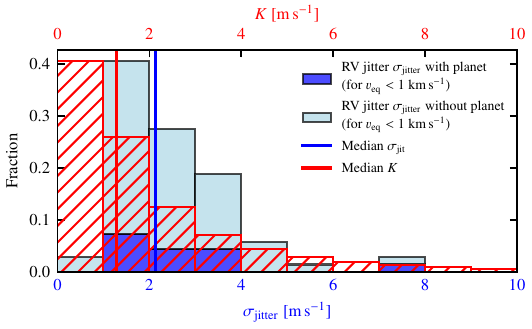}
      \caption{RV semi-amplitude distribution of model planet distribution. The observed jitter in CARMENES of stars with and without known planets (blue and light blue bars) is compared to the RV semi-amplitudes of the model planet population (red hatched bars). The median RV semi-amplitude (red line) lies at \SI{1.3}{\meter\per\second}, and the median RV jitter of stars with and without planets (blue line) is \SI{2.1}{\meter\per\second}.}
      \label{fig:hist_hidden_planets}
    \end{figure}

    To compute an expected RV semi-amplitude, we randomly draw from the log-uniform distribution of orbital periods, and from a power-law distribution  of minimum mass fitted by \citet{ribas2023}. We also randomly draw a stellar mass from our sample of slowly rotating stars with $v_{\mathrm{eq}}<\SI{1}{\kilo\meter\per\second}$. From period, minimum mass, and stellar mass, we then compute the RV semi-amplitude. We repeat the experiment for $10^6$ iterations using a Markov chain Monte Carlo (MCMC) approach. The resulting distribution is shown in Fig.~\ref{fig:hist_hidden_planets}. We overplot a histogram of jitter in slow rotating stars with $v_{\mathrm{eq}}<\SI{1}{\kilo\meter\per\second}$, where the jitter is not dominated by the modulation of the stellar rotation. The amplitudes of the modeled planet population are not sufficient to explain the observed jitter. However, we also see that the RV semi-amplitudes of approximately a fourth of the model planets have semi-amplitudes within \SIrange{1}{2}{\meter\per\second}. The median semi-amplitude lies at \SI{1.3}{\meter\per\second}, corresponding to a jitter of \SI{0.9}{\meter\per\second} (under the assumption of circular orbits and evenly sampled RV time series). Thus, hidden planets likely contribute to the observed jitter floor.
    
\clearpage

\section{High-jitter stars}
    \label{sec:high-jitter stars}
    \paragraph{J01033+623} is the second component in a wide binary companion. The primary component is J01026+623, which is part of the CARMENES GTO program. The stars are separated by \SI{293.1 \pm 0.1}{\arcsec} \citep{dorda2011}. Due to the large separation, the binary companion is not expected to induce short-term RV variations. J01033+623 is also a young disk star and exhibits a strong magnetic field \citep{contreras2024,shulyak2019}. The M5.0-type star is a known active star \citep{2018A&A...614A..76J,talor2018}.
    
    \paragraph{J04198+425} is one of the latest spectral type stars in the CARMENES sample (M8.5 V) and exhibits large RV jitter of \SI{232 \pm 43}{\meter\per\second} for its equatorial rotation velocity of \SI{5.70 \pm 0.57}{\kilo\meter\per\second}. The rotation period of \SI{0.99}{\day} this star is derived from TESS data by \citet{2024A&A...684A...9S} and is marked as secure. Since the star is situated within a crowded field of stars, it cannot be excluded that the photometric modulation seen in the TESS data is produced by neighboring star. The projected rotation velocity, however, points to even slower rotation with \SI{3.6\pm 2.3}{\kilo\meter\per\second}.
    
    \paragraph{J06574+740} is an M4.0 V spectral type star. The light curve of the star shows a beating signal as is discussed by \citet{2024A&A...684A...9S}. They conclude that star could be an unresolved binary as indicated by astrometric excess noise or could be exhibiting strong differential rotation. The star exhibits a moderately large average magnetic field strength of \SI{3330 \pm 830}{\gauss}. Furthermore, the distance of the star is disputed, which affects the radius determination of the star as discussed in \citep{schweitzer2019}.

    \paragraph{YZ CMi (J07446+035)} has been of high interest for studies of stellar activity in the past \citep[see][and references therein]{baroch2020}. The connection between its RV signal and its stellar rotation is well-established. For example, \citet{baroch2020} use a single spot model to fit chromatic RVs from CARMENES as well as photometric data. Similar analyses have been performed by \citet{biscz2022} and \citet{ikuta2023}. Furthermore, \citet{2021A&A...652A..28L} and \citet{schoefer2022} analyzed time series of various spectroscopic activity indicators from CARMENES data and showed that they exhibit a \SI{2.8}{\day} periodicity coinciding with the rotation period of the star.

    \paragraph{J09449$-$123 and J17338+169} are among the fastest rotating stars in CARMENES. The two stars have very large average magnetic fields of \SI{4860 \pm 990}{\gauss} and \SI{8130 \pm 940}{\gauss}. Additionally, J09449$-$123 is part of Argus, which has an estimated age of \SI{40}{\mega\year} and thus it is a very young star \citep{malo2013}. J17338+169 is also likely young, as it is part of the young disk \cite{contreras2024}, and $\mathrm{H\alpha}$ active \citet{2018A&A...614A..76J}. Both stars have been classified as RV-loud \citep{talor2018}. The RV time series of both stars show a long-term trend. The slope is \SI{-1.9 \pm 2.5}{\meter\per\second\per\day} for J09449$-$123 and \SI{1.1 \pm 1.1}{\meter\per\second\per\day} for J17338+169. The trends are not significant, therefore the stars have not been flagged and are included in our sample. In both cases, the number of data points is low.

    \paragraph{J13536+776} is an M4.0-type star at the upper edge of the prediction interval in Fig.~\ref{fig:jitter_vrot_all}. Its jitter value has a large uncertainty with \SI{95 \pm 34}{\meter\per\second}. Inspection of its RV time series reveals and outlier measurement, which likely causes the large uncertainty and the excess RV jitter. The outlier RV measurement has been taken during a flaring event, which has previously been reported by \cite{fuhrmeister2023}.

    \paragraph{J19422$-$207} is a M5.1-type star with a large average magnetic field strength of \SI{3840 \pm 350}{\gauss}. It is part of the young disk \citep{contreras2024}. The star has a similar equatorial rotation velocity (\SI{8.33 \pm 0.28}{\kilo\meter\per\second}) and jitter (\SI{90 \pm 12}{\meter\per\second}) as the series of high-jitter stars with concentrated distributions of magnetic filling factors. The filling factor distribution of J19422$-$207 does not reveal a dominating component and we compute a low filling factor index $z$ of \num{1.54}.
    
    \paragraph{EV Lac (J22468+443)} is another well-known active star. It has an equatorial rotation velocity of \SI{4}{\kilo\meter\per\second} and is thus located at the lower end of the velocity range in this group of outliers. As for YZ~CMi, EV~Lac has been investigated with star spot mapping \citep{ikuta2023}. \citet{jeffers2022} perform a detailed analysis of the CARMENES RVs and activity in this star. Using low-resolution Doppler imaging, they propose a complex spot pattern and demonstrate a strong correlation of the center-of-light variations, or activity-induced RVs from starspots, with the measured RVs. Interestingly, they find significant signals in the periodograms of multiple activity indicators at half of the rotation period instead of the full rotation period as seen in most CARMENES stars \citep{2021A&A...652A..28L}. Signals at half of the stellar period are associated with spot-induced variations, whereas variations induced by convective red- or blueshift are associated with periodogram peaks at the full rotation period \citet{baroch2020}. The activity signal in EV Lac shows signs of evolution according to \citet{jeffers2022}. 

    The magnetic fields of EV~Lac and YZ~CMi have been under frequent investigation \citep{reiners2007,morin2008,see2019,afram2019,shulyak2019,reiners2022,cristofari2023}. Both stars are in the regime of magnetic field saturation \citep{reiners2022}. However, YZ~CMi shows a dipolar axisymmetric field geometry,  while EV~Lac has a non-axisymmetric field geometry \citep{morin2008}.  \\

    \paragraph{The three stars J08536$-$034, J18131+260 and J23548+385} are less well-studied members of the CARMENES sample. J08536$-$034 (GJ~3517) is an M9.0V star, and the coolest and least massive star in the CARMENES sample. It is also the star with the largest RV jitter in the group of stars with increased jitter and has a large internal error of $\sim$\SI{60}{\meter\per\second} due to its faintness and low S/N spectra. Additionally, it has the lowest Rossby number in the CARMENES sample. J18131+260 has been mentioned by \citet{baran2011} as a flaring star with an unusually long flare decay time. 

    \clearpage

    \begin{figure*}
        \centering
        \captionsetup{justification=raggedright,singlelinecheck=false}
        \begin{tabular}{cc}
            \includegraphics[width=0.5\linewidth]{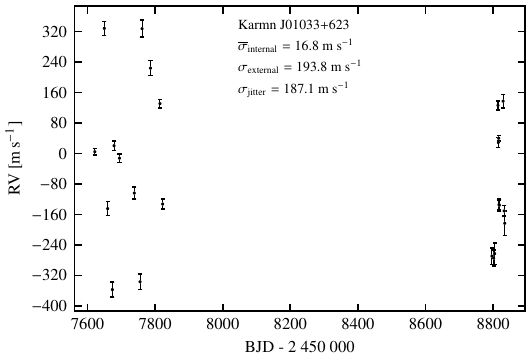} &
            \includegraphics[width=0.5\linewidth]{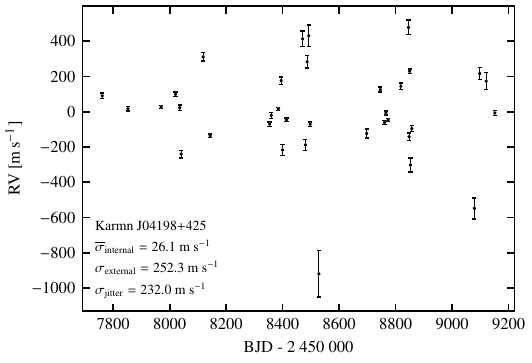} \\
            \includegraphics[width=0.5\linewidth]{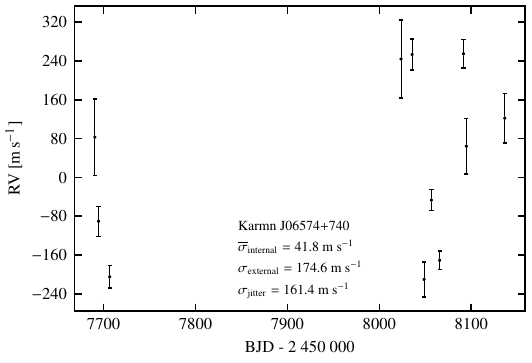} &
            \includegraphics[width=0.5\linewidth]{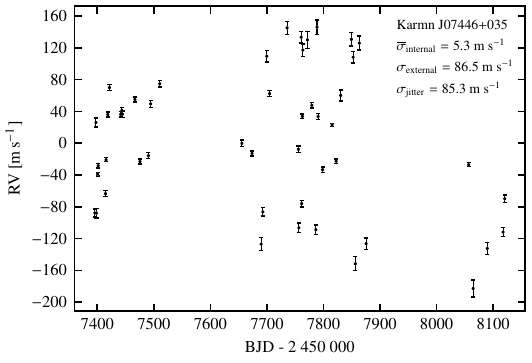} \\
            \includegraphics[width=0.5\linewidth]{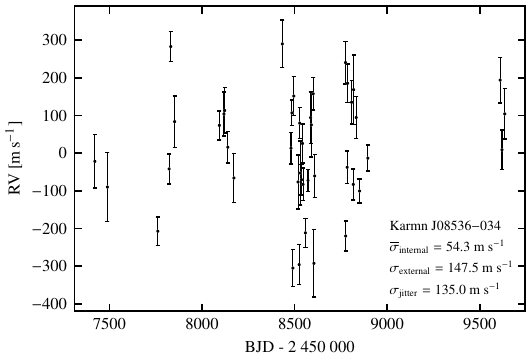} &
            \includegraphics[width=0.5\linewidth]{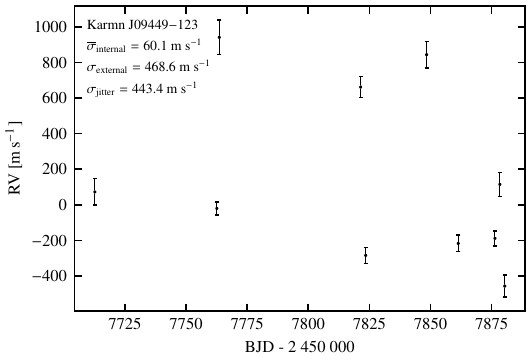} \\
        \end{tabular}
            \caption{RV time series of high-jitter stars.}
    \end{figure*}
    \clearpage

    \begin{figure*}
        \centering
        \ContinuedFloat 
        \captionsetup{justification=raggedright,singlelinecheck=false}
        \begin{tabular}{cc}
            \includegraphics[width=0.5\linewidth]{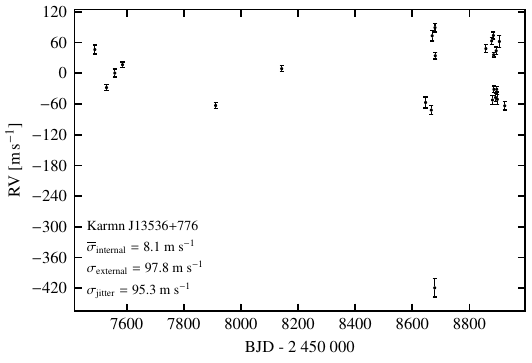} &
            \includegraphics[width=0.5\linewidth]{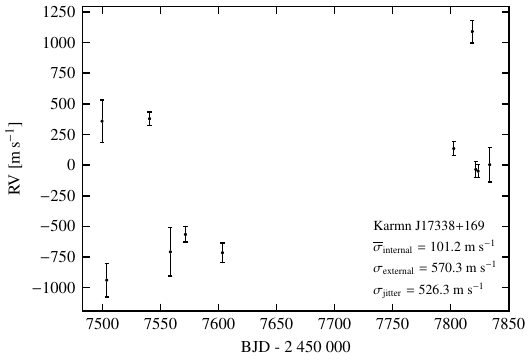} \\
            \includegraphics[width=0.5\linewidth]{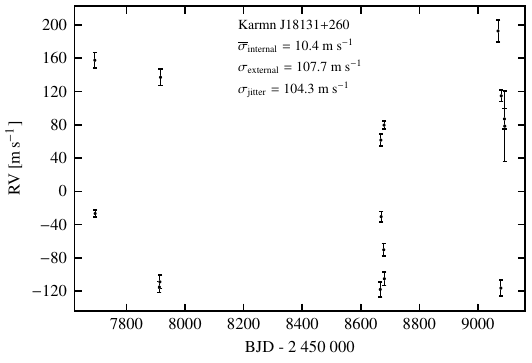} &
            \includegraphics[width=0.5\linewidth]{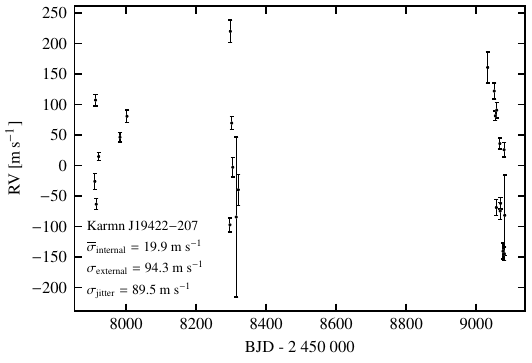} \\
            \includegraphics[width=0.5\linewidth]{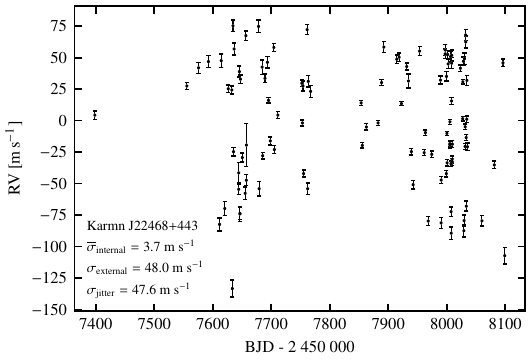} &
            \includegraphics[width=0.5\linewidth]{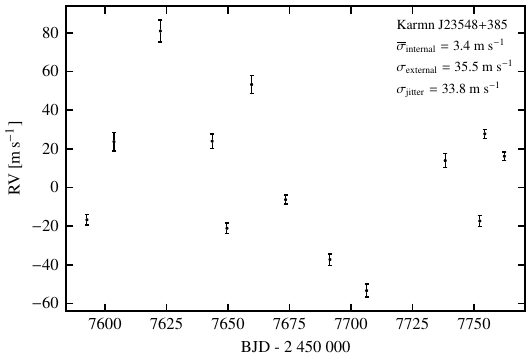} \\
        \end{tabular}
        \caption{RV time series of high-jitter stars (continued).}
    \end{figure*}
    
    \clearpage

\section{Additional figures}
    \label{sec:additional_figures}

    \begin{figure}[ht!]
        \center
        \includegraphics[width=1\linewidth]{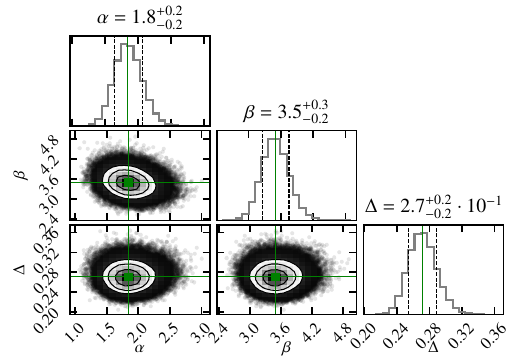}
        \caption{MCMC posterior distribution of jitter-rotation relation fit (Eq.~(\ref{eq:jitter_vrot})).}
        \label{fig:corner_jitter_rotation_relation}
    \end{figure}
    
    \begin{figure}[ht!]
        \center
        \includegraphics[width=1\linewidth]{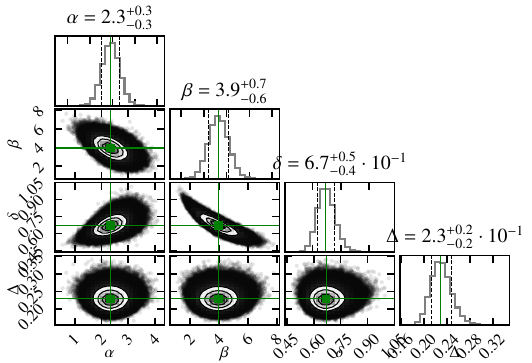}
        \caption{MCMC posterior distribution of jitter-rotation-magnetic field relation fit (Eq.~(\ref{eq:jitter_bvrot})).}
        \label{fig:corner_jitter_rotation_bfield_relation}
    \end{figure}

    \begin{figure}[ht!]
      \center
      \includegraphics[width=1\linewidth]{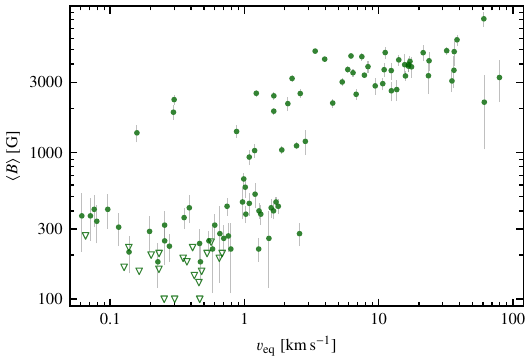}
      \caption{Average magnetic field as a function of equatorial rotation velocity.}
      \label{fig:vrot_bfield}
    \end{figure}

    \begin{figure}[!ht]
      \center
      \includegraphics[width=1\linewidth]{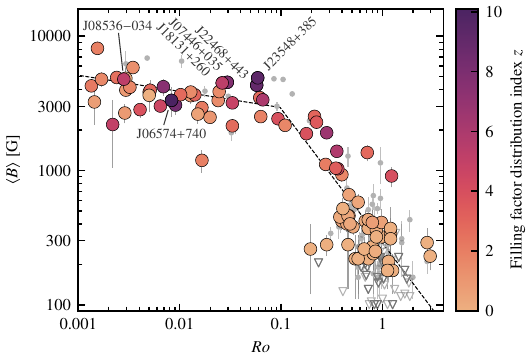}
      \caption{Mean magnetic field as a function of Rossby number. CARMENES targets removed from the jitter sample are shown in grey. The dashed lines indicate relations from \citet{2024A&A...684A...9S}.}
      \label{fig:ro_bfield}
    \end{figure}

\section{Additional tables}

    \clearpage

    \onecolumn
    {            
        \footnotesize
        \begin{longtable}{@{}l l c c c S[table-format=3.2(2)e2]@{}}
            \caption{CARMENES M dwarfs excluded from the jitter sample. The table shows CARMENES stars with at least 10 observations, that were flagged because of a known planet (P), planet candidate (P?), known multiplicity (M), or significant ($5\sigma$) trend (T). \label{tab:excluded_stars}} \\

            \toprule
            \toprule
            Karmn & Name & Flags & {Planet} & {Multiplicity} & {Trend} \\
                  &      &       & {Reference} & {Reference} &         \\
                  &      &       &             &             & {[\si{\centi\meter\per\second\per\day}]} \\
            \midrule
            
            \endfirsthead

            \caption[]{continued.} \\

            \toprule
            \toprule
            Karmn & Name & Flags & {Planet} & {Multiplicity} & {Trend} \\
                  &      &       & {Reference} & {Reference} &         \\
                  &      &       &             &             & {[\si{\centi\meter\per\second\per\day}]} \\
            \midrule
            \endhead
    
            \bottomrule
            \endfoot
                    
              J00067$-$075 &                  GJ 1002 &    P &                          SM23 &               &                \\
               J00162+198W &                   EZ Psc &    M &                            &            Bar18 &                \\
                J00183+440 &                   GX And &    P &               How14, Tri18 &               &                \\
                J00184+440 &                   GQ And &    T &                            &               &       2.73(46) \\
                J00403+612 &  2MASS J00402129+6112490 &    P &                         GA23b &               &                \\
                J01026+623 &                BD+61 195 &    P &                         Per19 &               &                \\
                J01056+284 &                  GJ 1029 &    M &                            &     Bar18, Win20 &                \\
                J01066+192 &          LSPM J0106+1913 &    P &                         Cha22 &               &                \\
              J01125$-$169 &                   YZ Cet &    P &                         AD17a &               &                \\
                J01518+644 &                G 244-037 &    T &                            &               &      -2.36(41) \\
                J02002+130 &                   TZ Ari &    P &                  Fen20, Qui22 &               &                \\
                J02222+478 &                BD+47 612 &    P &                         Hob18 &               &                \\
             J02489$-$145E &          PM J02489-1432E &   P? &                            &               &                \\
             J02489$-$145W &          PM J02489-1432W &    P &                         Kos21 &               &                \\
                J02530+168 &         Teegarden's Star &    P &                         Zec19 &               &                \\  
                J02573+765 &                 G 245-61 &    P &                         Sot21 &               &                \\
                J03133+047 &                   CD Cet &    P &                         Bau20 &               &                \\
              J04167$-$120 &                LP 714-47 &    P &                         Dre20 &               &                \\
                J04311+589 &                 G 175-34 &    M &                            &           Stra77 &                \\
                J04343+430 &           PM J04343+4302 &    P &                         Blu21 &               &                \\
              J04406$-$128 &                 TOI-2457 &   P? &                            &               &                \\
                J04429+189 &                HD 285968 &    P &           End08, For09, Tri18 &               &                \\
                J04520+064 &                Wolf 1539 &   PT &                         How10 &               &       29.5(15) \\
              J04538$-$177 &                   GJ 180 &    P &                  Fen20, Tuo14 &               &                \\
              J05314$-$036 &                 HD 36395 &    T &                            &               &        7.8(14) \\
                J05337+019 &                 V371 Ori &    M &                            &     Rei12, Bar21 &                \\
                J05532+242 &                  Ross 59 &    M &                            &            Bar18 &                \\
              J06105$-$218 &                HD  42581 &   PT &                  Fen20, Tuo14 &               &       7.32(76) \\
                J06371+175 &                HD 260655 &    P &                         Luq22 &               &                \\
              J06396$-$210 &               LP 780-032 &    T &                            &               &       -137(17) \\
                J06548+332 &                 Wolf 294 &    P &                  Rib23, Sto20 &               &                \\
              J07001$-$190 &  2MASS J07000682-1901235 &    M &                            &            Bar21 &                \\
                J07274+052 &            Luyten's Star &    P &                         AD17b &               &                \\
              J07361$-$031 &               BD-02 2198 &   MT &                            &            Bar21 &      -2101(63) \\
                J08023+033 &                G 50-16 A &    P &                         Kem20 &               &                \\
              J08409$-$234 &               LP 844-008 &    P &                  Fen20, Joh07 &               &                \\
                J08413+594 &               LP 090-018 &    P &                Mor19, Rib23 &               &                \\
                J09033+056 &               NLTT 20861 &    T &                            &               &       -584(32) \\
                J09140+196 &               LP 427-016 &   MT &                            &            Bar21 &      236.9(28) \\
                J09144+526 &                 HD 79211 &   PT &                          GA20 &               &      -10.1(11) \\
              J09360$-$216 &                   GJ 357 &    P &                         Luq19 &               &                \\
                J09561+627 &                BD+63 869 &    P &                  Fen20, Tuo19 &               &                \\
                J10023+480 &               BD+48 1829 &    P &                         Hob19 &               &                \\
                J10088+692 &          TYC 4384-1735-1 &    P &                         Blu20 &               &                \\
              J10182$-$204 &               NLTT 23956 &    M &                            &            Bar18 &                \\
              J10185$-$117 &               LP  729-54 &    P &                         Now20 &               &                \\
                J10196+198 &                   AD Leo &   P? &                         Tuo18 &               &                \\
                J10289+008 &               BD+01 2447 &    P &                         Ama21 &               &                \\
                J10354+694 &               LP 037-179 &    M &                            &            Bar18 &                \\
                J10504+331 &                G 119-037 &    M &                            &            Bar21 &                \\
                J10564+070 &                   CN Leo &    P &                         Tuo19 &               &                \\
                J11033+359 &            Lalande 21185 &   PT &           But17, Dia19, Sto20 &               &       0.61(12) \\
                J11044+304 &          LSPM J1104+3027 &   P? &                            &               &                \\
                J11055+435 &                   WX UMa &    T &                            &               &       -6.1(12) \\
                J11126+189 &               StKM 1-928 &    T &                            &               &      -5.08(95) \\
                J11302+076 &                    K2-18 &    P &               Clo17, Sar18 &               &                \\
                J11417+427 &                Ross 1003 &    P &           Hag10, Tri18, Tri20 &               &                \\
                J11421+267 &                 Ross 905 &    P &          But04, Tri18 &               &                \\
                J11423+230 &                LP 375-23 &   P? &                            &               &                \\
                J11474+667 &    1RXS J114728.8+664405 &    T &                            &               &      618.4(94) \\
                J11476+786 &                   GJ 445 &    T &                            &               &       3.42(60) \\
                J11477+008 &                   FI Vir &    P &                         Bon18 &               &                \\
                J11509+483 &                  GJ 1151 &    P &                  Bla23, Mah21 &               &                \\
               J12123+544S &                HD 238090 &    P &                 Sto20 &               &                \\
                J12156+526 &               StKM 2-809 &    T &                            &               &   1.59(24)e+03 \\
                J12230+640 &                 Ross 690 &   PT &                         End22 &               &      -11.0(21) \\
                J12388+116 &                 Wolf 433 &    P &                         Fen20 &               &                \\
                J12479+097 &                 Wolf 437 &    P &                         Tri21 &               &                \\
                J13209+342 &               BD+35 2439 &    T &                            &               &       24.9(18) \\
                J13229+244 &                Ross 1020 &    P &                         Luq18 &               &                \\
                J13255+688 &  2MASS J13253177+6850106 &  PMT &                         GA22 &               &       -293(11) \\
                J13299+102 &               BD+11 2576 &    P &                         Dam22 &               &                \\
              J13591$-$198 &               LP 799-007 &   P? &                            &               &                \\
              J14010$-$026 &                HD 122303 &    P &                         SM17a &               &                \\
                J14155+046 &                  GJ 1182 &    M &                            &            Bar18 &                \\
              J14342$-$125 &                   HN Lib &    P &                          GA23a &               &                \\
              J15194$-$077 &                   HO Lib &    P &           Bon05, May09, Udr07 &               &                \\
                J15412+759 &                   UU UMi &    M &                            &     Bar18, Bar21 &                \\
              J15474$-$108 &               LP 743-031 &   MT &                            &            Bar21 &   2.92(24)e+03 \\
                J15583+354 &                G 180-018 &    P &                         Kem22 &               &                \\
               J16167+672S &                HD 147379 &    P &                  Rei18, Rib23 &               &                \\
                J16254+543 &                   GJ 625 &    P &                         SM17b &               &                \\
              J16303$-$126 &                V2306 Oph &    P &                         Wri16 &               &                \\
                J16581+257 &               BD+25 3173 &    P &                         Joh10 &               &                \\
                J17033+514 &                G 203-042 &    P &                         Gor23 &               &                \\
                J17071+215 &                 Ross 863 &   P? &                            &               &                \\
                J17355+616 &               BD+61 1678 &    P &                         Pin19 &               &                \\
                J17364+683 &                BD+68 946 &    P &                         Bur14 &               &                \\
                J17378+185 &               BD+18 3421 &    P &                  Aff19, Lal19 &               &                \\
                J17578+046 &           Barnard's Star &    P &                         Rib18 &               &                \\
                J18346+401 &               LP 229-017 &    T &                            &               &      -5.11(49) \\
                J18353+457 &               BD+45 2743 &    P &                          GA21 &               &                \\
                J18363+136 &                 Ross 149 &    T &                            &               &        6.0(11) \\
              J18409$-$133 &               BD-13 5069 &    P &                         Gor23 &               &                \\
               J18427+596N &                HD 173739 &    T &                            &               &       23.2(30) \\
               J18427+596S &                HD 173740 &    T &                            &               &      -24.8(23) \\
              J18498$-$238 &                V1216 Sgr &   P? &                            &               &                \\
                J18580+059 &               BD+05 3993 &    P &                          TP21 &               &                \\
                J19025+754 &          LSPM J1902+7525 &   P? &                            &               &                \\
               J19169+051N &                V1428 Aql &    P &                         Kam18 &               &                \\
               J19206+731S &  2MASS J19204172+7311434 &   PM &                         Cad22 &            Gai20 &                \\
                J20198+229 &              LP  395-8 A &    M &                            &            Bar18 &                \\
                J20260+585 &                Wolf 1069 &    P &                         Kos23 &               &                \\
                J20450+444 &               BD+44 3567 &    P &                         Pal23 &               &                \\
              J20451$-$313 &                   AU Mic &    P &                 Mart21, Pla20 &               &                \\
             J20556$-$140N &                 GJ 810 A &    M &                            &            Bar18 &                \\
                J21164+025 &          LSPM J2116+0234 &    P &                         Lal19 &               &                \\
                J21221+229 &           TYC 2187-512-1 &    P &                         Qui22 &               &                \\
                J21466+668 &                G 264-012 &    P &                         Ama21 &               &                \\
                J21474+627 &           TYC 4266-736-1 &    P &                         Esp22 &               &                \\
              J22096$-$046 &               BD-05 5715 &   PT &          But06, Mon14 &               &      -29.6(45) \\
                J22102+587 &         UCAC4 744-073158 &   P? &                            &               &                \\
                J22125+085 &                Wolf 1014 &    T &                            &               &       -6.9(11) \\
              J22137$-$176 &               LP 819-052 &    P &                         Luq18 &               &                \\
                J22252+594 &                G 232-070 &    P &                         Nag19 &               &                \\
                J22298+414 &                G 215-050 &   P? &                            &               &                \\
              J22532$-$142 &                   IL Aqr &    P &  Marc01, Marc98, Riv05, Riv10 &               &                \\
                J22565+165 &                HD 216899 &    T &                            &               &       1.42(16) \\
              J23064$-$050 &              2MUCD 12171 &    P &                  Gil16, Gil17 &               &                \\
                J23113+085 &               NLTT 56083 &    M &                            &            Gai20 &                \\
              J23556$-$061 &                   GJ 912 &   MT &                            &            Bar21 &       -723(39) \\
                J23585+076 &                Wolf 1051 &   MT &                            &            Bar21 &  -2.94(49)e+03 \\
                 
        \end{longtable}

        \tablebib{AD17a: \citet{astudillo2017a}, AD17b: \citet{astudillo2017b}, Aff19: \citet{2019A&A...622A.193A}, Ama21: \citet{2021A&A...650A.188A}, Bar18: \citet{baroch2018}, Bar21: \citet{2021A&A...653A..49B}, Bau20: \citet{bauer2020}, Blu20: \citet{2020A&A...639A.132B}, Blu21: \citet{2020A&A...639A.132B}, Bon05: \citet{2005A&A...443L..15B}, Bon18: \citet{2018A&A...613A..25B}, Bur14: \citet{2014ApJ...789..114B}, But04: \citet{2004ApJ...617..580B}, But06: \citet{2006PASP..118.1685B}, But17: \citet{2017AJ....153..208B}, Cad22: \citet{2022AJ....164...96C}, Cha22: \citet{chaturvedi2022}, Dam22: \citet{2022A&A...666A.187D}, Dia19: \citet{2019A&A...625A..17D}, Dre20: \citet{2020A&A...644A.127D}, End08: \citet{2008ApJ...673.1165E}, End22: \citet{2022AJ....164..238E}, Esp22: \citet{2022AJ....163..133E}, Fen20: \citet{feng2020}, For09: \citet{2008ApJ...673.1165E}, For09: \citet{2008psa..conf..191F}, GA20: \citet{2020A&A...637A..93G}, GA21: \citet{2021A&A...649A.157G}, GA22: \citet{2022A&A...658A.138G}, GA23a: \citet{2023A&A...675A.141G}, GA23b: \citet{2023A&A...675A.177G},  Gil16: \citet{2016Natur.533..221G}, Gil17: \citet{2017Natur.542..456G}, Gor23: \citet{2023A&A...680A..28G}, Gai20: \citet{2020yCat.1350....0G}, Hag10: \citet{2010ApJ...715..271H}, Hob18: \citet{hobson2018}, Hob19: \citet{2019A&A...625A..18H}, How10: \citet{2010ApJ...721.1467H}, How14: \citet{howard2014}, Joh07: \citet{2007ApJ...670..833J}, Joh10: \citet{2010PASP..122..149J}, Kam18: \citet{2018A&A...618A.115K}, Kem20: \citet{2020A&A...642A.236K}, Kem22: \citet{2022A&A...659A..17K}, Kos21: \citet{kossakowski2021}, Kos23: \citet{2023A&A...670A..84K}, Lal19: \citet{2019A&A...627A.116L}, Luq18: \citet{2018A&A...620A.171L}, Luq19: \citet{2019A&A...628A..39L}, Luq22: \citet{2022A&A...664A.199L}, Mah21: \citet{2021ApJ...919L...9M}, Marc98: \citet{1998ApJ...505L.147M}, Marc01: \citet{2001ApJ...556..296M}, Mart21: \citet{2021A&A...649A.177M}, May09: \citet{2009A&A...507..487M}, Mon14: \citet{2014ApJ...781...28M}, Mor19: \citet{2019Sci...365.1441M}, Nag19: \citet{2019A&A...622A.153N}, Now20: \citet{2020A&A...642A.173N}, Pal23: \citet{2023A&A...678A..80P}, Per19: \citet{perger2019}, Pin19: \citet{2019A&A...625A.126P}, Pla20: \citet{2020Natur.582..497P}, Qui22: \citet{2022A&A...663A..48Q}, Qui22: \citet{2022A&A...663A..48Q}, Rei12: \citet{2012AJ....143...93R}, Rei18: \citet{2018A&A...609L...5R}, Rib18: \citet{2018Natur.563..365R}, Rib23: \citet{ribas2023}, Riv05: \citet{2005ApJ...634..625R}, Riv10: \citet{2010ApJ...719..890R}, SM17a: \citet{2017A&A...597A.108S}, SM17b: \citet{2017A&A...605A..92S}, SM23: \citet{mascareno2023}, Sot21: \citet{2021A&A...649A.144S}, Sto20: \citet{2020A&A...643A.112S}, TP21: \citet{2021A&A...648A..20T}, Tuo14: \citet{2014MNRAS.441.1545T}, Tuo18: \citet{2018AJ....155..192T}, Tuo19: \citet{2019A&A...628A..39L}, Tri18: \citet{trifonov2018}, Tri20: \citet{2020A&A...638A..16T}, Tri21: \citet{2021Sci...371.1038T}, Udr07: \citet{2007A&A...469L..43U}, Wri16: \citet{2016ApJ...817L..20W}, Win20: \citet{winters2020}, Zec19: \citet{zechmeister2019}.} \\
    }

    \clearpage

    {
    \tiny
    
    \begin{landscape}

        \begin{longtable}{
            @{}
            l
            l
            S[table-format=1.4(2),separate-uncertainty = true]
            S[table-format=2.4(2),separate-uncertainty = true]
            c
            S[table-format=1.4(2),separate-uncertainty = true]
            S[table-format=>2.2(2),separate-uncertainty = true]
            c
            S[table-format=1.2]
            S[table-format=>3(3),separate-uncertainty = true]
            S[table-format=2.1]
            S[table-format=3.0]
            S[table-format=4.3(2),separate-uncertainty = true]
            @{}
            }

            \caption{\label{table:sample}Stellar properties and jitter of the \num{239} targets in the CARMENES VIS jitter sample.\tablefootmark{a}} \\

            \toprule
            \toprule
            Karmn & Name  & {$R$}           & {$P_\mathrm{rot}$} & {Ref} & {$v_\mathrm{eq}$}             & {$v \sin i$}                  & Ref & $\log Ro$ & {$\langle B \rangle$}    & {$z$} & $N_{\mathrm{RV}}$ & $\sigma_{\mathrm{jitter}}$\\
                  &       &  [\si{\rsol}]   & [\si{\day}]        &       & [\si{\kilo\meter\per\second}] & [\si{\kilo\meter\per\second}] &     &           & [\si{\gauss}]            &       &                   & [\si{\meter\per\second}]  \\
            \midrule
            
            \endfirsthead

            \caption[]{continued.} \\
            
            \toprule
            \toprule
            Karmn & Name  & {$R$}           & {$P_\mathrm{rot}$} & {Ref} & {$v_\mathrm{eq}$}             & {$v \sin i$}                  & Ref & $\log Ro$ & {$\langle B \rangle$}    & {$z$} & $N_{\mathrm{RV}}$ & $\sigma_{\mathrm{jitter}}$\\
                  &       &  [\si{\rsol}]   & [\si{\day}]        &       & [\si{\kilo\meter\per\second}] & [\si{\kilo\meter\per\second}] &     &           & [\si{\gauss}]            &       &                   & [\si{\meter\per\second}]  \\
            \midrule
            
            \endhead
            
        \bottomrule
        \endfoot
                J00051+457 &               BD+44 4548 &  0.4888(44) &     &           &     &  <=2.0 &     Rei18 &     &     &       &   51 &      2.42(33) \\
                J00158+135 &                    GJ 12 &  0.2688(83) &     &           &     &     &        &     &     &       &   17 &      2.01(54) \\
               J00162+198E &               LP 404-062 &  0.2796(51) &     105(10) &      DA19 &   0.135(13) &  <=2.0 &     Rei18 &   0.10 &     &       &   22 &      2.26(43) \\
              J00286$-$066 &                  GJ 1012 &  0.3490(59) &     &           &     &  <=2.0 &     Rei18 &     &     &       &   50 &      2.01(27) \\
                J00389+306 &                Wolf 1056 &  0.4143(41) &    50.2(27) &      DA19 &   0.418(23) &  <=2.0 &     Rei18 &  -0.04 &     &       &   57 &      1.71(21) \\
                J00570+450 &                G 172-030 &  0.3358(44) &     &           &     &  <=2.0 &     Rei18 &     &     &       &   23 &      1.95(45) \\
                J01013+613 &                    GJ 47 &  0.3706(40) &    34.7(14) &      SM18 &   0.540(22) &  <=2.0 &     Rei18 &  -0.18 &     &       &   10 &      1.95(80) \\
                J01019+541 &                G 218-020 &  0.1560(14) &  0.280(5) &      DA19 &   28.18(56) &    30.6(31) &     Rei18 &  -2.76 &     &       &   20 &        34(18) \\
                J01025+716 &                 Ross 318 &  0.4636(70) &    50.5(27) &     Sha24 &   0.464(26) &  <=2.0 &     Rei18 &  -0.09 &     &       &  114 &      1.99(19) \\
                J01033+623 &                  V388 Cas &  0.2433(78) &  1.020(5) &     Sha24 &   12.07(39) &    10.5(15) &     Rei18 &  -2.04 &     &       &   23 &       187(24) \\
              J01048$-$181 &                  GJ 1028 &  0.1529(40) &     143(18) &     New18 &  0.0540(68) &  <=2.0 &     Rei18 &  -0.01 &     &       &  109 &      3.60(33) \\
                J01078+128 &                   G 2-21 &   0.506(15) &     &           &     &     &        &     &     &       &   47 &      4.56(58) \\
              J01339$-$176 &               LP 768-113 &  0.2767(56) &   7.750(95) &     Sha24 &   1.806(43) &  <=2.0 &     Rei18 &  -1.02 &     &       &   26 &       9.6(19) \\
              J01352$-$072 &             Barta 161 12 &   0.718(16) &   0.70(5) &     Sha24 &    51.9(39) &    59.8(69) &     Rei18 &  -2.13 &     &       &   11 &       270(80) \\
                J01433+043 &                    GJ 70 &  0.3991(47) &   14.65(30) &     Sha24 &   1.378(32) &  <=2.0 &     Rei18 &  -0.53 &     &       &   37 &      1.70(25) \\
                J01550+379 &           LSR J0155+3758 &  0.2071(67) &     &           &     &     &        &     &     &       &   19 &       5.8(28) \\
                J02015+637 &                G 244-047 &  0.4215(56) &     &           &     &  <=2.0 &     Rei18 &     &     &       &   49 &      2.41(32) \\
                J02022+103 &               LP 469-067 &  0.1487(53) &     &           &     &    4.5 &     Jen09 &     &     &       &   12 &    \text{$(1.5 \pm 2.4)\,i$\tablefootmark{b}} \\
                J02070+496 &                G 173-037 &  0.3284(43) &   7.200(83) &      DA19 &   2.307(40) &  <=2.0 &     Rei18 &  -0.95 &     &       &   53 &      3.67(44) \\
                J02088+494 &                G 173-039 &   0.417(31) &  0.750(5) &     Sha24 &    28.1(21) &    24.1(24) &     Rei18 &  -1.98 &     &       &   17 &        75(14) \\
                J02123+035 &                BD+02 348 &  0.4658(38) &     &           &     &  <=2.0 &     Rei18 &     &     &       &   64 &      1.30(30) \\
                J02336+249 &                   GJ 102 &  0.2272(33) &   3.00(5) &     Sha24 &   3.832(85) &     3.0(15) &     Rei18 &  -1.50 &     &       &   30 &       5.8(10) \\
                J02358+202 &                BD+19 381 &  0.4836(52) &     &           &     &  <=2.0 &     Rei18 &     &     &       &   43 &      1.55(37) \\
                J02362+068 &                   BX Cet &  0.2688(73) &     &           &     &  <=2.0 &     Rei18 &     &     &       &   50 &      2.38(31) \\
                J02442+255 &                   VX Ari &  0.3491(43) &    38.7(17) &      DA19 &   0.456(21) &  <=2.0 &     Rei18 &  -0.21 &     &       &   51 &      1.53(24) \\
                J02465+164 &               LP 411-006 &   0.127(13) &     &           &     &    6.20(80) &     Jen09 &     &     &       &   17 &      53.1(87) \\
                J02519+224 &                  RBS 365 &   0.611(14) &  0.860(5) &      DA19 &   35.94(87) &    27.2(27) &     Rei18 &  -2.01 &     &       &   14 &       112(11) \\
               J02565+554W &                 Ross 364 &  0.6096(35) &    51.2(28) &      SM18 &   0.602(33) &  <=2.0 &     Rei18 &   0.09 &     &     0 &   10 &      1.59(89) \\
                J03090+100 &                  GJ 1055 &  0.1780(60) &     &           &     &    4.70(70) &     Jen09 &     &     &       &   10 &       3.5(22) \\
                J03142+286 &               LP 299-036 &  0.1295(45) &     &           &     &     &        &     &     &       &   12 &       176(45) \\
                J03181+382 &                HD 275122 &  0.5690(47) &    77.2(58) &      DA19 &   0.373(28) &  <=2.0 &     Rei22 &   0.22 &  <=180 &     0 &   56 &      1.52(22) \\
                J03213+799 &                   GJ 133 &  0.4048(33) &     &           &     &  <=2.0 &     Rei22 &     &    320(110) &     0 &   34 &      2.19(33) \\
              J03217$-$066 &                 G 77-046 &  0.4510(44) &   21.00(57) &     Sha24 &   1.087(31) &  <=2.0 &     Rei22 &  -0.40 &    930(110) &   1.7 &   18 &      3.51(83) \\
              J03230+420 &                  GJ 1059 &  0.1673(56) &     &           &     &     &        &     &     &       &   16 &       7.9(32) \\
               J03463+262 &                 HD 23453 &  0.5727(38) &     &           &     &  <=2.0 &     Rei22 &     &     470(60) &   0.3 &   49 &      3.37(51) \\
                J03473+086 &                LTT 11262 &  0.1693(55) &     &           &     &     &        &     &     &       &   11 &       4.9(15) \\
              J03473$-$019 &                 G 80-021 &   0.494(38) &   3.880(28) &     Sha24 &    6.45(50) &     6.0(20) &     Rei22 &  -1.21 &   3490(200) &   2.7 &   11 &      23.0(52) \\
                J03531+625 &                 Ross 567 &  0.3530(47) &     &           &     &  <=2.0 &     Rei22 &     &     260(70) &     0 &   43 &      3.39(39) \\
              J04153$-$076 &              omi02 Eri C &   0.263(10) &     &           &     &     2.2(20) &     Rei22 &     &   3020(140) &   2.8 &   50 &      7.24(57) \\
                J04198+425 &           LSR J0419+4233 &   0.112(11) &  0.9900(50) &     Sha24 &    5.70(57) &     3.6(23) &     Rei18 &  -2.17 &     &       &   35 &       232(43) \\
                J04225+105 &          LSPM J0422+1031 &  0.4180(60) &     &           &     &  <=2.0 &     Rei22 &     &     190(70) &     0 &   19 &       1.2(11) \\
                J04290+219 &                BD+21 652 &  0.6647(90) &   25.40(79) &      DA19 &   1.324(45) &  <=2.0 &     Rei22 &  -0.14 &     380(60) &   0.4 &  161 &      3.65(24) \\
                J04376+528 &                BD+52 857 &  0.5466(58) &   15.47(33) &     Laf21 &   1.787(42) &  <=2.0 &     Rei22 &  -0.40 &     430(50) &   0.5 &  123 &      4.13(32) \\
              J04376$-$110 &                BD-11 916 &  0.4511(41) &     &           &     &  <=2.0 &     Rei22 &     &    215 &     0 &   48 &      1.81(29) \\
                J04429+214 &  2MASS J04425586+2128230 &  0.3668(53) &    47.8(25) &      DA19 &   0.388(21) &  <=2.0 &     Rei22 &  -0.17 &     420(90) &   0.2 &   15 &      2.64(57) \\
                J04588+498 &               BD+49 1280 &  0.5934(42) &   17.46(41) &     Sha24 &   1.719(42) &  <=2.0 &     Rei22 &  -0.33 &     460(40) &   0.7 &   56 &      8.35(64) \\
                J05019+011 &    1RXS J050156.7+010845 &   0.652(15) &  2.0800(90) &     Sha24 &   15.85(36) &     7.0(20) &     Rei22 &  -1.62 &   3320(250) &   1.5 &   19 &      88.5(83) \\
              J05033$-$173 &               LP 776-046 &  0.2685(43) &     &           &     &  <=2.0 &     Rei22 &     &     270(70) &     0 &   74 &      3.97(35) \\
                J05062+046 &          RX J0506.2+0439 &   0.620(15) &  0.8900(50) &     Sha24 &   35.22(85) &    29.0(29) &     Rei22 &  -2.04 &   3070(490) &   8.4 &   12 &        55(16) \\
              J05084$-$210 &  2MASS J05082729-2101444 &     &  0.2800(50) &     Sha24 &     &    25.2(25) &     Rei18 &  -2.65 &     &       &   36 &   920(200) \\
                J05127+196 &                   GJ 192 &  0.4199(32) &    32.7(12) &     Sha24 &   0.650(25) &  <=2.0 &     Rei22 &  -0.21 &  <=190 &     0 &   51 &      1.67(29) \\
                J05280+096 &                  Ross 41 &  0.2321(37) &   17.20(40) &      DA19 &   0.683(19) &  <=2.0 &     Rei22 &  -0.63 &    205 &     0 &   17 &      1.61(47) \\
                J05348+138 &                  Ross 46 &  0.3442(56) &     &           &     &  <=2.0 &     Rei22 &     &     340(60) &   0.3 &   22 &       1.7(11) \\
              J05360$-$076 &                Wolf 1457 &  0.3209(55) &     &           &     &  <=2.0 &     Rei22 &     &    105 &     0 &   37 &      1.06(42) \\
                J05365+113 &                V2689 Ori &  0.5680(44) &   11.75(20) &     Sha24 &   2.446(46) &  <=2.0 &     Rei22 &  -0.56 &    1110(60) &   2.1 &  126 &     10.77(56) \\
                J05366+112 &           PM J05366+1117 &  0.2802(40) &   5.440(50) &     Sha24 &   2.606(44) &     3.4(20) &     Rei22 &  -1.20 &   2520(180) &   1.9 &   15 &       7.9(16) \\
                J05394+406 &           LSR J0539+4038 &   0.115(11) &     &           &     &     5.3(20) &     Rei22 &     &   1830(380) &   9.4 &   18 &        30(18) \\
                J05415+534 &                HD 233153 &  0.5439(42) &   17.39(40) &     Sha24 &   1.582(39) &  <=2.0 &     Rei22 &  -0.40 &     420(60) &   0.5 &   95 &      3.68(30) \\
                J05421+124 &                V1352 Ori &  0.2387(55) &     &           &     &  <=2.0 &     Rei22 &     &  <=140 &     0 &   49 &      1.48(27) \\
                J06000+027 &                 G 99-049 &  0.2433(29) &  1.8100(70) &     Sha24 &   6.801(84) &     6.2(20) &     Rei22 &  -1.70 &   2490(220) &   1.1 &   14 &      18.7(30) \\
                J06011+595 &                G 192-013 &  0.2594(53) &    95.1(84) &     Don23 &   0.138(13) &  <=2.0 &     Rei22 &   0.06 &     210(60) &     0 &   79 &      2.04(21) \\
                J06024+498 &                G 192-015 &  0.1477(32) &     105(10) &      DA19 &  0.0712(70) &  <=2.0 &     Rei22 &  -0.14 &    370(120) &   0.1 &  127 &      3.29(35) \\
                J06103+821 &                   GJ 226 &  0.4094(47) &    44.6(22) &      DA19 &   0.464(23) &  <=2.0 &     Rei22 &  -0.08 &  <=100 &     0 &   57 &      3.17(30) \\
                J06246+234 &                  Ross 64 &  0.1808(27) &     &           &     &  <=2.0 &     Rei22 &     &    620(140) &     0 &   10 &      1.74(67) \\
                J06318+414 &               LP 205-044 &   0.364(87) &   0.300(50) &     Sha24 &      61(18) &    54.0(54) &     Rei22 &  -2.66 &  2200(1150) &     5 &   37 &       179(18) \\
                J06421+035 &                G 108-021 &  0.3732(59) &    83.4(67) &      DA19 &   0.226(18) &  <=2.0 &     Rei22 &   0.10 &     180(60) &     0 &   19 &      1.25(49) \\
                J06574+740 &  2MASS J06572616+7405265 &   0.284(16) &  0.6100(50) &     Sha24 &    23.5(13) &    32.0(32) &     Rei22 &  -2.08 &   3330(830) &  14.2 &   11 &       161(23) \\
                J06594+193 &                  GJ 1093 &  0.1343(33) &     &           &     &  <=2.0 &     Rei22 &     &    390(120) &   0.6 &   27 &      1.99(47) \\
                J07033+346 &               LP 255-011 &  0.2619(42) &    8.00(10) &     Sha24 &   1.656(33) &     2.4(20) &     Rei22 &  -1.03 &   2430(160) &   3.9 &   14 &       8.5(15) \\
                J07044+682 &                   GJ 258 &  0.3960(49) &     &           &     &  <=2.0 &     Rei22 &     &    135 &     0 &   19 &      3.12(60) \\
              J07051$-$101 &  2MASS J07051194-1007528 &  0.1959(63) &     &           &     &     &        &     &     &       &   12 &   520(100) \\
              J07287$-$032 &                  GJ 1097 &  0.3745(46) &     115(12) &     Sha24 &   0.164(17) &  <=2.0 &     Rei22 &   0.24 &    155 &     0 &   25 &      2.42(35) \\
               J07319+362N &                V* BL Lyn &  0.3972(24) &   16.40(36) &      DA19 &   1.225(28) &  <=2.0 &     Rei22 &  -0.67 &   2530(140) &     3 &   52 &      4.79(41) \\
                J07393+021 &               BD+02 1729 &  0.5827(37) &   11.45(19) &     Sha24 &   2.575(46) &  <=2.0 &     Rei22 &  -0.55 &     280(50) &   0.1 &   48 &      2.96(36) \\
              J07403$-$174 &               LP 783-002 &   0.124(12) &     &           &     &  <=2.0 &     Rei22 &     &    280(170) &     0 &   50 &      3.49(57) \\
                J07446+035 &                   YZ CMi &   0.342(86) &   2.780(15) &     Sha24 &     6.2(16) &     4.6(20) &     Rei22 &  -1.53 &   4540(150) &   7.7 &   49 &      85.3(65) \\
                J07472+503 &  2MASS J07471385+5020386 &   0.280(10) &  1.3200(50) &     Sha24 &   10.74(39) &    10.3(20) &     Rei22 &  -1.78 &   2940(290) &   2.1 &   15 &      17.8(31) \\
                J07558+833 &                  GJ 1101 &  0.2725(92) &  1.1100(50) &     Sha24 &   12.42(42) &    11.8(20) &     Rei22 &  -1.89 &   3610(480) &   0.8 &   11 &        66(12) \\
                J07582+413 &                  GJ 1105 &  0.2651(44) &     &           &     &  <=2.0 &     Rei22 &     &  <=120 &     0 &   28 &      1.46(49) \\
                J08119+087 &                 Ross 619 &  0.1621(24) &     &           &     &  <=2.0 &     Rei22 &     &     920(90) &   1.3 &   14 &      0.36(66) \\
              J08126$-$215 &                   GJ 300 &  0.2715(72) &     &           &     &  <=2.0 &     Rei22 &     &  <=200 &     0 &   18 &      3.01(78) \\
                J08161+013 &                  GJ 2066 &  0.4375(46) &    40.7(18) &      DA19 &   0.544(25) &  <=2.0 &     Rei22 &  -0.11 &     250(40) &   0.1 &   72 &      1.81(22) \\
                J08293+039 &  2MASS J08292191+0355092 &  0.4734(46) &     &           &     &  <=2.0 &     Rei22 &     &     410(80) &   0.1 &   14 &      1.92(87) \\
                J08298+267 &                   DX Cnc &   0.124(12) &  0.4600(50) &      DA19 &    13.7(14) &    11.5(20) &     Rei22 &  -2.54 &   2680(440) &   0.9 &   34 &      52.8(54) \\
                J08315+730 &               LP 035-219 &  0.2854(72) &     105(10) &      DA19 &   0.138(14) &  <=2.0 &     Rei22 &   0.16 &    225 &     0 &   19 &      1.71(46) \\
                J08358+680 &                G 234-037 &  0.3625(34) &     &           &     &  <=2.0 &     Rei22 &     &     580(80) &   0.6 &   11 &      1.60(59) \\
                J08526+283 &                rho Cnc B &  0.2694(61) &     &           &     &  <=2.0 &     Rei22 &     &    295 &     0 &   11 &      2.53(87) \\
              J08536$-$034 &               LP 666-009 &   0.102(10) &  0.4600(50) &     Sha24 &    11.2(11) &     8.4(20) &     Rei22 &  -2.55 &   4790(470) &   5.3 &   43 &       135(16) \\
                J08599+729 &               LP 036-098 &  0.2080(67) &     &           &     &     &        &     &     &       &   17 &       3.1(29) \\
                J09003+218 &               LP 368-128 &   0.123(12) &  0.4400(50) &     New16 &    14.1(14) &    13.0(20) &     Rei22 &  -2.87 &   4270(370) &   2.2 &   23 &        43(17) \\
                J09028+680 &               LP 060-179 &  0.2752(70) &     &           &     &  <=2.0 &     Rei22 &     &    225 &     0 &   10 &      2.88(62) \\
                J09029+716 &          LSPM J0902+7138 &   0.546(16) &     &           &     &     &        &     &     &       &   14 &       6.3(19) \\
                J09133+688 &                G 234-57A &   0.588(61) &   10.40(16) &      SM18 &    2.86(30) &  <=2.0 &     Rei18 &  -0.69 &     &       &   16 &       5.8(11) \\
                J09143+526 &                 HD 79210 &  0.5711(58) &   17.54(41) &     Sha24 &   1.647(42) &  <=2.0 &     Rei22 &  -0.35 &     400(50) &   0.4 &   69 &      6.63(63) \\
                J09307+003 &                  GJ 1125 &  0.3000(48) &     &           &     &  <=2.0 &     Rei22 &     &     180(50) &     0 &   21 &      3.37(43) \\
                J09411+132 &                  Ross 85 &  0.4494(39) &     &           &     &  <=2.0 &     Rei22 &     &     390(70) &   0.1 &   47 &      2.65(37) \\
                J09423+559 &                   GJ 363 &  0.3733(63) &    74.3(54) &      DA19 &   0.254(19) &  <=2.0 &     Rei22 &  -0.01 &     320(80) &     0 &   10 &      3.37(88) \\
                J09425+700 &                   GJ 360 &  0.4945(67) &   21.00(57) &      DA19 &   1.191(36) &     2.1(20) &     Rei22 &  -0.44 &   1030(110) &   4.3 &   48 &      5.37(47) \\
                J09428+700 &                   GJ 362 &  0.4204(74) &   24.33(74) &     Sha24 &   0.874(31) &  <=2.0 &     Rei22 &  -0.45 &   1390(140) &   5.8 &   50 &      7.05(49) \\
              J09447$-$182 &                  GJ 1129 &  0.2753(51) &     &           &     &  <=2.0 &     Rei22 &     &    155 &     0 &   11 &   \text{$(0.92 \pm 0.15)\,i$\tablefootmark{b}} \\
              J09449$-$123 &                G 161-071 &   0.318(82) &  0.4400(50) &     Sha24 &    36.6(94) &    35.3(35) &     Rei22 &  -2.53 &   4860(990) &   0.1 &   10 &       443(80) \\
                J09468+760 &               BD+76 3952 &  0.5271(51) &    50.6(27) &     Sha24 &   0.527(29) &  <=2.0 &     Rei22 &   0.08 &    205 &     0 &   26 &      1.83(45) \\
              J09511$-$123 &               BD-11 2741 &  0.5044(37) &     &           &     &  <=2.0 &     Rei22 &     &     150(50) &     0 &   24 &      1.81(38) \\
                J09535+355 &                 Wolf 327 &   0.441(13) &     &           &     &     &        &     &     &       &   22 &      3.28(51) \\
                J09597+472 &                G 146-005 &  0.2988(94) &     &           &     &     &        &     &     &       &   11 &       3.6(19) \\
                J10087+355 &                 Wolf 346 &  0.3069(91) &     &           &     &     &        &     &     &       &   16 &      1.84(92) \\
              J10122$-$037 &                   AN Sex &  0.4941(54) &   23.00(67) &     Sha24 &   1.087(34) &  <=2.0 &     Rei22 &  -0.42 &     450(80) &   0.3 &   75 &      5.12(42) \\
              J10167$-$119 &                   GJ 386 &  0.4738(55) &     &           &     &  <=2.0 &     Rei22 &     &  <=100 &     0 &   19 &      1.96(44) \\
                J10238+438 &               LP 212-062 &  0.2504(88) &     &           &     &     &        &     &     &       &   12 &       5.5(27) \\
              J10251$-$102 &               BD-09 3070 &  0.5072(41) &   25.05(78) &     Sha24 &   1.024(33) &  <=2.0 &     Rei22 &  -0.29 &     380(50) &   0.3 &   28 &      2.77(62) \\
              J10350$-$094 &               LP 670-017 &  0.3792(45) &     &           &     &  <=2.0 &     Rei22 &     &  <=260 &     0 &   16 &      1.48(71) \\
                J10416+376 &                  GJ 1134 &  0.2217(37) &     &           &     &  <=2.0 &     Rei22 &     &  <=410 &     0 &   11 &      2.25(74) \\
              J10482$-$113 &               LP 731-058 &   0.119(12) &   1.500(50) &     Mor10 &    4.00(42) &     2.1(15) &     Rei18 &  -2.43 &     &       &   73 &      3.48(65) \\
                J10508+068 &                   EE Leo &  0.2551(63) &    64.0(42) &      DA19 &   0.202(14) &  <=2.0 &     Rei22 &  -0.17 &  <=200 &     0 &   52 &      1.42(31) \\
              J10584$-$107 &               LP 731-076 &  0.2056(30) &     &           &     &     2.8(20) &     Rei22 &     &   4100(170) &   4.3 &   48 &      8.78(79) \\
                J11000+228 &                 Ross 104 &   0.373(89) &    53.2(30) &     Sha24 &   0.355(87) &  <=2.0 &     Rei22 &  -0.06 &     360(60) &   0.2 &   60 &      0.89(32) \\
                J11026+219 &                   DS Leo &   0.535(23) &   14.26(28) &     Sha24 &   1.899(91) &     2.3(20) &     Rei22 &  -0.46 &    1040(60) &   3.4 &   52 &      11.4(12) \\
                J11054+435 &              BD+44 2051A &  0.3843(46) &     &           &     &  <=2.0 &     Rei22 &     &  <=100 &     0 &  112 &      2.90(28) \\
                J11108+479 &                G 176-015 &  0.2778(87) &     &           &     &     &        &     &     &       &   12 &       1.8(18) \\
               J11110+304W &               HD  97101B &  0.5406(59) &     &           &     &  <=2.0 &     Rei22 &     &     280(70) &     0 &   51 &      2.45(28) \\
              J11201$-$104 &               LP 733-099 &  0.5072(10) &   5.640(54) &     Sha24 &   4.550(44) &     4.6(20) &     Rei22 &  -0.96 &   2170(160) &   1.7 &   24 &      14.7(25) \\
              J11306$-$080 &               LP 672-042 &  0.3459(56) &     &           &     &  <=2.0 &     Rei22 &     &     380(90) &   0.1 &   15 &      1.35(44) \\
              J11467$-$140 &                   GJ 443 &  0.5002(59) &     &           &     &  <=2.0 &     Rei22 &     &     270(80) &     0 &   14 &      1.58(60) \\
                J11511+352 &               BD+36 2219 &  0.4454(52) &   22.80(66) &      DA19 &   0.988(31) &  <=2.0 &     Rei22 &  -0.33 &     660(60) &   0.5 &  109 &      1.90(21) \\
              J12100$-$150 &               LP 734-032 &  0.3612(75) &    79.3(61) &     Sha24 &   0.230(18) &  <=2.0 &     Rei22 &  -0.05 &  <=160 &     0 &   57 &      1.38(23) \\
              J12111$-$199 &                 LTT 4562 &  0.3491(37) &     &           &     &  <=2.0 &     Rei22 &     &  <=210 &     0 &   18 &      1.47(73) \\
                J12189+111 &                   GL Vir &  0.1619(34) &  0.4900(50) &      DA19 &   16.71(39) &    13.0(20) &     Rei22 &  -2.53 &   3970(460) &   0.9 &   12 &      40.0(97) \\
              J12248$-$182 &                 Ross 695 &  0.2798(49) &     &           &     &  <=2.0 &     Rei22 &     &     290(70) &     0 &   14 & \text{$(0.42 \pm 0.72)\,i$\tablefootmark{b}}  \\
                J12312+086 &               BD+09 2636 &  0.5493(37) &     &           &     &  <=2.0 &     Rei22 &     &     360(40) &   0.3 &   47 &      3.95(44) \\
              J12373$-$208 &               LP 795-038 &  0.4003(85) &     &           &     &  <=2.0 &     Rei22 &     &    290(150) &     0 &   18 &      1.81(74) \\
                J13005+056 &                   FN Vir &  0.2086(51) &   0.600(50) &     Sha24 &    17.6(15) &    16.4(20) &     Rei22 &  -2.30 &   3830(620) &   1.5 &   11 &      30.0(97) \\
                J13102+477 &                G 177-025 &  0.1719(23) &    29.1(10) &     Sha24 &   0.299(11) &  <=2.0 &     Rei22 &  -0.65 &   2290(170) &   3.5 &   36 &      4.22(69) \\
             J13283$-$023W &                Ross 486A &  0.4200(68) &    46.4(23) &     Rae20 &   0.458(24) &  <=2.0 &     Rei22 &  -0.14 &  <=130 &     0 &   10 &       3.2(11) \\
                J13427+332 &                Ross 1015 &  0.2689(44) &     &           &     &  <=2.0 &     Rei22 &     &     240(80) &     0 &   18 &      2.19(53) \\
                J13450+176 &               BD+18 2776 &  0.4888(27) &     &           &     &  <=2.0 &     Rei22 &     &  <=100 &     0 &   31 &      1.48(41) \\
                J13457+148 &                HD 119850 &  0.4859(77) &    52.3(29) &      SM15 &   0.470(27) &  <=2.0 &     Rei22 &   0.05 &     180(50) &     0 &  241 &      2.64(20) \\
              J13458$-$179 &               LP 798-034 &  0.3345(57) &     &           &     &  <=2.0 &     Rei18 &     &     &       &   10 &      2.58(70) \\
                J13536+776 &          RX J1353.6+7737 &   0.303(16) &  1.2300(50) &     Sha24 &   12.45(65) &     9.7(20) &     Rei22 &  -1.82 &   2630(400) &   0.6 &   24 &        95(34) \\
                J13582+125 &                 Ross 837 &  0.2221(36) &     &           &     &  <=2.0 &     Rei18 &     &     &       &   10 &       2.8(10) \\
                J14082+805 &                BD+81 465 &  0.5881(45) &     &           &     &  <=2.0 &     Rei22 &     &     400(60) &   0.4 &   37 &      3.52(39) \\
                J14173+454 &          RX J1417.3+4525 &  0.2809(43) &  0.3700(50) &     Sha24 &   38.42(79) &    14.9(20) &     Rei22 &  -2.46 &   5850(500) &   1.3 &   12 &       137(26) \\
                J14251+518 &                tet Boo B &  0.3958(50) &     &           &     &  <=2.0 &     Rei22 &     &  <=120 &     0 &   12 &      2.42(51) \\
               J14257+236E &              BD+24 2733B &  0.5891(39) &     &           &     &  <=2.0 &     Rei22 &     &     270(50) &     0 &   49 &      1.52(34) \\
               J14257+236W &              BD+24 2733A &  0.6055(44) &     111(11) &      DA19 &   0.276(28) &  <=2.0 &     Rei22 &   0.47 &     230(50) &     0 &   63 &      2.75(37) \\
              J14307$-$086 &               BD-07 3856 &  0.6334(44) &     &           &     &  <=2.0 &     Rei22 &     &     210(40) &     0 &   92 &      2.19(21) \\
                J14321+081 &               LP 560-035 &  0.1417(77) &  0.7600(50) &     Pas23 &    9.44(52) &     6.8(20) &     Rei22 &  -2.39 &   2840(370) &   3.9 &   55 &      40.9(37) \\
                J14524+123 &                  G 66-37 &  0.4980(66) &   26.03(83) &     Sha24 &   0.968(33) &  <=2.0 &     Rei22 &  -0.34 &    460(110) &   0.4 &   26 &      4.47(60) \\
                J14544+355 &                Ross 1041 &  0.3535(60) &     &           &     &  <=2.0 &     Rei22 &     &    255 &     0 &   33 &      2.60(39) \\
                J15013+055 &                   G 15-2 &  0.3386(43) &     &           &     &  <=2.0 &     Rei22 &     &    440(120) &   0.3 &   19 &      3.09(44) \\
                J15095+031 &                Ross 1047 &  0.4266(42) &     &           &     &  <=2.0 &     Rei22 &     &  <=100 &     0 &   16 &      2.13(41) \\
                J15100+193 &                G 136-072 &   0.320(10) &     &           &     &    3.0 &    Jeff18 &     &     &       &   11 &      4.37(95) \\
                J15218+209 &                   OT Ser &    0.52(11) &   3.350(21) &     Sha24 &     7.8(17) &     4.1(20) &     Rei22 &  -1.18 &   3360(100) &     6 &   51 &      33.1(31) \\
                J15305+094 &               NLTT 40406 &  0.1274(13) &   0.300(50) &      DA19 &    21.5(36) &    13.4(20) &     Rei22 &  -2.77 &   4780(600) &   1.6 &   14 &      36.0(72) \\
                J15499+796 &               LP 022-420 &  0.2988(35) &  0.1900(50) &     Sha24 &    79.6(23) &    31.0(31) &     Rei22 &  -2.84 &  3240(1040) &   0.7 &   14 &       166(24) \\
              J15598$-$082 &               BD-07 4156 &  0.4745(39) &   20.00(52) &      SM18 &   1.200(33) &  <=2.0 &     Rei22 &  -0.39 &    520(100) &     0 &   24 &      3.35(45) \\
                J16028+205 &                   GJ 609 &  0.2487(51) &     &           &     &  <=2.0 &     Rei22 &     &     210(70) &     0 &   52 &      1.81(19) \\
               J16167+672N &                   EW Dra &  0.4507(84) &    40.4(18) &     Don23 &   0.564(28) &  <=2.0 &     Rei22 &  -0.17 &    245 &     0 &  100 &      2.00(23) \\
                J16313+408 &                G 180-060 &  0.1727(46) &  0.5100(50) &     Sha24 &   17.13(49) &     7.4(20) &     Rei22 &  -2.49 &   4170(380) &   1.7 &   11 &        98(22) \\
                J16327+126 &                  GJ 1203 &  0.3991(49) &     &           &     &  <=2.0 &     Rei22 &     &    235 &     0 &   13 &      2.03(68) \\
                J16462+164 &               LP 446-006 &  0.4155(50) &     &           &     &  <=2.0 &     Rei22 &     &    195 &     0 &   16 &      1.60(39) \\
             J16554$-$083N &                 Wolf 629 &  0.2172(43) &     &           &     &  <=2.0 &     Rei22 &     &     260(70) &     0 &   30 &      1.56(56) \\
              J16555$-$083 &                     VB 8 &   0.115(12) &  1.0900(50) &     Laf21 &    5.36(54) &     4.0(20) &     Rei22 &  -2.19 &   3020(200) &   4.6 &  120 &      19.3(13) \\
              J16570$-$043 &               LP 686-027 &  0.2631(26) &  1.2100(50) &     Kir12 &   11.00(12) &     9.7(20) &     Rei22 &  -1.85 &   3660(380) &   3.1 &   13 &      13.7(29) \\
              J17052$-$050 &                 Wolf 636 &  0.4758(39) &    50.2(27) &      DA19 &   0.479(26) &  <=2.0 &     Rei22 &   0.01 &    155 &     0 &   48 &      2.08(29) \\
                J17115+384 &                 Wolf 654 &  0.3736(76) &    62.6(40) &      DA19 &   0.302(20) &  <=2.0 &     Rei22 &  -0.05 &  <=100 &     0 &   72 &      1.33(24) \\
                J17166+080 &                  GJ 2128 &  0.3978(43) &     &           &     &  <=2.0 &     Rei18 &     &     &       &   18 &      1.07(62) \\
                J17198+417 &                   GJ 671 &  0.3568(42) &    71.5(51) &      DA19 &   0.252(18) &  <=2.0 &     Rei22 &   0.10 &  <=100 &     0 &   23 &      1.54(37) \\
                J17303+055 &               BD+05 3409 &  0.5195(30) &    34.6(14) &     Sha24 &   0.760(31) &  <=2.0 &     Rei22 &  -0.10 &     270(60) &   0.1 &   54 &      2.50(34) \\
                J17338+169 &    1RXS J173353.5+165515 &   0.325(86) &  0.2700(50) &     Sha24 &      61(16) &    38.0(38) &     Rei22 &  -2.81 &   8130(940) &     2 &   11 &   530(110) \\
                J17481+159 &  2MASS J17481125+1558465 &   0.368(11) &     &           &     &     &        &     &     &       &   30 &      3.07(81) \\
                J17542+073 &                  GJ 1222 &  0.3060(57) &     &           &     &  <=2.0 &     Rei22 &     &    280(150) &     0 &   12 &       3.6(13) \\
                J17578+465 &                G 204-039 &  0.3929(53) &    30.3(11) &      DA19 &   0.656(25) &  <=2.0 &     Rei22 &  -0.34 &    280(150) &     0 &   26 &      3.06(40) \\
                J18012+355 &                G 182-034 &   0.400(12) &     &           &     &     &        &     &     &       &   39 &      2.88(57) \\
                J18022+642 &               LP 071-082 &  0.1780(49) &  0.2800(50) &     Sha24 &    32.2(11) &    12.9(20) &     Rei22 &  -2.62 &   4930(310) &   2.6 &   25 &      41.1(49) \\
                J18027+375 &                  GJ 1223 &  0.1601(39) &     124(14) &     New16 &  0.0654(73) &  <=2.0 &     Rei22 &  -0.07 &  <=270 &     0 &  114 &      1.95(31) \\
              J18051$-$030 &                HD 165222 &  0.4410(34) &     &           &     &  <=2.0 &     Rei22 &     &     270(50) &   0.1 &   54 &      1.62(27) \\
              J18075$-$159 &                  GJ 1224 &  0.1733(24) &   3.870(27) &      DA19 &   2.265(35) &  <=2.0 &     Rei22 &  -1.48 &   3190(160) &   4.8 &   16 &       3.8(12) \\
                J18131+260 &                LP 390-16 &   0.338(13) &   2.280(11) &     Sha24 &    7.50(29) &     6.3(20) &     Rei22 &  -1.58 &   4490(270) &   5.3 &   16 &       104(11) \\
                J18165+048 &                 G 140-51 &  0.1925(43) &     &           &     &  <=2.0 &     Rei22 &     &  <=130 &     0 &   51 &      4.08(55) \\
                J18174+483 &          TYC 3529-1437-1 &    0.52(11) &   15.83(34) &     Sha24 &    1.65(36) &     2.2(20) &     Rei22 &  -0.55 &   1910(100) &   6.6 &   69 &     10.82(85) \\
               J18180+387E &                 G 204-58 &  0.2922(34) &     &           &     &  <=2.0 &     Rei22 &     &  <=270 &     0 &   20 &      1.55(69) \\
                J18189+661 &                LP 71-165 &  0.1608(46) &  0.5200(50) &     Sha24 &   15.65(47) &    14.5(20) &     Rei22 &  -2.29 &   3960(740) &   2.9 &   12 &      30.1(66) \\
                J18221+063 &                 Ross 136 &  0.2932(49) &     &           &     &  <=2.0 &     Rei22 &     &    215 &     0 &   16 &  \text{$(0.64 \pm 0.68)\,i$\tablefootmark{b}} \\
                J18224+620 &                  GJ 1227 &  0.1772(46) &     &           &     &  <=2.0 &     Rei22 &     &    500(220) &   0.7 &   60 &      2.80(36) \\
                J18319+406 &                G 205-028 &  0.3467(50) &    50.2(27) &      DA19 &   0.349(19) &  <=2.0 &     Rei22 &  -0.15 &  <=190 &     0 &   23 &      2.36(34) \\
                J18356+329 &           LSR J1835+3259 &   0.101(10) &  0.1200(50) &     Sha24 &    42.6(46) &    49.2(49) &     Rei18 &  -3.16 &     &       &   54 &       191(20) \\
                J18419+318 &                 Ross 145 &  0.3434(40) &     &           &     &  <=2.0 &     Rei22 &     &    105 &     0 &   28 &      2.71(45) \\
              J18480$-$145 &                G 155-042 &  0.3663(61) &     &           &     &  <=2.0 &     Rei22 &     &    155 &     0 &   18 &      2.98(49) \\
                J18482+076 &                G 141-036 &  0.1552(38) &   2.760(15) &      DA19 &   2.845(72) &     3.6(20) &     Rei22 &  -1.78 &   1190(230) &     2 &   52 &      6.51(73) \\
                J19070+208 &                 Ross 730 &  0.3198(39) &     &           &     &  <=2.0 &     Rei22 &     &  <=110 &     0 &   43 &      0.59(50) \\
                J19072+208 &                HD 349726 &  0.3131(35) &     &           &     &  <=2.0 &     Rei22 &     &  <=110 &     0 &   51 &      1.65(25) \\
                J19084+322 &                G 207-019 &  0.3382(49) &    74.1(54) &      DA19 &   0.231(17) &  <=2.0 &     Rei22 &   0.00 &    205 &     0 &   30 &      0.99(36) \\
                J19098+176 &                  GJ 1232 &  0.2007(42) &    80.1(62) &      DA19 &   0.127(10) &  <=2.0 &     Rei22 &  -0.16 &    165 &     0 &   31 &      2.68(44) \\
               J19169+051S &                V1298 Aql &   0.109(11) &     &           &     &     3.7(20) &     Rei22 &     &   1210(180) &  22.3 &   50 &       9.5(14) \\
                J19216+208 &                  GJ 1235 &  0.2085(45) &     133(15) &      DA19 &  0.0793(93) &  <=2.0 &     Rei22 &   0.08 &    340(100) &   0.3 &   25 &      2.14(52) \\
                J19242+755 &                  GJ 1238 &  0.1342(69) &     &           &     &     &        &     &     &       &  203 &      2.17(30) \\
                J19251+283 &                 Ross 164 &   0.354(94) &     &           &     &  <=2.0 &     Rei22 &     &     250(80) &     0 &   28 &       4.1(15) \\
                J19255+096 &          LSPM J1925+0938 &  0.1749(71) &     &           &     &    34.7(35) &     Rei18 &     &     &       &   76 &       103(23) \\
                J19346+045 &               BD+04 4157 &  0.5501(66) &   21.79(60) &     Sha24 &   1.277(39) &  <=2.0 &     Rei22 &  -0.14 &     220(40) &     0 &   50 &      4.29(42) \\
              J19422$-$207 &  2MASS J19421282-2045477 &  0.2208(74) &  1.3400(50) &     Sha24 &    8.34(28) &     6.7(20) &     Rei22 &  -1.89 &   3840(350) &   1.5 &   24 &        90(12) \\
                J19511+464 &                G 208-042 &  0.2773(69) &  0.5900(50) &     Sha24 &   23.78(62) &    24.6(25) &     Rei22 &  -2.16 &   4210(620) &   7.5 &   13 &        38(12) \\
              J19573$-$125 &              HD 188807 B &  0.2179(73) &     &           &     &     &        &     &     &       &   11 &  1150(250) \\
              J20093$-$012 &  2MASS J20091824-0113377 &  0.1596(87) &  1.3700(50) &     Sha24 &    5.89(32) &     5.5(20) &     Rei22 &  -1.99 &   3670(190) &   2.8 &   12 &      33.8(55) \\
                J20305+654 &                   GJ 793 &  0.3752(51) &    32.8(13) &      DA19 &   0.579(24) &     2.7(20) &     Rei22 &  -0.27 &    220(110) &     0 &   51 &      1.90(49) \\
                J20336+617 &                  GJ 1254 &  0.3791(73) &   12.60(23) &      DA19 &   1.522(40) &  <=2.0 &     Rei22 &  -0.71 &    260(140) &     0 &   52 &      1.58(26) \\
                J20405+154 &                  GJ 1256 &  0.2007(44) &     106(10) &      DA19 &  0.0958(95) &  <=2.0 &     Rei22 &  -0.01 &    410(110) &   0.6 &   27 &      2.43(50) \\
              J20525$-$169 &               LP 816-060 &  0.2318(55) &     &           &     &  <=2.0 &     Rei22 &     &     380(80) &   0.4 &   53 &      1.73(27) \\
                J20533+621 &                HD 199305 &  0.5232(47) &     &           &     &  <=2.0 &     Rei22 &     &     350(80) &     0 &  153 &      1.59(16) \\
             J20556$-$140S &                 GJ 810 B &  0.1632(24) &     135(16) &     New18 &  0.0613(72) &  <=2.0 &     Rei22 &   0.08 &    370(160) &     0 &   53 &      3.13(48) \\
              J20567$-$104 &                 Wolf 896 &  0.4596(71) &     &           &     &  <=2.0 &     Rei22 &     &     290(50) &   0.2 &   20 &      3.38(58) \\
              J21019$-$063 &                 Wolf 906 &  0.4513(52) &    49.3(26) &     Sha24 &   0.463(25) &  <=2.0 &     Rei22 &  -0.09 &     240(60) &   0.1 &   67 &      3.26(32) \\
                J21152+257 &               LP 397-041 &  0.5445(78) &    34.8(14) &      DA19 &   0.792(34) &  <=2.0 &     Rei22 &  -0.24 &    220(110) &     0 &   25 &       4.6(11) \\
                J21348+515 &                 Wolf 926 &  0.4412(59) &    54.3(31) &      DA19 &   0.411(24) &  <=2.0 &     Rei22 &  -0.02 &    225 &     0 &   69 &      2.44(23) \\
                J21463+382 &          LSPM J2146+3813 &  0.2144(28) &    93.9(83) &     Don23 &   0.116(10) &  <=2.0 &     Rei22 &   0.08 &     310(80) &     0 &   52 &      1.70(30) \\
              J21466$-$001 &                 Wolf 940 &  0.2806(48) &     &           &     &  <=2.0 &     Rei22 &     &    235 &     0 &   24 &      3.24(42) \\
                J22012+283 &                 V374 Peg &  0.3242(60) &  0.4500(50) &      DA19 &   36.45(78) &    36.4(36) &     Rei22 &  -2.30 &   3630(880) &     0 &   30 &        97(14) \\
              J22020$-$194 &               LP 819-017 &  0.3314(46) &     &           &     &  <=2.0 &     Rei22 &     &    165 &     0 &   12 &      0.70(96) \\
                J22021+014 &               BD+00 4810 &  0.5601(51) &   22.06(62) &     Sha24 &   1.285(38) &  <=2.0 &     Rei22 &  -0.32 &     400(50) &   0.3 &   75 &      3.24(34) \\
                J22057+656 &               G 264-18 A &  0.4675(46) &     120(13) &     Laf21 &   0.196(21) &  <=2.0 &     Rei22 &   0.44 &     290(80) &     0 &   89 &      1.86(21) \\
                J22060+393 &                G 189-001 &   0.396(12) &     &           &     &     &        &     &     &       &   29 &      2.55(42) \\
                J22114+409 &    1RXS J221124.3+410000 &  0.1754(34) &    30.0(11) &      DA19 &   0.296(12) &  <=2.0 &     Rei22 &  -0.75 &   1880(210) &   3.9 &   52 &      3.87(58) \\
                J22115+184 &                 Ross 271 &  0.5031(58) &    36.3(15) &      DA19 &   0.701(30) &  <=2.0 &     Rei22 &  -0.18 &     260(60) &     0 &   67 &      5.19(38) \\
              J22231$-$176 &               LP 820-012 &  0.1903(23) &   4.570(37) &     Sha24 &   2.107(31) &     2.3(20) &     Rei22 &  -1.48 &   2150(240) &   2.8 &   15 &      14.2(31) \\
                J22330+093 &               BD+08 4887 &  0.4490(43) &    37.8(16) &     Sha24 &   0.601(26) &  <=2.0 &     Rei22 &  -0.06 &     320(60) &     0 &   81 &      3.09(28) \\
                J22468+443 &                   EV Lac &  0.3408(63) &   4.350(34) &     Sha24 &   3.964(79) &     4.1(20) &     Rei22 &  -1.24 &   4320(110) &     9 &  105 &      47.6(25) \\
              J22503$-$070 &               BD-07 5871 &  0.5395(30) &     &           &     &  <=2.0 &     Rei22 &     &     440(50) &   0.2 &   53 &      1.95(45) \\
                J22518+317 &                   GT Peg &   0.537(34) &  1.6300(58) &      DA19 &    16.7(11) &    12.7(20) &     Rei22 &  -1.61 &   3870(340) &   1.5 &   12 &        69(21) \\
                J22526+750 &               NLTT 55174 &  0.2871(93) &     &           &     &  <=2.0 &     Fou18 &     &     &       &   16 &      2.08(62) \\
                J22559+178 &              StKM 1-2065 &  0.5412(36) &   27.00(89) &      SM18 &   1.014(34) &  <=2.0 &     Rei22 &  -0.24 &     580(90) &   0.4 &   11 &    \text{$(0.93 \pm 0.37)\,i$\tablefootmark{b}}  \\
                J23216+172 &               LP 462-027 &   0.376(11) &    74.7(55) &      DA19 &   0.254(20) &  <=2.0 &     Rei22 &  -0.07 &     250(80) &   0.2 &   64 &      2.28(26) \\
                J23245+578 &               BD+57 2735 &  0.5351(40) &    36.5(15) &     Sha24 &   0.742(31) &  <=2.0 &     Rei22 &  -0.14 &     430(60) &   0.3 &   60 &      4.06(31) \\
                J23340+001 &                Wolf 1039 &  0.4339(53) &     &           &     &  <=2.0 &     Rei22 &     &  <=190 &     0 &   42 &      2.20(34) \\
              J23351$-$023 &                  GJ 1286 &  0.1304(32) &     178(26) &     Don23 &  0.0371(55) &  <=2.0 &     Rei22 &   0.09 &    910(160) &     4 &   69 &     \text{$(0.67 \pm 0.59)\,i$\tablefootmark{b}}  \\
              J23381$-$162 &                G 273-093 &  0.3646(46) &     &           &     &  <=2.0 &     Rei22 &     &  <=160 &     0 &   54 &      1.56(23) \\
                J23419+441 &                   HH And &  0.1591(41) &     106(10) &      DA19 &  0.0759(76) &  <=2.0 &     Rei22 &  -0.10 &    410(100) &   0.7 &   98 &      2.14(22) \\
                J23431+365 &                  GJ 1289 &  0.2301(42) &    73.7(53) &     Don23 &   0.158(12) &  <=2.0 &     Rei22 &  -0.15 &   1360(190) &   2.9 &   40 &      2.20(36) \\
                J23492+024 &                   BR Psc &  0.4172(56) &    49.9(27) &      SM18 &   0.423(23) &  <=2.0 &     Rei22 &  -0.04 &    145 &     0 &  469 &     1.634(98) \\
              J23505$-$095 &               LP 763-012 &  0.2965(61) &     &           &     &  <=2.0 &     Rei22 &     &    255 &     0 &   71 &      2.83(45) \\
                J23548+385 &          RX J2354.8+3831 &  0.3195(32) &   4.800(50) &     Sha24 &   3.367(49) &     3.6(20) &     Rei22 &  -1.23 &   4900(130) &   9.3 &   13 &      33.8(64) \\
            
        \end{longtable}

        \tablefoot{
            \tablefoottext{a}{Listed are: CARMENES star identifier, additional identifier, stellar radius, stellar rotation period, equatorial rotation velocity, projected equatorial rotation velocity, average magnetic field strength, magnetic field component distribution index, number of RV measurements, and RV jitter.}
            \tablefoottext{b}{The letter $i$ stands for an imaginary jitter value, i.e. a negative jitter variance $\sigma_{\mathrm{jitter}}^2$. A negative jitter variance results from overestimated internal RV uncertainties.}
        }

        \tablebib{
            DA19: \citet{diezalonso2019}, 
            Don23: \citet{2023MNRAS.525.2015D}, 
            Jeff18: \citet{2018A&A...614A..76J}, 
            Jen09: \citet{2009ApJ...704..975J}, 
            Kir12: \citet{2012AcA....62...67K}, 
            Laf21: \citet{2021A&A...652A..28L}, 
            Mor10: \citet{2010MNRAS.407.2269M}, 
            New16: \citet{2016ApJ...821...93N}, 
            New18: \citet{2018AJ....156..217N}, 
            Pas23: \citet{2023AJ....166...16P}, 
            Rae20: \citet{2020A&A...637A..22R}, 
            Rei18: \citet{reiners2018}, 
            Rei22: \citet{reiners2022},
            Sha24: \citet{2024A&A...684A...9S}, 
            SM15: \citet{2015MNRAS.452.2745S}, 
            SM18: \citet{2018A&A...612A..89S}.
        }
    
    \end{landscape}
    }

    \clearpage

    \onecolumn

    \begin{table*}
        \sisetup{detect-weight,mode=text}
        
        \centering
        \caption{Fitted relations between RV jitter, stellar rotation velocity and average magnetic field.\tablefootmark{a}}
        \label{tab:fit_results}
        \begin{tabularx}{\linewidth}{@{}X S[round-mode = figures,round-precision = 1] S[round-mode = figures,round-precision = 3] S[round-mode = figures,round-precision = 3]  S[round-mode = figures,round-precision = 3]@{}} 
            
            \toprule
            \toprule
            {${\sigma}_{\mathrm{jitter}}$} &  {Number of parameters}  & {$\ln \mathcal{L} $} &     {AIC} &        {BIC} \\
            \midrule
                                                                                                     $\alpha$                          &        1 & -90.6500 &  185.3000 &  190.277273 \\
                                                                                                     ${\alpha\cdot {v_{\mathrm{eq}}}}$ &        1 & -33.9600 &   71.9200 &   76.897273 \\
                                                                                                  ${\alpha\cdot {\langle B \rangle }}$ &        1 & -33.9300 &   71.8600 &   76.837273 \\
                                                                           ${\alpha\cdot {v_{\mathrm{eq}}}\cdot {\langle B \rangle }}$ &        1 & -52.9300 &  109.8600 &  114.837273 \\
                                                                           
                                                                                               ${\alpha\cdot {v_{\mathrm{eq}}}^\beta}$ &        2 & -19.5700 &   45.1400 &   52.605909 \\
                                                                     ${\alpha\cdot {{v_{\mathrm{eq}}}^\beta\cdot \langle B \rangle }}$ &        2 & -18.7800 &   43.5600 &   51.025909 \\
                                                                     ${\alpha\cdot {{v_{\mathrm{eq}}}\cdot \langle B \rangle }^\beta}$ &        2 &  -6.2770 &   18.5540 &   26.019909 \\
                                                                                ${\sqrt{\alpha^2+{(\beta\cdot {v_{\mathrm{eq}}})}^2}}$ &        2 &  -3.4670 &   12.9340 &   20.399909 \\
                                                                             ${\sqrt{\alpha^2+{(\beta\cdot {\langle B \rangle })}^2}}$ &        2 & -33.9200 &   73.8400 &   81.305909 \\
                                                      ${\sqrt{\alpha^2+{(\beta\cdot {v_{\mathrm{eq}}}\cdot {\langle B \rangle })}^2}}$ &        2 &  -4.0860 &   14.1720 &   21.637909 \\
                                                      
                                                                         ${\sqrt{\alpha^2+{(\beta\cdot {v_{\mathrm{eq}}}^\gamma)}^2}}$ &        3 &  -3.4210 &   14.8420 &   24.796545 \\
                                                                      ${\sqrt{\alpha^2+{(\beta\cdot {\langle B \rangle }^\gamma)}^2}}$ &        3 &    -23.9 &      55.7 &        65.7 \\
                             $\boldsymbol{{\sqrt{\alpha^2+{(\beta\cdot {({v_{\mathrm{eq}}}\cdot {\langle B \rangle })}^\gamma)}^2}}}$  &     \B 3 & \B 7.4030 &  \B -6.8060  & \B  3.148545 \\
                                                 ${\sqrt{\alpha^2+{(\beta\cdot v_{\mathrm{eq}}\cdot {\langle B \rangle }^\gamma)}^2}}$ &        3 &   1.7690 &    4.4620 &   14.416545 \\
                                                 ${\sqrt{\alpha^2+{(\beta\cdot {v_{\mathrm{eq}}}^\gamma\cdot \langle B \rangle )}^2}}$ &        3 &   4.8120 &   -1.6240 &    8.330545 \\
                                                    ${\sqrt{\alpha^2+{({v_{\mathrm{eq}}}^\beta\cdot {\langle B \rangle }^\gamma)}^2}}$ &        3 &  -2.5110 &   13.0220 &   22.976545 \\
                                            $ {\sqrt{\alpha^2+{(\beta\cdot v_{\mathrm{eq}})}^2+{(\gamma\cdot \langle B \rangle )}^2}}$ &        3 &  -0.6316 &    9.2632 &   19.217745 \\
                                                                 $ {\alpha+\beta\cdot v_{\mathrm{eq}}+\gamma\cdot \langle B \rangle }$ &        3 &  -0.9671 &    9.9342 &   19.888745 \\
     ${{\alpha+\beta\cdot {({v_{\mathrm{eq}}}\cdot \langle B \rangle )}+\gamma\cdot {({v_{\mathrm{eq}}}\cdot \langle B \rangle )}^2}}$ &        3 &   1.8760 &    4.2480 &   14.202545 \\            
     
                                        ${\sqrt{\alpha^2+{(\beta\cdot {v_{\mathrm{eq}}}^\gamma\cdot {\langle B \rangle }^\delta)}^2}}$ &        4 &   7.5950 &   -5.1900 &    7.253182 \\
            \bottomrule
    
        \end{tabularx}
        \tablefoot{
            \tablefoottext{a}{The bold type indicates the model that has been adopted.}
        }    

    \end{table*}
      
\end{appendix}

\end{document}